\documentclass[prd,aps,superscriptaddress,tightenlines,nofootinbib,floatfix,preprintnumbers,11pt,longbibliography]{revtex4-1}
\usepackage{epsfig}
\usepackage{amsfonts}
\usepackage{amsmath}
\usepackage{amssymb}
\usepackage{dsfont}
\usepackage{hyperref}
\usepackage{xspace}
\usepackage{slashed}
\hypersetup{
    colorlinks=true,       
    linkcolor=blue,          
    citecolor=blue,        
    filecolor=blue,      
    urlcolor=blue           
}
\usepackage{pifont}
\usepackage{color}
\usepackage[dvipsnames]{xcolor}
\usepackage{graphicx}  %
\usepackage{bm}  %
\usepackage{comment}
\usepackage{graphics}
\usepackage{float}
\usepackage{caption}
\usepackage{ragged2e}
\usepackage{tikz} 
\usetikzlibrary{shapes,arrows,positioning,automata,backgrounds,calc,er,patterns}
\usepackage{tikz-feynman}
\tikzfeynmanset{compat=1.0.0}
\usepackage{multirow}

\newcommand{\isovec}[1]{{\vec{#1}}} 
\newcommand{\spacevec}[1]{{\mathbf #1}}

\newcommand{\simle}{\hspace*{0.2em}\raisebox{0.5ex}{$<$}
     \hspace{-0.8em}\raisebox{-0.3em}{$\sim$}\hspace*{0.2em}}

\newcommand{\slashpartial}{\slash\hspace{-0.5em}\partial}
\newcommand{\slashD}{\slash\hspace{-0.7em}D}
\newcommand{\slashA}{\slash\hspace{-0.7em}A}

\newcommand{\vL}{\ensuremath{\mathcal{L}}}

\newcommand{\cq}{\mathcal Q}

\newcommand{\dslash}[1]{#1 \llap{/\kern-0.5pt}}
\newcommand{\Dslash}[1]{#1 \llap{/\kern+1.5pt}}
\newcommand{\DDslash}[1]{#1 \llap{/\kern+2.3pt}}
\newcommand{\dslashh}[1]{#1 \llap{/\kern+1pt}}
\newcommand{\abs}[1]{|#1|}

\newcommand{\boldtau}{\mbox{$\isovec \tau$}}

\newcommand{\boldsigma}{\mbox{\boldmath $\sigma$}}

\newcommand{\bea}{\begin{eqnarray}}
\newcommand{\eea}{\end{eqnarray}}
\newcommand{\be}{\begin{equation}}
\newcommand{\ee}{\end{equation}}
\newcommand{\bma}{\begin{pmatrix}}
\newcommand{\ema}{\end{pmatrix}}
\newcommand{\nn}{\nonumber}

\newcommand{\NLDBD}{$0 \nu \beta \beta$}

\newcommand{\graham}[1]{{\color{purple}#1}}

\begin{document}
\preprint{LA-UR-25-30400}

\title{
\vspace*{0.5cm}
\huge
Three-nucleon lepton-number-violating potentials in chiral EFT and their matrix elements in light nuclei 
\vspace*{.5cm}
}

\vspace*{.5cm}

\author{Graham Chambers-Wall}
\affiliation{Department of Physics, Washington University in Saint Louis, Saint Louis, MO 63130, USA}

\author{Justin Lieffers}
\affiliation{Department of Physics, University of Arizona, Tucson, AZ 85721, USA}

\author{Garrett B. King}
\affiliation{Theoretical Division, Los Alamos National Laboratory, Los Alamos, 
NM 87545, USA}

\author{Emanuele Mereghetti}
\affiliation{Theoretical Division, Los Alamos National Laboratory, Los Alamos, 
NM 87545, USA}

\author{Saori Pastore}
\affiliation{Department of Physics, Washington University in Saint Louis, Saint Louis, MO 63130, USA}
\affiliation{McDonnell Center for the Space Sciences at Washington University in St. Louis, MO 63130, USA}

\author{Maria Piarulli}
\affiliation{Department of Physics, Washington University in Saint Louis, Saint Louis, MO 63130, USA}
\affiliation{McDonnell Center for the Space Sciences at Washington University in St. Louis, MO 63130, USA}

\author{R. B. Wiringa}
\affiliation{Physics Division, Argonne National Laboratory, Argonne, IL 60439, USA}

\begin{abstract}
    We derive the three-nucleon neutrinoless double beta decay potential in a $\Delta$-full chiral effective field theory through next-to-next-to-next-to leading order in Weinberg's power counting. The matrix elements of the resulting operators are computed in light nuclei using Variational Monte Carlo with wave functions constructed from the Norfolk family of nuclear interactions. We find that three-nucleon corrections induce a modest quenching of the total nuclear matrix elements. We discuss model dependencies and the potential impact of these corrections on the sensitivity of experimental programs to probe lepton number violating parameters. These results provide a benchmark for many-body methods capable of reaching heavier nuclei of experimental interest.
\end{abstract}

\maketitle

\tableofcontents

\newpage

\section{Introduction}

The oscillation experiments of the last decade \cite{Ahmad:2001an,Eguchi:2002dm} have definitively proved that neutrinos have masses, which are constrained 
to be at least six orders of magnitude smaller than those of their charged lepton partners \cite{KATRIN:2021uub,KATRIN:2024cdt}.  
As the only electromagnetically neutral elementary fermions in the Standard Model (SM), 
neutrinos could have a Majorana or Dirac mass terms. Both terms  are forbidden in the minimal version of the SM, the first by the fact that left-handed neutrinos $\nu_L$ are charged under $SU(2)_L$, while the second requires the introduction of a new right handed field $\nu_R$. Therefore, neutrino masses point to physics beyond the SM, and  
understanding their Majorana or Dirac nature could have profound implications for other open problems in the field, such as the origin of the matter-antimatter asymmetry in the Universe \cite{Fukugita:1986hr}.

Neutrinoless double beta decay ($0\nu\beta\beta$)---a process in which two neutrons inside a nucleus turn into two protons, with the emission of two electrons and zero neutrinos---violates lepton number by two units and provides the most sensitive probe of the Majorana nature of active neutrinos. 
The present generation of experiments put very strong limits on the  $0\nu\beta\beta$
half-life  $T_{1/2}^{0\nu}$
of $^{76}$Ge \cite{GERDA:2020xhi,Majorana:2019nbd,LEGEND:2025jwu},
$^{82}$Se \cite{Arnold:2018tmo,CUPID:2019gpc}
$^{100}$Mo \cite{NEMO-3:2015jgm,Augier:2022znx,AMoRE:2024loj}, $^{130}$Te \cite{CUORE:2019yfd},  $^{136}$Xe  \cite{KamLAND-Zen:2016pfg,EXO-200:2019rkq,KamLAND-Zen:2024eml},
at the level of  $10^{24}$--$10^{26}$ yr.
In the minimal scenario, in which neutrino masses are induced by a dimension-five gauge-invariant, lepton-number-violating (LNV) effective operator~\cite{Weinberg:1979sa}, these can be translated into constraints on the effective neutrino mass $m_{\beta\beta}= \sum_i m_i U^2_{ei}$,  which combines neutrino masses, $m_i$, and mixing parameters, $U_{ei}$. This quantity is currently limited to fall in the range of $0.03-0.5$ eV, depending on the adopted value of the nuclear matrix (NMEs). The next generation of experiments will improve the limits on the half-lives by two orders of magnitude with the goal of completely covering the values of $m_{\beta\beta}$ predicted in the inverted hierarchy of neutrino masses~\cite{Adams:2022jwx,Agostini:2022zub}. 

The relation between the half-life of a given isotope and $m_{\beta\beta}$, or 
between $T_{1/2}^{0\nu}$ and fundamental LNV parameters for more general LNV mechanisms, depends on the NMEs of the $0\nu\beta\beta$ transition operators. 
In the case of light-Majorana-neutrino exchange,
the transition operator  (often called ``neutrino potential'') has a leading  two-nucleon component, which receives long- and pion-range contributions~\cite{Haxton:1984ggj,Simkovic:1999re}.
An investigation in both pionless and chiral Effective Field Theory (EFT)~\cite{Cirigliano:2017tvr,Cirigliano:2018hja,Cirigliano:2019vdj}  revealed that, in order to obtain LNV scattering amplitudes and matrix elements independent of ultraviolet regulators, the neutrino potential also has a leading order (LO) short-range component that acts on two neutrons or protons in the $^1S_0$ nucleon-nucleon channel. While the long- and pion-range components are determined by single-nucleon parameters (such as the axial coupling $g_A$), the short-range operator is a genuine two-nucleon  
effect, which comes with a coupling, $g_\nu^{\rm NN}$, which cannot be determined by single nucleon couplings and form factors. 
The best determination of $g_{\nu}^{\rm NN}$ is currently obtained by modeling the generalized forward
Compton scattering amplitude $n(p_1)n(p_2)W^+(k) \rightarrow p(p^\prime_1)p(p^\prime_2)W^-(k)$ \cite{Cirigliano:2020dmx,Cirigliano:2021qko}. The resulting value of $g_{\nu}^{\rm NN}$, which is in rough agreement with the expectations from naive isospin  
\cite{Cirigliano:2019vdj}
and from large $N_c$ arguments \cite{Richardson:2021xiu}, induces a
 significant corrections both in the light nuclei used for benchmarking {\it ab initio} calculations \cite{Pastore:2017ofx,Cirigliano:2019vdj,Weiss:2021rig} and in realistic $0\nu\beta\beta$ candidates \cite{Jokiniemi:2021qqv,Wirth:2021pij,Belley:2023btr,Belley:2023lec}.
Future lattice QCD calculations of the LNV $nn \rightarrow pp e e$ scattering amplitude will hopefully put the determination of the two-body neutrino potential on even firmer grounds
\cite{Davoudi:2020gxs,Davoudi:2021noh,Davoudi:2024ukx}.

To carry out a robust assessment of the theoretical uncertainties associated with $0\nu\beta\beta$ NMEs, it is necessary to construct the neutrino potential beyond leading order \cite{Cirigliano:2022rmf}.
Various next-to-next-to-leading (N$^2$LO) corrections to the $0\nu\beta\beta$ amplitude have been considered in the literature, including from the nucleon axial and vector form factors \cite{Haxton:1984ggj,Simkovic:1999re}, 
from pion-neutrino loops \cite{Cirigliano:2017djv}, and from corrections to the closure approximation \cite{Cirigliano:2017djv}.  Their evaluation in  light \cite{Pastore:2017ofx} and heavy nuclei \cite{Castillo:2024jfj} suggests that these corrections follow the power counting of chiral EFT. 
Here we focus on three-body corrections to the neutrino potential, which, as we will see, also start contributing at N$^2$LO. Most of the three-body  corrections arising at the first two non-trivial orders can be thought of as arising from connecting a two-body vector or axial current with a one-body current via a neutrino propagator, and a subset of these contributions were studied in Refs.
\cite{Menendez:2011qq,Wang:2018htk}.
A long-standing problem in the calculation
of $\beta$ decay rates is the systematic 
overprediction of the matrix element of the Gamow-Teller
$\spacevec{\sigma} \isovec{\tau}$ operator obtained in phenomenological models, such as the nuclear shell model \cite{Chou:1993zz,Wildenthal:1983zz,Martinez-Pinedo:1996zvt}. 
This problem is usually addressed by ``quenching'' $g_A$, that is by reducing its value from the physical nucleon axial coupling, $g_A = 1.27$, to an effective coupling $g_{A}^{\rm eff} = q g_A$, with $q$ ranging from $q \approx 0.9$ in light nuclei to $q\sim 0.75$ for $A\sim 50$ \cite{Chou:1993zz,Wildenthal:1983zz,Martinez-Pinedo:1996zvt}. 
The quenching is even more pronounced in the isotopes of interest for $0\nu\beta\beta$ experiments \cite{Caurier:2011gi,Engel:2016xgb}.  
For example, in order to reproduce the $2\nu\beta\beta$ decay rate of $^{76}$Ge and $^{136}$Xe,
quenching factors in the ranges 
$q_{^{76}\rm Ge}= \left[0.55,0.67\right]$ \cite{Jokiniemi:2022ayc} and
$q_{^{136}\rm Xe}= \left[0.42,0.57\right]$ \cite{CoelloPerez:2018ghg} are needed.
As the $0\nu\beta\beta$ rate
is proportional to $g_A^4$, 
the naive application of these quenching factors to $0\nu\beta\beta$ matrix elements would increase the half-life by a factor between 5 and 30,
seriously limiting the chances of the next generation of $0\nu\beta\beta$ experiments to probe interesting portions of the LNV parameter space. 
\textit{Ab initio} calculations of the Gamow-Teller matrix elements indicate that the quenching of $g_A$ is associate with limitations in the treatment of two-nucleon dynamics, particularly two-nucleon correlations and currents~\cite{Gysbers:2019uyb,Pastore:2017uwc,King:2020wmp}. Here, we want to investigate their effect 
at moderate momentum transfer, as it applies in the case of $0\nu\beta\beta$ decay. 

The paper is organized as follows. In Section \ref{Sec:2}, we detail the chiral Lagrangian and weak currents needed to construct LNV potentials and in Section \ref{Sec:3} we review the neutrino potentials in the two-nucleon sector. 
In Section \ref{Sec:4} we provide the three-nucleon neutrino potentials
and discuss the difference of our work with the literature. In particular, 
in Sections \ref{Sec:4a} and
\ref{Sec:4b} 
we construct the three-nucleon potential 
at N$^2$LO  and N$^3$LO in Weinberg's power counting,
while in Section \ref{Sec:4c}
we consider a N$^4$LO contribution induced by the short-range operator $g_{\nu}^{\rm NN}$. 
The potentials in this sections are derived in momentum space. Section \ref{coordspace} schematically discusses the coordinate space expressions, which are given in detail in Appendix \ref{sec.A}. 
The difference of our work with the literature is discussed in Section \ref{Sec:4d}.
In Section \ref{Sec:5}, we review the quantum Monte Carlo computational method in \ref{Sec:5a} and present matrix elements and their densities for three transitions, $^6\mathrm{He}\rightarrow {^6}\mathrm{Be}$, $^8\mathrm{He}\rightarrow {^8}\mathrm{Be}$, and ${^{12}}\mathrm{Be}\rightarrow {}^{12}\mathrm{C}$, whose total isospin changes by either 0 ($A=6$) or 2 ($A=8,12$), in \ref{Sec:5b}; finally, we conclude in Section \ref{Sec:6}.





\section{Electroweak and LNV chiral Lagrangians}
\label{Sec:2}

In this section, we review the strong and electroweak chiral Lagrangians that will be used in the calculation. The starting point is the quark-level QCD Lagrangian, complemented by charged leptons, Majorana neutrinos, electromagnetic and  weak interactions 
\begin{eqnarray}\label{quark}
\mathcal L_{\rm QCD} &=& -\frac{1}{2} \textrm{Tr} \left[G^{\mu \nu} G_{\mu \nu}\right]
- \frac{1}{4} F_{\mu \nu} F^{\mu \nu} +  \bar \nu_L i \slashpartial \nu_L  - \frac{m_{\beta\beta}}{2} \nu^T_{eL} C \nu_{eL}
+ \bar{\ell} \left(  i \slashpartial + e\, Q_e\, \slashA - m_{\ell}  \right)\ell \nonumber \\
& & + \bar q_L  i \slashD q_L + \bar q_R  i \slashD q_R -  \bar q_L M q_R - \bar q_R M q_L 
+ \bar q_L \gamma^\mu \left(l_\mu + \hat{l}_\mu \right) q_L  
+ \bar q_R \gamma^\mu \left(r_\mu + \hat{r}_\mu\right) q_R  
\, , 
\end{eqnarray}
where $A_\mu$ is the photon field,  $G_{\mu\nu}$ and $F_{\mu\nu}$ are the gluon and photon field strengths,  $C= i \gamma^2 \gamma^0$ is the charge conjugation matrix, and the effective electron neutrino Majorana mass is $m_{\beta\beta} = \sum m_{i} U_{e i}^2$, with $m_i$ denoting the eigenvalues of the neutrino mass matrix and $U_{ei}$ the elements of the Pontecorvo-Maki-Nakagawa-Sakata neutrino mixing matrix. The electron and muon fields  are denoted by $\ell$, with $e >0$ being the proton charge and $Q_e = -1$.  The quark doublet $q = (u\; d)^T$ has associated mass $M = {\rm diag} (m_u, m_d)$, and 
\begin{equation}
    q_{R, L} = \frac{1\pm \gamma_5}{2} q, \qquad 
    D_\mu q = (\partial_\mu - i g_s t^a G^a_\mu) q, \end{equation}
where $t^a$ are $SU(3)$ generators and  $g_s$ is the QCD coupling constant.    
We coupled the quarks to photons and leptons by defining the external currents as
\begin{eqnarray}
l_\mu = \frac{e}{2} A_\mu \, \tau^3 
- 2 \sqrt{2}\, G_F \left[V_{ud} \, \bar e_L \gamma_\mu  \nu_L \, \tau^+  + {\rm h.c.}\right]\,,  
& & \qquad \hat{l}_\mu = \frac{e}{6} A_\mu\,, 
\\
r_\mu = \frac{e}{2} A_\mu \, \tau^3\,,  
& & \qquad \hat{r}_\mu = \frac{e}{6} A_\mu\,,
\end{eqnarray}
where $\tau^a$ denote Pauli matrices, $G_F$ is the Fermi constant and $V_{ud}$  the $u$-$d$ element of the Cabibbo-Kobayashi-Maskawa quark mixing matrix.
We neglect weak neutral currents, which do not play an important role in this discussion. 

In the absence of electromagnetic and weak interactions, $e=G_F=0$, and in the massless limit, $m_u=m_d=0$,
the Lagrangian in Eq.~\eqref{quark} is invariant under independent $SU(2)$ rotations of left- and right-handed quarks, which form the chiral group $SU(2)_L \otimes SU(2)_R$.  Chiral symmetry is spontaneously broken down to the isovector subgroup $SU(2)_V$, leading to the emergence of pions as  Goldstone bosons. The symmetry is also explicitly broken by the quark masses and their electromagnetic and weak interactions, but the breaking is small and can be accounted for perturbatively. 
The approximate chiral symmetry of the QCD Lagrangian and its spontaneous breaking imply that pions interact weakly between themselves and with nucleons.  This observation can be formalized into an EFT for pions \cite{Weinberg:1978kz,Gasser:1983yg,Gasser:1984gg,Weinberg:1996kr}
and nucleons \cite{Jenkins:1990jv,Bernard:1995dp}.
The chiral Lagrangian also provides the foundation for the chiral expansion of nuclear potential and currents \cite{Weinberg:1990rz,Weinberg:1991um,Weinberg:1996kr,Ordonez:1992xp,Ordonez:1993tn,Ordonez:1995rz}.
Here we briefly summarize the terms in the pion, pion-nucleon, and nucleon-nucleon Lagrangian that we will need for the derivation of three-nucleon operators.
We start in Section \ref{Sec:pwc} by summarizing the power counting rules for few-body forces.
In Section  \ref{Sec:Lag1}
we present interactions between hadrons and one  weak current, which will give rise to long-range contributions to the LNV potentials. In Section \ref{Sec:Lag2} we construct operators induced by hard neutrino exchanges, which are not resolved in chiral EFT and lead to short-range contributions.

\subsection{Power counting}
\label{Sec:pwc}

Chiral perturbation theory ($\chi$PT) \cite{Weinberg:1978kz,Gasser:1983yg,Gasser:1984gg,Weinberg:1996kr} and its generalization to multi-nucleon systems, usually called chiral EFT \cite{Weinberg:1990rz,Weinberg:1991um},
allow one to organize contributions to meson and nucleon scattering amplitudes, nuclear potentials and electroweak currents in an expansion in powers of $\epsilon_\chi= p/\Lambda_\chi$, 
where $p$ is a low-momentum scale in the system and  $\Lambda_\chi \sim 4 \pi F_\pi \sim 1$ GeV
is the intrinsic mass scale of QCD.
Interactions in the chiral Lagrangian are organized according to their chiral index  $\chi = d + n/2-2$ \footnote{The chiral index is traditionally denoted by $\Delta$, here we use a different symbol not to induce confusion with the nucleon-$\Delta$ mass splitting. } \cite{Weinberg:1978kz,Manohar:1983md},
where $d$ counts the number of derivatives, insertions of the pion mass $m_\pi$ or of the nucleon-$\Delta$ mass splitting $\Delta$, and $n$ counts the number of baryon fields \cite{Weinberg:1978kz}. For LNV interactions, it is convenient to assume $G_F \, \bar \ell \, \Gamma \ell^\prime  = \mathcal O(p)$, so that the power counting rules discussed below can be immediately applied to LNV transition operators. In practice, we will consider at most two powers of $G_F$.

In systems with $A\le 1$, all momenta and energies are of order $p$
and for $p \ll \Lambda_\chi$
the  perturbative expansion of the $\chi$PT Lagrangian implies a similar perturbative organization of scattering amplitudes, with 
loops and insertions of subleading terms in the $\chi$PT Lagrangian causing suppression by powers of $\epsilon_\chi$. 
For systems with $A\ge 2$, 
the energy scale  $p^2/2m_N$ becomes relevant, and  the class of ``reducible'' diagrams, 
in which the intermediate state consists purely of propagating nucleons, is enhanced by factors of 
$m_N/p$ with respect to the $\chi$PT
power counting and need to be resummed \cite{Weinberg:1990rz,Weinberg:1991um}, leading to the appearance of bound states. 
Diagrams whose intermediate states contain interacting nucleons and pions --``irreducible''-- do not suffer from this infrared enhancement, and they   
follow the $\chi$PT power counting~\cite{Weinberg:1990rz,Weinberg:1991um}. 
Reducible diagrams are then obtained by gluing irreducible diagrams with intermediate states consisting of $A$ free-nucleon propagators,
which is equivalent to solving the Schr\"{o}dinger or Lippmann-Schwinger equation with a potential $V$ given by the sum of irreducible diagrams.   
In the case of two-body potentials, using topological identities for connected graphs, Weinberg showed that 
a diagram with $L$ loops and $V_i$ vertices with chiral index $\chi_i$
gives a correction of $\mathcal O(Q^\delta)$ to the potential \cite{Weinberg:1978kz,Weinberg:1990rz,Weinberg:1991um},
with 
\begin{equation}
    \delta =  2L + \sum V_i \chi_i.
\end{equation}
With this counting, the isospin-symmetric strong potential receives $\delta = 0$
corrections from one-pion-exchange and contact interactions in the $^3S_1$ and $^1S_0$ channel, which will be discussed in the following section. 
The electromagnetic potential  gets contributions of $\mathcal O(e^2 p^{-2})$ from Coulomb exchange and from the pion mass splitting and, in this power counting, $\mathcal O(e^2 p^0)$ from short-range four-nucleon operators in the $^1S_0$  channel. 

For three- and few-nucleon potentials, there is some ambiguity in the counting of the suppression induced by having 
more connections between nucleon lines. 
Weinberg assigned  
to each disconnected nucleon line a factor of $(4\pi)^2 p^{-4}$, from the four-dimensional energy-momentum conservation delta function
\cite{Weinberg:1991um},
implying that the presence of one less separately connected piece has the same cost of an irreducible pion loop
\begin{equation}\label{eq:nuWPC}
    \delta_{\, \rm Weinberg} = 2 (A-1-C) + 2L + \sum V_i \chi_i,
\end{equation}
where $C$ is the number of separately connected pieces in one diagram.
Friar, on the other hand, assigned the factor 
$(4\pi) p^{-4}$ to the delta function
\cite{Friar:1996zw},
which is compatible with the non-relativistic nature of bound state loops and results in an index 
\begin{equation}
    \delta_{\, \rm Friar} =  (A-1-C) + 2L + \sum V_i \chi_i.
\end{equation}
These two different counting schemes have implications for the strong three-body force, which, in chiral EFT with explicit $\Delta$s, contributes at NLO ($\mathcal O(\epsilon_\chi)$) in Friar's power counting (FPC), and N$^2$LO ($\mathcal O(\epsilon^2_\chi)$)
in Weinberg's power counting (WPC).
Similarly, as we will discuss, LNV three-body potential are suppressed by  
$\epsilon_\chi$ in FPC 
and $\epsilon^2_\chi$ in WPC. The difference is even more important for four-body potentials, which appear at N$^2$LO in FPC and N$^4$LO in WPC.
We will use WPC nomenclature, as this is widely used in the literature. However, we stress that the size of three-body corrections should be confirmed by explicit calculations (see also Ref. \cite{Hammer:2019poc} for more discussion of different power counting schemes in chiral EFT).

Finally, the definition of the chiral index $\chi$ for the operators in the $\chi$PT Lagrangian,
\begin{equation}\label{chi}
\chi = d + n/2-2,
\end{equation}
is based on naive dimensional analysis (NDA) \cite{Manohar:1983md}. 
While this assumption works in the 
meson and single nucleon sectors, NDA is known to fail in the two-nucleon system, where 
renormalization requires the promotion of certain nucleon-nucleon operators  \cite{Kaplan:1996xu,Kaplan:1998tg,Kaplan:1998we,Nogga:2005hy,Long:2011xw,Long:2012ve,Cirigliano:2018yza}. In all cases in which the enhancement factors are known, we will incorporate them 
in our power counting, by modifying the chiral index $\chi$. Renormalization studies in the three- and four-nucleon sectors are not as advanced \cite{Song:2016ale},
and thus, in these sectors, we will mostly rely  on WPC.

Before writing down in detail 
the chiral Lagrangian, we establish our notation.  
In the case of LNV, 
we define the Hamiltonian mediating $0\nu\beta\beta$ decays in terms of a ``neutrino potential" $V_\nu$
\begin{equation}
    H_{\rm LNV} =  2 G_F^2 V_{ud}^2 m_{\beta\beta} \bar e_L C \bar e_L^T  V_{\nu}.
\end{equation}
The neutrino potential has a leading two-nucleon component and can be expanded in contributions involving progressively more nucleons
\begin{eqnarray}
    V_\nu = V_{\nu,\, 2} + V_{\nu,\, 3} + V_{\nu,\, 4} + \ldots =\sum_{a\neq b} V^{(a,b)}_{\nu}  + \sum_{a\neq b \neq c} V_{\nu}^{(a,b,c)} +
\sum_{a\neq b \neq c \neq d} V_{\nu}^{(a,b,c,d)}
+ \ldots.
\end{eqnarray}
Each component of the neutrino potential has a chiral expansion
\begin{equation}
    V^{}_{\nu,\, j} = \sum_\delta V^{(\delta)}_{\nu,\, j}.
\end{equation}
In the two body sector, $j=2$, the leading order neutrino potential has $\delta = 0$,
and the first subleading corrections appear at $\delta = 2$ \cite{Cirigliano:2017tvr,Cirigliano:2019jig}. 
We will derive three-body terms up to N$^3$LO ($\delta = 3$) in WPC, where
we indicate with N$^\delta$LO terms that are suppressed by $\delta$ powers of $\epsilon_\chi$. Notice that this differs from the standard nomenclature in the strong part of the nucleon-nucleon potential (see e.g. \cite{Entem:2017gor}), where N$^\delta$LO denotes terms suppressed by $\epsilon_\chi^{\delta+1}$ compared to the leading.

\subsection{Chiral Lagrangian with electromagnetic and weak currents}
\label{Sec:Lag1}

At leading order (LO) in chiral perturbation theory, the mesonic representation  of Eq. \eqref{quark} is given by
\begin{equation}\label{Eq:LagPi}
\mathcal L^{(0)}_{\pi}  = \frac{F_0^2}{4} \mathrm{Tr}\left[(D_\mu U)^\dagger (D^\mu U)\right]+ \frac{F_0^2}{4} \mathrm{Tr}\left[U^\dagger \chi + U \chi^\dagger\right]\,,
\end{equation}
where the superscript $(0)$ denotes the chiral index  defined in Eq. \eqref{chi}. Here $\chi = 2 B M$, with $M$ the quark mass matrix, $B = 2.8$ GeV 
and the matrix $U$, which encodes the pion fields, is given by
\begin{align}
U=u^2=
\exp\left(\frac{i\boldsymbol{\pi}\cdot\boldsymbol{\tau}}{F_0}\right)
\end{align}
with $F_0$ the pion decay constant in the chiral limit. 
In our normalization, the physical pion decay constant is $F_\pi =92.3$ MeV.  At the order we are working, we can set $F_0 = F_\pi$, and we will identify them in what follows.
The chiral covariant derivative $D_\mu U$ is defined as
\begin{equation}
D_\mu U = \partial_\mu U - i (l_\mu + \hat{l}_\mu) U + i U (r_\mu + \hat{r}_\mu)\,.
\end{equation}
Expanding out the pion fields, Eq. \eqref{Eq:LagPi} yields the LO interactions of pions with axial and vector currents.

The leading order baryon Lagrangian is given by
\begin{equation}\label{LagN}
    \mathcal L^{(0)}_{\pi N} = \bar N_v i v \cdot \mathcal D N_v  + g_A \bar N_v S \cdot u N_v  - \bar{T}_{v\, i}^\mu (i v \cdot \mathcal D^{i j} - \Delta \delta^{ij} + g_1 S\cdot u^{i j}) T_{v\, \mu j}  + h_A \left( \bar{T}_{v\, i}^\mu w^i_\mu N + \bar N w^i_\mu T_{v\, i}^\mu  \right)
\end{equation}
where $v^\mu$ and $S^\mu$
are the nucleon velocity and spin, $v^\mu = (1, \bf 0)$ and $S^\mu = (0, \boldsymbol{\sigma}/2)$ 
in the nucleon rest frame. 
$N_v$ and $T_{v\, i}^\mu$ denote a heavy nucleon
and heavy $\Delta$ fields \cite{Jenkins:1990jv}. For ease of notation, we will drop the subscript $v$ in what follows. 
For the $\Delta$, we follow the conventions of Ref. \cite{Siemens:2020vop},
and the indices $i,j$ denote isospin indices.
$g_A$, $g_1$ and $h_A$ are the nucleon, $\Delta$, and nucleon-$\Delta$ axial couplings. $g_A$ is well determined, $g_A = 1.27$, while
for $h_A$ and $g_1$, we will use large $N_c$ relations \cite{Siemens:2016jwj},
which yield
$h_A = 1.40 \pm 0.05$,
$g_1 = 2.32 \pm 0.26$\footnote{See Ref. \cite{Siemens:2016jwj} for a discussion of the uncertainty range on $h_A$ and $g_1$.}. The nucleon and $\Delta$ covariant derivative are given by 
\begin{eqnarray}
\mathcal D_\mu N &=& (\partial_\mu + \Gamma_\mu)N  \nonumber \\
\mathcal D^{i j}_\mu T_{\nu j} &=& \left(\partial_\mu \delta^{i j} + \Gamma_\mu \delta^{i j} - i \epsilon^{i j k} \, {\rm Tr} (\tau^k \Gamma_\mu)\right)\, T_{\nu j}  \nonumber \\
\Gamma_\mu &=& \frac{1}{2}\left[u^\dagger\left(\partial_\mu -i(l_\mu + 3\hat{l}_\mu)\right)u+u\left(\partial_\mu - i (r_\mu + 3\hat{r}_\mu) \right)u^\dagger\right]\,, 
\end{eqnarray}
while $u_\mu$ is defined by
\begin{eqnarray}
u_\mu &=& -i\left[u^\dagger\left(\partial_\mu -i (l_\mu +\hat{l}_\mu)\right)u-u\left(\partial_\mu - i (r_\mu + \hat{r}_\mu)\right)u^\dagger\right]\,. \end{eqnarray}
The objects entering $\Delta$ and $\Delta$-nucleon axial interactions are related to $u_\mu$ by
\begin{eqnarray}
w^i_\mu = \frac{1}{2} {\rm Tr} (\tau^i u_\mu), \qquad
u_\mu^{i j} =  \xi^{i l}_{3/2} u_\mu \, \xi^{l j}_{3/2}, \, \qquad  
\xi^{i j}_{3/2} = \frac{2}{3} \delta^{i j} - \frac{i}{3} \epsilon^{i j k}\tau^k.
\end{eqnarray}
The propagator for the $\Delta$ is the following \cite{Krebs:2018D3f}:
\begin{eqnarray}
D_{\mu\nu}^{ij}(p)=-i \frac{P_{\mu\nu}^{3/2}\xi_{3/2}^{ij}}{v\cdot p - \Delta}
\end{eqnarray}
where  the spin projector  $P^{3/2}_{\mu\nu}$ is given by
\begin{align}
P^{3/2}_{\mu\nu} &= g_{\mu\nu}-v_{\mu}v_{\nu} + \frac{4}{3}S_{\mu}S_{\nu} = - \delta_{\mu i} \delta_{\nu j}\left(\frac{2}{3}\delta_{ij}-\frac{i}{3}\epsilon_{ijk}\sigma^{k} \right). 
\end{align}
The  NLO nucleon Lagrangian is given by \cite{Bernard:1995dp}
\begin{eqnarray}
    \mathcal L^{(1)}_{\pi N} &=& \bar N \Bigg[ \frac{1}{2 m_N} \left( (v \cdot \mathcal D)^2 - \mathcal D^2 \right) - i \frac{g_A}{2 m_N} \left\{ S \cdot \mathcal D, v \cdot u\right\} + c_1 \textrm{Tr}(\chi_+)
    +  \left(c_2 - \frac{g^2_A}{8 m_N}\right) (v\cdot u)^2 \nn \\
    & & + \frac{c_3}{2} {\rm Tr}( u \cdot u ) + \left(c_4 + \frac{1}{4 m_N} \right) \left[S^\mu, S^\nu\right] u_\mu u_\nu  + c_5 \tilde\chi_+
    \nn \\ & & - \frac{i}{4 m_N} \left[S^\mu,S^\nu\right]
    \left( (1+\kappa_1) f^+_{\mu\nu} + \frac{1}{2} (\kappa_0 - \kappa_1) \textrm{Tr}\left(f^+_{\mu\nu}\right)\right)
    \Bigg] N, \label{Eq:LagPiN_nlo}
\end{eqnarray}
where 
\begin{eqnarray}
    \chi_{\pm} &=& u^\dagger \chi u^\dagger \pm u \chi^\dagger u\,, \qquad \tilde \chi_{\pm} = \chi_\pm  - \frac{1}{2} \textrm{Tr}(\chi_\pm), \qquad
    f^+_{\mu \nu} = u^{\dagger} F_{L\, \mu \nu} u + u F_{R\, \mu \nu} u^{\dagger}.
\end{eqnarray}
The field strengths are given by
\begin{equation}
    F_{L}^{\mu\nu} = \partial^\mu l^\nu -\partial^\mu l^\nu - i [l^\mu, l^\nu], \qquad  
    F_{R}^{\mu\nu} = \partial^\mu r^\nu -\partial^\mu r^\nu - i [r^\mu, r^\nu].
\end{equation}
The first term in Eq. \eqref{Eq:LagPiN_nlo}
gives a correction to the nucleon propagator, while the second is a recoil correction to the pion-nucleon coupling.
In the construction of the LNV potentials, we will count powers of the nucleon mass as  $m_N = \mathcal O(\Lambda_\chi^2/Q)$, using the power counting often adopted for the nucleon-nucleon potentials \cite{Epelbaum:2008ga} or two-body weak and electromagnetic currents \cite{Kolling:2011mt,Krebs:2016rqz} (see Ref. \cite{Hammer:2019poc} for more discussion). In this counting, the first two terms in Eq. \eqref{Eq:LagPiN_nlo} are effectively neglected.
$c_1$, $c_2$, $c_3$ and $c_4$ induce couplings of the nucleon to two pions
and to one pion and one axial current.
The three-nucleon force contribution from the $\Delta$-field is exactly reproduced at N$^2$LO in the
EFT without explicit $\Delta$-fields through the resonance saturation of the LECs $c_{3,4}$, which show up in the subleading $\pi\pi$NN interactions. They are related as the following: $c_3^\Delta = -2c_4^\Delta = -4h^2_A/(9\Delta)$ \cite{Epelbaum:2007sq}.
$c_5$ arises in the presence of strong isospin breaking, $m_u \neq m_d$, and contributes to the neutron-proton mass difference. It also gives rise to interactions with two pions. Finally the last term represents the magnetic moment of the neutron and proton, which, in the presence of weak interactions, also leads to weak magnetism.
For the lengthy NLO $\Delta$ and nucleon-$\Delta$ Lagrangians, we refer to Refs. \cite{Hemmert:1996rw,Hemmert:1997ye,Siemens:2020vop}.
The NLO $\Delta$-$N$ Lagrangian contains recoil terms, proportional to $1/m_N$, which we neglect. In addition it contains a $\Delta$-nucleon transition magnetic moment, proportional to the LEC $b_1$, and two $\Delta N \pi\pi$ couplings, proportional to $b_{4,5}$. These also induce couplings of the axial current to one pion, a nucleon and a $\Delta$. As we discuss later, $b_{1,4,5}$ do not contribute to the three-body LNV potential at tree level.
The only NLO single-baryon LECs we need are therefore $c_{1,3,4}$.
The values of $c_1$, $c_3$ and $c_4$ that we use in the three-body neutrino potential are given in 
Tab. \ref{tab:lecs}
and are taken from Ref. \cite{Krebs:2007rh}. These values are chosen to be consistent with those implemented in the Norfolk interactions \cite{Piarulli:2016vel}. A more up-to-date analysis of pion-nucleon scattering yields \cite{Siemens:2016jwj}
\begin{equation}\label{eq:c34M}
c_1 = -0.74(2) \text{ GeV}^{-1}, \qquad  
    c_3 = -0.65(22) \text{ GeV}^{-1}, \qquad c_4 = 0.96(11) \text{ GeV}^{-1},
\end{equation}
where we report the values obtained from the NLO fits within a 
$\Delta$-full chiral theory 
\cite{Siemens:2016jwj},
compatible with the order at which the LECs appear in our calculation. 
We will use the difference between 
Eq. \eqref{eq:c34M} and the values in Table \ref{tab:lecs} to give a rough assessment of the uncertainty in the three-body contribution to $0\nu\beta\beta$, even though a consistent analysis would require to refit the LECs in the three-nucleon force and in the axial two-body current as we change $c_{1,3,4}$.

In the two nucleon sector, in WPC there are two S-wave interactions \cite{Weinberg:1991um} 
\begin{eqnarray}
\mathcal L^{(0)}_{NN} = - \frac{C_S}{2}\bar N N \, \bar N N - \frac{C_T}{2}\bar N\boldsigma N \cdot \bar N \boldsigma N.  
\end{eqnarray}
For the study of $0\nu\beta\beta$ is convenient to work in a partial wave basis,
\begin{eqnarray}
\mathcal L^{(0)}_{NN} = - C^{^3S_1}_0 \left( N^T P^i_{^3 S_1} N\right)^{\dagger} N^T P^i_{^3 S_1} N -
 C^{^1S_0}_0 \left( N^T P^a_{^1 S_0} N\right)^{\dagger}
    \left( N^T P^a_{^1 S_0} N\right), 
\end{eqnarray}
with the projectors $P^i_{^3S_1}$ and $P^a_{^1S_0}$ defined as 
\begin{eqnarray}
P^i_{^3S_1} = \frac{1}{\sqrt{8}} \sigma_2 \tau_2 \sigma^i, \qquad
P^a_{^1S_0} = \frac{1}{\sqrt{8}} \sigma_2 \tau_2 \tau^a,
\end{eqnarray}
and $i$ ($a$) denote a spin (isospin) index.
The coefficients $C^{^1S_0}_0$ and $C^{^3S_1}_0$
are related to $C_{S,T}$ by
\begin{eqnarray}
C^{^1S_0}_0 = C_S - 3 C_T, \qquad
C^{^3S_1}_0 = C_S + C_T.
\end{eqnarray}
Ref. \cite{Kaplan:1996xu} showed that renormalization requires 
$C_0^{^1S_0}$  to be mass dependent.
At the operator level, this is accomplished by \begin{equation}\label{eq:C0D2}
     C_0^{^1S_0} \rightarrow C^{^1S_0}_0 + D^{^1S_0}_2 {\rm Tr}[\chi_+].
\end{equation}
Additional deficiencies of WPC can be cured by promoting short-range operators in the $P$-wave to leading order \cite{Nogga:2005hy}. 
At NLO, one also has a coupling of a pion (or an axial current) to two-nucleons, which is usually expressed as \cite{Friar:1998zt,Cohen:1995cc}
\begin{eqnarray}\label{eq:LagCD}
\mathcal L^{(1)}_{N N\pi} 
& =&  -\frac{c_D}{4 \Lambda_\chi F_\pi^2} 
\bar N \sigma^i  \tau^a N \, \bar N N \left(\frac{\partial_i \pi^a}{F_\pi} - l_i^a + r_i^a + \ldots \right),
\end{eqnarray}
with $c_D = \mathcal O(1)$
and the antisymmetry of the nuclear wavefunction has been used to reduce the spin/isospin structures   \cite{Epelbaum:2002vt}. $c_D$ induces 
a three-body force and the leading contact contribution to the two-body
axial current.

\begin{table}[]
    \centering
    \begin{tabular}{c c c c c}
    \hline\hline
     $g_A$ & $h_A$ & $c_1$ & $c_3$ & $c_4$\\
     \hline
     1.27 & 1.40 & -0.57 & -0.79 & 1.33\\
     \hline\hline
    \end{tabular}
    \caption{Summary of low-energy-constants used in the three-body LNV potential. The LECs $g_A$ and $h_A$ are adimensional, and $c_1$, $c_3$, and $c_4$ are given in GeV$^{-1}$.}
    \label{tab:lecs}
\end{table}

\subsection{Short-range  LNV operators}
\label{Sec:Lag2}

In addition to the long-range effects discussed above, short-range operators are generated by the insertion of two 
currents connected by the exchange of hard  neutrinos. 
For the derivation of the LNV potentials at N$^3$LO we need to construct the LNV chiral Lagrangians in the 
$\pi\pi$, $\pi N$ and $N\!N$ sectors. 
As isospin-breaking electromagnetic operators induced by hard photon exchange have a very similar structure to LNV operators \cite{Cirigliano:2019vdj}, we will also mention short-range CIB operators, which have been constructed, for example, in Refs. \cite{VanKolck:1993ee, Fettes:1998ud, Muller:1999ww, Gasser:1983yg,Epelbaum:2007sq}. 

To construct operators that transform like two insertions of the weak and 
electromagnetic currents, we introduce the spurion fields for the left- and right-handed currents
\begin{eqnarray}
\mathcal Q_L  &=& u^\dagger Q_L u\,, 
\qquad  
\mathcal Q_R  = u Q_R u^\dagger\,,
\end{eqnarray}
where $u$
incorporates the pion fields.
Under left- and right-handed chiral rotations $L$ and $R$, respectively, 
the meson $u$ and nucleon $N$ fields transform 
as $u \rightarrow L u K^{\dagger} = K u R^\dagger$ and $N \rightarrow K N$,
where $K$ is an $SU(2)$ matrix that depends nonlinearly on the pion field.
The spurions $Q_{L,R}$ transform like currents,
\begin{eqnarray}
Q_L & \rightarrow& L Q_L L^\dagger\,, 
\qquad 
\,\,\, Q_R \rightarrow R Q_R R^\dagger\,, \\
\mathcal Q_L  
&\rightarrow& K \mathcal Q_L K^\dagger\,, 
\qquad 
\mathcal Q_R  
\rightarrow K \mathcal Q_R K^\dagger\,.
\end{eqnarray}
One then writes the most general Lagrangian involving $\mathcal Q_{L,R}$
that is invariant under chiral symmetry.
The way in which weak and electromagnetic currents break the symmetry is recovered by taking
$Q\to Q^{\rm w}$ 
or $Q\to Q^{\rm em}$,
with
\begin{equation}
Q^{\rm w}_L = \tau^+\,,\qquad Q^{\rm w}_R = 0\,, \qquad
Q^{\rm em}_L = Q^{\rm em}_R = \tau_3/2\,.
\end{equation}
Because two insertions of $\cq^{\rm w}_L$ give rise to isospin 2 ($T=2$) interactions in the 
\NLDBD\ case, in the electromagnetic case we will also focus on $T=2$ operators. 

In the mesonic sector, the only operator that can be constructed with
two insertions of $Q_{L,R}$ and no derivatives is the $T=2$ interaction 
\begin{equation}
\mathcal L^{(0)}_{e^2,\, \pi} = 
 Z e^2 F_\pi^4 \, 
\textrm{Tr} [\mathcal Q^{\rm em}_L \mathcal Q^{\rm em}_R ]\,,
\label{CIBmeson0}
\end{equation}
where, at LO in $\chi$PT, $Z$ is related to the pion-mass (squared) splitting by 
\begin{equation}
Z e^2 F_\pi^2 = \frac{1}{2}\delta m^2_\pi  = \frac{1}{2} \left( m_{\pi^\pm}^2 - m_{\pi^0}^2 \right)\,.
\end{equation}
There is no interaction of the type $Q_{L}^2$ that would lead to $|\Delta L|=2$, where $L$ here denotes the lepton number. 
The first such interaction contains two chiral-covariant derivatives 
of the pion field,
and it is given by \cite{VanKolck:1993ee,Gasser:2002am,Cirigliano:2017tvr}
\begin{eqnarray}
 \mathcal L^{(2)}_{e^2,\, \pi}  &=&  
-e^2 F_\pi^2 \, \kappa_3 \left[ \textrm{Tr}( \mathcal Q^{\rm em}_L u^\mu) 
\, \textrm{Tr}( \mathcal Q^{\rm em}_L u_\mu )
- \frac{1}{3}\,\textrm{Tr} \left(\mathcal Q^{\rm em }_L\mathcal Q^{\rm em }_L\right) 
\, \textrm{Tr} \left( u^\mu u_\mu\right) +  (L \rightarrow R)\right] \,,
\nonumber\\
\mathcal L^{(2)}_{G_F^2\, \pi}   &= & 
\left(2 \sqrt{2}\, G_F V_{ud}\right)^2  m_{\beta \beta} \, 
\bar e_L C\bar e_L^T \, \frac{}{} \frac{5g_\nu^{\pi \pi}}{3(16\pi)^2} F_\pi^2 
 \left[   
\textrm{Tr}(\mathcal Q^{\rm w}_L u^\mu) 
\, \textrm{Tr}( \mathcal Q^{\rm w}_L u_\mu )
- \frac{1}{3}\, \textrm{Tr}\left(\mathcal Q^{\rm w}_L\mathcal Q^{\rm w}_L\right)
\, \textrm{Tr} \left( u^\mu u_\mu\right)
\right]  \nonumber \\ & &  + {\rm h.c.}\,,
\label{eq:gnupipi} 
\end{eqnarray}
where we used the notation of Ref. \cite{Gasser:2002am} for the electromagnetic operator
\footnote{Differently from Ref. \cite{Gasser:2002am}, we subtracted the trace 
part of the $\kappa_3$ operator to isolate the $T=2$ representation. This shift
in the $T=0$ part can be absorbed in a redefinition of the isospin-invariant 
operator $\kappa_1$ defined in Ref. \cite{Gasser:2002am}.}.
$g_{\nu}^{\pi\pi}$ is a LEC of $\mathcal O(1)$, so that the operator in 
Eq. \eqref{eq:gnupipi} contributes to the neutrino potential at N$^2$LO, 
together with the pion-neutrino loops discussed in 
Ref. \cite{Cirigliano:2017tvr}. 
The factors of $e^2$ and $(2 \sqrt{2} G_F V_{ud})^2
m_{\beta \beta} \, \bar e_L C\bar e_L^T$ appear due to two insertions of electromagnetic
and weak currents, respectively. 
This allows us to identify \cite{Cirigliano:2017tvr}
\begin{equation}
g_{\nu}^{\pi\pi}= - \frac{3}{5} \left(16\pi\right)^2 \kappa_3\,. 
\end{equation}
The model estimate of Ref. \cite{Ananthanarayan:2004qk} for $\kappa_3$ gives
$g_{\nu}^{\pi\pi}(\mu = m_\rho) = -7.6$, in agreement with two recent LQCD 
extraction that found $g_{\nu}^{\pi\pi}(\mu = m_\rho)= -10.89 \pm 0.79$ \cite{Feng:2018pdq,
Tuo:2019bue}
and $g_{\nu}^{\pi\pi}(\mu = m_\rho)= -10.78 \pm 0.52$
\cite{Detmold:2020jqv}. Eq. \eqref{eq:gnupipi} only contains $LL$ operators.  The  $LR$ mesonic operators that are relevant for electromagnetic 
are given in Ref. \cite{Gasser:2002am}. They give N$^2$LO corrections to the two-body potential,
which amount to a renormalization of the pion mass splitting,
but they do not induce three-body effects at the order we need.

In the single baryon sector, the lowest order operator one can build with the insertion of two currents are, in the notation of Refs. \cite{Gasser:2002am,Epelbaum:2007sq}
\begin{eqnarray}
\label{eq:f1}
\mathcal L^{(1)}_{e^2,\, N\pi} &=& e^2 F_\pi^2 f_1  \textrm{Tr} (\mathcal Q^{\rm em}_R \mathcal Q^{\rm em}_L) \bar N N + \frac{e^2 F_\pi^2}{2} f_3 \textrm{Tr} (\mathcal Q^{\rm em}_R \mathcal Q^{\rm em}_R +  \mathcal Q^{\rm em}_L \mathcal Q^{\rm em}_L) \bar N N \nonumber \\
& & - \frac{e^2 F_\pi^2}{4} \left( f^{\Delta}_4 + f^{\Delta}_5 \right) T^\mu_i \left( 
\textrm{Tr}(\tau^i \mathcal Q^{\rm em}_L)
\textrm{Tr}(\tau^j \mathcal Q^{\rm em}_L) + \textrm{Tr}(\tau^i \mathcal Q^{\rm em}_R)
\textrm{Tr}(\tau^j \mathcal Q^{\rm em}_R)  \right) T^\nu_j g_{\mu \nu} \nonumber \\
& & - \frac{e^2 F_\pi^2}{4} \left( f^{\Delta}_4 - f^{\Delta}_5 \right) T^\mu_i \left( 
\textrm{Tr}(\tau^i \mathcal Q^{\rm em}_L)
\textrm{Tr}(\tau^j \mathcal Q^{\rm em}_R) + \textrm{Tr}(\tau^i \mathcal Q^{\rm em}_R)
\textrm{Tr}(\tau^j \mathcal Q^{\rm em}_L)  \right) T^\nu_j g_{\mu \nu}. 
\end{eqnarray}
The LECs $f_1$ and $f_3$ scale as $\mathcal O(\Lambda^{-1}_\chi)$. They are calculated in the perturbative chiral quark model to be of the following sizes: 
$f_1 = -(4.6 \pm 0.4)$ GeV$^{-1}$
and $f_3 = 4.2 \pm 0.4$ GeV$^{-1}$ in Ref. \cite{Lyubovitskij:2002}. The second operator has isospin 0, and is just an electromagnetic correction to the average nucleon mass.  In particular, $f_3$ vanishes for $\mathcal Q_L = \mathcal Q_L^{\rm w}$ and does not contribute to $0\nu\beta\beta$.
The first operator, which appears only in the electromagnetic case, has a $T=2$ component that contributes to CIB three-body forces at N$^3$LO. 
The operator $f_4^{\Delta}$ contributes to the $T=2$ $\Delta$ mass splitting \cite{Epelbaum:2007sq}
\begin{equation}
    m_{\Delta^{++}} - m_{\Delta^+} - m_{\Delta^0} + m_{\Delta^-} = \delta m^{(2)}_\Delta= - \frac{4}{3} e^2 F_\pi^2 f_4^{\Delta}.
\end{equation}
This mass splitting is unfortunately not known very well. Epelbaum and collaborators estimate \cite{Epelbaum:2007sq}
\begin{equation}
    \delta m^{(2)}_\Delta = -(1.7 \pm 2.7)\, {\rm MeV}, \qquad f_4^{\Delta} =  1.6 \pm 2.6 \, {\rm GeV}^{-1}\,, 
\end{equation}
in rough agreement with 
power-counting expectations.
When $\mathcal Q_L \rightarrow \mathcal Q^{\rm w}_L$ and 
$e^2 \rightarrow (2 \sqrt{2} G_F V_{ud})^2 \allowbreak
m_{\beta \beta} \, \bar e_L C\bar e_L^T$,  
$f_4^{\Delta}  + f_5^{\Delta}$  contributes to $0\nu\beta\beta$.
This operator induces a momentum independent $\Delta^0 \rightarrow \Delta^{++} e^- e^-$ interaction, which has no equivalent in the nucleon sector.

In the single-nucleon sector, 
the lowest-order $|\Delta L|=2$ interaction involves one derivative. Focusing on terms with only $\mathcal Q^{\rm em}_L$ ($\mathcal Q^{\rm w}_L$) or $\mathcal Q^{\rm em}_R$, one can write 
\cite{VanKolck:1993ee,vanKolck:1996rm,Gasser:2002am,Cirigliano:2017tvr}
\footnote{We again subtracted the trace terms compared to the $O_4$ and $O_5$ 
operators in Ref. \cite{Gasser:2002am}, such that the operators in 
Eq. \eqref{eq:CT1em} 
have $T=2$. These redefinitions would be absorbed by shifting the couplings of 
the $O_1$ and $O_2$ operators of Ref. \cite{Gasser:2002am}.}
\begin{eqnarray} 
\vL_{e^2,\, \pi N}^{(2)}&=& e^2 F_\pi^2 \, \frac{g_4 + g_5}{4} 
\left[{\rm Tr}\left(u_\mu \mathcal Q^{\rm em}_L \right)
\bar N S^\mu \mathcal Q^{\rm em}_L N 
-\frac{1}{3} \,\textrm{Tr}\left(\mathcal Q^{\rm em}_L\,\mathcal Q^{\rm em}_L\right) 
\bar N S^\mu u_\mu N
+ (L \rightarrow R) 
\right] \,, 
\nonumber \\
\vL_{G_F^2,\, \pi N}^{(2)}  &=&
\left(2\sqrt{2} G_F V_{ud}\right)^2  m_{\beta \beta} \, \bar e_L C\bar e_L^T
\, \frac{ g_A  g_\nu^{\pi N}}{4 (4\pi)^2} 
\nonumber \\
&&
\times \left[{\rm Tr}\left(u_\mu \mathcal Q^{\rm w}_L \right) 
\bar N S^\mu \mathcal Q^{\rm w}_L N
- \frac{1}{3} \,\textrm{Tr}\left(\mathcal Q^{\rm w }_L\mathcal Q^{\rm w }_L\right) 
\bar N S^\mu u_\mu N \right] + {\rm h.c.} \,,
\label{eq:CT1em}
\end{eqnarray}
where the LEC $g_{\nu}^{\pi N}=\mathcal O(1)$
is related to the electromagnetic LEC $g_4+g_5$ by \cite{Cirigliano:2017tvr} 
\begin{equation}
g_\nu^{\pi N}= \left(4\pi F_\pi\right)^2 \, \frac{g_4 + g_5}{g_A}
\equiv -\frac{2}{g_A} \left( \frac{4\pi}{e} \right)^2 \bar{\beta}_{10}\,. 
\label{gnupiN}
\end{equation}
The electromagnetic interactions induce CIB in the pion-nucleon couplings, but
at the moment there exist no good estimates besides NDA.
There is only a bound $\bar{\beta}_{10}=5(18)\cdot 10^{-3}$ \cite{vanKolck:1996rm}
extracted from 
the Nijmegen partial-wave analysis \cite{Stoks:1993tb,vanKolck:1997fu} 
of $N\!N$ scattering, which translates to $|g_\nu^{\pi N}|\simle 61$.
This introduces a source of uncertainty at N$^2$LO in the chiral expansion
of the neutrino potential.
Also in this case we did not write down the $LR$ operators, as they are given in 
\cite{Gasser:2002am}.

We  now come to the $N\!N$ sector, where the failure of WPC requires 
$|\Delta L|=2$ contact interactions  at LO.
The associated electromagnetic operators were constructed in Ref. \cite{Walzl:2000cx}. 
In the 
$m_{u,d} \rightarrow 0$ limit, 
there are only two rank-2 isospin operators with two insertions of $Q_{L,R}$ 
\cite{VanKolck:1993ee,Cirigliano:2017tvr},
\begin{eqnarray}
\mathcal L^{(0)}_{e^2,\, NN} 
&= & \frac{e^2}{4} 
\left\{ \bar N \mathcal Q^{\rm em}_L N \, 
\bar N \!\left({\cal C}_1\mathcal Q^{\rm em}_L
+{\cal C}_2\mathcal Q^{\rm em}_R\right) \! N
- \frac{1}{6} 
\textrm{Tr}\left[\mathcal Q^{\rm em}_L
\!\left({\cal C}_1\mathcal Q^{\rm em}_L+{\cal C}_2\mathcal Q^{\rm em}_R\right)\right]
\bar N \boldtau N \cdot \bar N \boldtau N\right\}
\nonumber\\ 
&&  + (L \rightarrow R)\,,
\nonumber\\
\mathcal L^{(0)}_{G_F^2,\, NN}  &=& \left(2\sqrt{2} G_F V_{ud}\right)^2  
m_{\beta \beta} 
\bar e_L C\bar e_L^T  \, \frac{g_\nu^{\rm NN}}{4} 
\left[ \bar N \mathcal Q^{\rm w}_L N \, \bar N \mathcal Q^{\rm w}_L N  
- \frac{1}{6} \textrm{Tr}\left(\mathcal Q^{\rm w}_L\,\mathcal Q^{\rm w}_L\right)
\bar N \boldtau N \cdot \bar N \boldtau N \right] \nonumber \\
&& + {\rm h.c.} \,.
\label{C12def}
\end{eqnarray}
As before, the LECs $g_\nu^{\rm NN}$ and ${\cal C}_1$ are related,
$g_\nu^{\rm NN}={\cal C}_1$.
The renormalization group equation of $\mathcal C_1$ and $\mathcal C_2$ implies that these operators acquire an implicit quark mass dependence, via the dependence of $C_0^{^1S_0}$ in Eq. \eqref{eq:C0D2}.
The full set of 
such $N\!N$ operators with up to two mass insertions is constructed in 
App. A of Ref. \cite{Cirigliano:2019vdj}.
In the isospin limit $m_u=m_d$, we can include quark-mass corrections by 
replacing ${\cal C}_1$ and ${\cal C}_2$ by the combinations
\begin{equation}
g_{\nu}^{\rm NN} = {\cal C}_1 = \sum_n c^{(1)}_{n} m_\pi^{2n}\,,   
\qquad
{\cal C}_2 = \sum_n c^{(2)}_n m_\pi^{2n}\,,
\end{equation}
where $c_n^{(1,2)}$ are the couplings of the electromagnetic operators with $n$ mass 
insertions.
The equality $g_{\nu}^{\rm NN} = {\cal C}_1$ relies only on isospin symmetry 
and is not spoiled by insertions of the average quark mass. 
Derivative operators in the $^1S_0$ channel, of the form \begin{eqnarray}
\mathcal L_{G_F^2,\, NN}^{(1)} &=&  \left(2\sqrt{2} G_F  V_{ud} \right)^2 
m_{\beta \beta} \,  \bar e_L C\bar e_L^T \, \frac{g_{2\, \nu}^{\rm NN} }{8}  
\nonumber \\ 
& & \times
\left[ (N^T  \overleftrightarrow{\boldsymbol\nabla}^{2} P_{^1S_0}^{\,+}   N) 
(N^T  P_{^1S_0}^{\,-} N)^{\dagger}  
+ (N^T P_{^1S_0}^{\,+} N)(N^T\overleftrightarrow{\boldsymbol\nabla}^{2} 
P_{^1S_0}^{\,-}N)^{\dagger}
\right] 
+ \textrm{h.c.},
\label{g2nudef}
\end{eqnarray}
contribute at N$^2$LO in the two-body sector. Here $\overleftrightarrow{\boldsymbol{\nabla}}$ denotes  $\overleftrightarrow{\boldsymbol{\nabla}} = \overrightarrow{\boldsymbol{\nabla}} - \overleftarrow{\boldsymbol{\nabla}}$.

At sufficiently high orders, light neutrino exchange will also induce short-range three-nucleon LNV interactions. 
Naively, one could write down non-derivative three-nucleon operators of the form
\begin{eqnarray}\label{eq:Lag_3b}
    \mathcal L^{(3)}_{G_F^2, NNN} &=& \left(2\sqrt{2} G_F V_{ud}\right)^2  
m_{\beta \beta} 
\bar e_L C\bar e_L^T  \, \frac{g_\nu^{\rm 3N}}{4}  \bar N N
\left[ \bar N \mathcal Q^{\rm w}_L N \, \bar N \mathcal Q^{\rm w}_L N  
- \frac{1}{6} \textrm{Tr}\left(\mathcal Q^{\rm w}_L\,\mathcal Q^{\rm w}_L\right)
\bar N \boldtau N \cdot \bar N \boldtau N \right] \nonumber \\
&& + {\rm h.c.} + \ldots,
\end{eqnarray}
with coefficients that, by naive dimensional analysis, would be expected to scale as
\begin{equation}
   \left. g_{\nu}^{\rm 3N} \right|_{\rm NDA} = \mathcal O\left(\frac{1}{F_\pi^2 \Lambda_\chi^3}\right).
\end{equation}
$g_{\nu}^{\mathrm{3N}}$ would then contribute to the $0\nu\beta\beta$ amplitude at N$^5$LO in WPC.
However, operators such as those in Eq. \eqref{eq:Lag_3b} involve three neutrons or three protons at the same space-time point, and thus can be shown to vanish  by symmetry (or, in the case of Eq. \eqref{eq:Lag_3b}, by a Fierz transformation).
Three-nucleon operators will thus involve derivatives,
implying further suppression in WPC.

\section{Two-nucleon sector}\label{Sec:3}

The derivation of the two-nucleon LNV  neutrino potential from a chiral EFT perspective has been carried out in Refs.~\cite{Cirigliano:2017tvr,Cirigliano:2018hja,Cirigliano:2019vdj}.
Here we limit ourselves to briefly reviewing its construction and refer the interested reader to the references above for in-depth discussions.

The Lagrangian in Eqs. \eqref{Eq:LagPi}, \eqref{LagN}, and \eqref{Eq:LagPiN_nlo}
gives rise to long-distance effects through couplings of 
photons and leptons to nucleons and pions. 
It induces the following one-body isovector  amplitude 
\begin{equation}
\mathcal A = \bar N  \left[\frac{l_\mu+r_\mu}{2}J_V^\mu+\frac{l_\mu-r_\mu}{2} J_A^\mu \right] N ,
\end{equation}
where  $J_V^\mu$ and $J_A^\mu$ are the single-nucleon vector and axial currents 
\bea \label{eq:currents1}
J^\mu_V  &=& g_V(\spacevec q^2) 
\left( v^\mu + \frac{p^\mu + p^{\prime \mu}}{2m_N} \right)
+ i g_M(\spacevec q^2)\,\epsilon^{\mu \nu \alpha \beta}\,
\frac{v_\alpha S_\beta q_\nu}{m_N}\,, 
\\
J^\mu_A  &=& - 2g_A(\spacevec q^2)  
\left(S^\mu  - \frac{S \cdot (p + p^\prime)}{2 m_N}\, v^\mu
+\frac{S \cdot q}{\spacevec q^2 + m_\pi^2}\,  q^\mu \right)\,. \label{eq:currents2}
\eea
Here $p$ and $p'$ stand for the momentum of the incoming neutron and outgoing 
proton, respectively, with momentum transfer defined as $q^\mu=(q^0,\, \spacevec q) =p^{\prime \mu}-p^\mu$.
The nucleon velocity and spin in the nucleon rest frame are given by $v^\mu = (1,\,\spacevec 0)$ and $S^\mu = (0,\,\boldsigma/2)$, respectively. 
Furthermore, $ \epsilon^{\mu \nu \alpha \beta}$ is the totally anti-symmetric 
tensor, with $\epsilon^{0123}=+1$. 

Up to NLO, the 
vector, axial, and magnetic nucleonic form factors are given by
\bea \label{eq:FF}
g_V(\spacevec q^2) &=& g_V = 1\, ,
\qquad g_A(\spacevec q^2) = g_A \simeq 1.27\,,
\qquad g_M(\spacevec q^2) = 1+\kappa_1\simeq 4.7\,,
\eea
where $\kappa_1\simeq 3.7$ is the nucleon isovector anomalous magnetic moment.
At N$^2$LO, $g_V$, $g_A$ and $g_M$ receive momentum-dependent corrections, while the induced pseudoscalar form factor deviates from the pion-pole form used  in Eq. \eqref{eq:currents2}. 
These effects are traditionally included in the neutrino potential  as form factors. In the calculation of the nuclear matrix elements, we follow the traditional approach and we use 
\begin{equation}
    g_V(\spacevec q^2) = \left(1 + \frac{\spacevec{q}^2}{\Lambda_V^2}\right)^{-2}, \qquad g_A(\spacevec q^2) =
    \left(1 + \frac{\spacevec{q}^2}{\Lambda_A^2}\right)^{-2}, \qquad g_M(\spacevec{q}^2) = (1+\kappa_1) g_V(\spacevec q^2)\, ,
\end{equation}
with $\Lambda_V = 850$ MeV and 
$\Lambda_A = 1040$ MeV.

Since we are interested in terms linear in the neutrino mass, the neutrino propagator reduces to a photon propagator in the Feynman gauge, and, as a result, the tree-level long-range neutrino potential, induced by the vector and axial currents in Eqs. \eqref{eq:currents1}  and \eqref{eq:currents2},  
and CIB photon-exchange potentials, which only receives contributions from the vector current, are very similar. 
As discussed in depth in Refs.~\cite{Cirigliano:2018hja,Cirigliano:2019vdj}, renormalization requires a LO counterterm for both the LNV and CIB potentials, captured by the $^1S_0$ couplings $g_\nu^{\rm NN}=\mathcal C_1$ and $\mathcal C_2$.

The LO neutrino potential is given by
\begin{eqnarray}\label{pot1}
V^{(0)}_{\nu,2} &=&  V_{\text{F}} + V_{\text{GT}} + V_{\text{T}}  + V_{\text{F}, S}
 \nn
\\
&=&  \sum_{i \neq j}\tau_i^+ \tau_j^+  \left\{ \frac{1}{\spacevec{q}^2} 
\left[
1- \frac{2g_A^2}{3} \boldsigma_{i} \cdot \boldsigma_{j} 
\left(1 + \frac{m_\pi^4}{2(\spacevec q^2 + m_\pi^2)^2}\right)  
- \frac{g_A^2 }{3} S_{ij} \left(1 - \frac{m_\pi^4}{(\spacevec q^2 + m_\pi^2)^2}
\right)
\right]  + 2 g_\nu^{\rm NN} \right\}, \nn \\
\end{eqnarray}
where the superscript $(0)$ denotes the chiral index, $\spacevec{q} = \spacevec{p}_1^\prime - \spacevec{p}_1 = \spacevec{p}_2 - \spacevec{p}_2^{\prime}$.
The Fermi (F), Gamow-Teller (GT) and Tensor (T) components are associated with the spin structures $\mathds{1}_i \mathds{1}_j$,
$\boldsigma_{i} \cdot \boldsigma_{j}$ and $S_{ij}$, respectively, with the tensor operator defined as
\begin{equation}
    S_{12}(\spacevec q) = \boldsigma_{1} \cdot \boldsigma_{2} - 3 \frac{ \boldsigma_{1} \cdot \spacevec{q} \,  \boldsigma_{2} \cdot \, \spacevec{q} }{\spacevec{q}^2}\label{tens1}\, .
\end{equation}
We split the Fermi piece in a long-distance part and a short-distance part, $V_{\text{F}, S}$, which is proportional to the LEC $g_{\nu}^{\rm NN}$.
In the CIB sector, electromagnetism causes a shift $\delta m_\pi$ in the charged pion masses, which also gives a LO long-range contribution. The LO CIB potential is 
\begin{eqnarray}
V^{(0)}_{{\rm CIB}} &=&  \frac{1}{2}\sum_{i\neq j}\left( \tau^{3}_i \tau^{3}_j - \frac{1}{3}\, \boldsymbol{\tau}_{i} \cdot \boldsymbol{\tau}_{j} 
\right) \Bigg\{ 
\frac{1}{\spacevec q^2} 
\left[ 1 - 
\frac{g_A^2}{3}  \frac{\delta m^2_\pi}{e^2 F_\pi^2} \,
\left(\boldsigma_{i} \cdot \boldsigma_{j}-S_{ij}\right)
\, \left(1-\frac{m_\pi^2}{\spacevec q^2 + m_\pi^2}\right)^2
\right]  \nonumber \\ & &  +  \left(\mathcal C_1 + \mathcal C_2\right) \Bigg\} \, .
\label{pot2} 
\end{eqnarray}

\begin{figure}[t]
\begin{center}
\begin{minipage}{0.25\textwidth}
    \scalebox{.75}{
    \begin{tikzpicture}[line width=1.2 pt]
    \begin{feynman}
    \vertex (a);
    \vertex [below=0.0 of a,dot] (adot) {};
    \vertex[below=2 of adot, dot] (b) {};
    \vertex[above left=1 of adot] (c);
    \vertex[above right=1 of adot] (d);
    \vertex[below left=1 of b] (e);
    \vertex[below right=1 of b] (f);

    \diagram*[]{
        (adot) -- [anti majorana] (b);
        (adot) -- [dashed, half right] (b); 
        (c) -- [dashed] (adot);
        (adot) -- [fermion] (d);
        (e) -- [dashed] (b);
        (b) -- [fermion] (f);
        };

    \vertex[below=1 of adot,square dot] (sq) {};
    \end{feynman}
\end{tikzpicture}}
\centering
\end{minipage}
\begin{minipage}{0.25\textwidth}
    \scalebox{.75}{
    \begin{tikzpicture}[line width=1.2 pt]
    \begin{feynman}
    \vertex (a);
    \vertex[right=1 of a, dot] (b) {};
    \vertex[right=1 of b,dot] (c) {};
    \vertex[right=1 of c, dot] (d) {};
    \vertex[right=1 of d] (e);
    \vertex[above=.75 of a] (f);
    \vertex[above=.75 of e] (g);
    \vertex[below=1 of b] (h);
    \vertex[below=1 of c] (i);
    \vertex[above=1 of c,dot] (top) {};
    \vertex[above right=1 of top] (electron);

    \diagram*[]{
        (a) -- [] (b);
        (d) -- [] (e);
        (top) -- [anti majorana, quarter right] (b);
        (b) -- [fermion] (f);
        (b) -- [] (c);
        (c) -- [] (d);
        (c) -- [dashed] (i);
        (top) -- [dashed,quarter left] (d);
        (top) -- [fermion] (electron);
        };

   \vertex[above left=1.03 of c, square dot,rotate=45] (square) {};
    \vertex[above=0.0 of b,dot] {};
    \vertex[above=0.0 of c,dot] {};
    \vertex[above=0.0 of d,dot] {};
    \end{feynman}
\end{tikzpicture}}
\centering
\end{minipage}
\begin{minipage}{0.25\textwidth}
    \scalebox{.75}{
    \begin{tikzpicture}[line width=1.2 pt]
    \begin{feynman}
    \vertex (a);
    \vertex[right=1 of a, dot] (b) {};
    \vertex[right=2 of b, dot] (c) {};
    \vertex[right=1 of c] (d);

    \vertex[below=2 of a] (e);
    \vertex[right=1 of e] (f);
    \vertex[right=2 of f] (g);
    \vertex[right=1 of g] (h);
    \vertex[right=1 of f] (ff);

    \vertex[below=2 of e] (i);
    \vertex[right=1 of i] (j);
    \vertex[right=2 of j] (k);
    \vertex[right=1 of k] (l);

    \vertex[below=1.2 of b] (m);

    \vertex[below=0.4 of c] (n);
    \vertex[below=1.6 of c] (o);

    \vertex[right=1 of b] (p);
    \vertex[below=2 of p] (q);
    \vertex[below=2 of q] (r);

    \vertex[right=1 of n] (s);
    \vertex[right=1 of o] (t);

    \vertex[below=1 of c] (cc);
    \vertex[above=.5 of h] (gg);
    \vertex[below=.5 of d] (dd);
    \vertex[below=2 of b,dot] (bb) {};
    \vertex[below=2 of c,dot] (cc) {}; 
    \vertex[right=1 of b] (b2);
    \vertex[above=2.5 of l] (m);
    \vertex[above=.5 of b2] (b3);
    
    \diagram*[]{
        (a) -- [] (b);
        (b) -- [] (c);
        (c) -- [] (d);
        
        (e) -- [] (f);
        (f) -- [] (g);
        (g) -- [] (h);


         (b) -- [anti majorana] (cc);
         (bb) -- [dashed] (c);
         (cc) -- [fermion] (m);
         (b) -- [fermion] (b3);
        };

    \vertex[above right=1.41 of bb, square dot,rotate=45] (dot2) {};

    \end{feynman}
\end{tikzpicture}}
\centering
\end{minipage}
\\
\vspace{20pt}
\begin{minipage}{0.25\textwidth}
    \hfill
\end{minipage}
\begin{minipage}{0.25\textwidth}
    \scalebox{.75}{
    \begin{tikzpicture}[line width=1.2 pt]
    \begin{feynman}
    \vertex (a);
    \vertex[right=1 of a, dot] (b) {};
    \vertex[right=1 of b,dot] (c) {};
    \vertex[right=1 of c, dot] (d) {};
    \vertex[right=1 of d] (e);
    \vertex[above=.75 of a] (f);
    \vertex[above=.75 of e] (g);
    \vertex[below=1 of b] (h);
    \vertex[below=1 of c] (i);

    \diagram*[]{
        (a) -- [] (b);
        (d) -- [] (e);
        (d) -- [anti majorana, half right] (b);
        (b) -- [fermion] (f);
        (d) -- [fermion] (g); 
        (b) -- [double] (c);
        (c) -- [double] (d);
        (c) -- [dashed] (i);
        };

    \vertex[above=.95 of c, square dot] (square) {};
    \vertex[above=0.0 of b,dot] {};
    \vertex[above=0.0 of c,dot] {};
    \vertex[above=0.0 of d,dot] {};
    \end{feynman}
\end{tikzpicture}}
\centering
\end{minipage}
\begin{minipage}{0.25\textwidth}
    \scalebox{.75}{
    \begin{tikzpicture}[line width=1.2 pt]
    \begin{feynman}
    \vertex (a);
    \vertex[right=1 of a, dot] (b) {};
    \vertex[right=2 of b, dot] (c) {};
    \vertex[right=1 of c] (d);

    \vertex[below=2 of a] (e);
    \vertex[right=1 of e] (f);
    \vertex[right=2 of f] (g);
    \vertex[right=1 of g] (h);
    \vertex[right=1 of f] (ff);

    \vertex[below=2 of e] (i);
    \vertex[right=1 of i] (j);
    \vertex[right=2 of j] (k);
    \vertex[right=1 of k] (l);

    \vertex[below=1.2 of b] (m);

    \vertex[below=0.4 of c] (n);
    \vertex[below=1.6 of c] (o);

    \vertex[right=1 of b] (p);
    \vertex[below=2 of p] (q);
    \vertex[below=2 of q] (r);

    \vertex[right=1 of n] (s);
    \vertex[right=1 of o] (t);

    \vertex[below=1 of c] (cc);
    \vertex[above=.5 of h] (gg);
    \vertex[below=.5 of d] (dd);
    \vertex[below=2 of b,dot] (bb) {};
    \vertex[below=2 of c,dot] (cc) {}; 
    \vertex[right=1 of b] (b2);
    \vertex[above=2.5 of l] (m);
    \vertex[above=.5 of b2] (b3);
    
    \diagram*[]{
        (a) -- [] (b);
        (b) -- [] (c);
        (c) -- [] (d);
        
        (e) -- [] (f);
        (f) -- [] (g);
        (g) -- [] (h);

         (b) -- [anti majorana] (cc);
         (bb) -- [dashed] (c);
         (cc) -- [fermion] (m);
         (b) -- [fermion] (b3);
         (bb) -- [double] (cc);
        };

    \vertex[above right=1.41 of bb, square dot,rotate=45] (dot2) {};

    \end{feynman}
\end{tikzpicture}}
\centering
\end{minipage}
\\
\vspace{20pt}
\begin{minipage}{0.25\textwidth}
    \hfill
\end{minipage}
\begin{minipage}{0.25\textwidth}
    \scalebox{.75}{
    \begin{tikzpicture}[line width=1.2 pt]
    \begin{feynman}
    \vertex (a);
    \vertex[right=1 of a, dot] (b) {};
    \vertex[right=1 of b] (c);
    \vertex[right=1 of c, dot] (d) {};
    \vertex[right=1 of d] (e);
    \vertex[above=.75 of a] (f);
    \vertex[above=.75 of e] (g);
    \vertex[below=1 of b] (h);
    
    \diagram*[]{
        (a) -- [] (e);
        (d) -- [anti majorana, half right] (b);
        (b) -- [fermion] (f);
        (d) -- [fermion] (g); 
        (b) -- [dashed] (h);
        };

    \vertex[above=.9 of c, square dot] (square) {};
        \vertex[above=-.0 of b, dot, minimum height=7] (dot) {};
        \vertex[above=-.0 of b, dot, white, minimum height=6] (dot) {};
        \vertex[above=-.0 of b, dot] (dot) {};
    \end{feynman}
\end{tikzpicture}}
\centering
\end{minipage}
\begin{minipage}{0.25\textwidth}
    \scalebox{.75}{
    \begin{tikzpicture}[line width=1.2 pt]
    \begin{feynman}
    \vertex (a);
    \vertex[right=1 of a] (b);
    \vertex[right=2 of b] (c);
    \vertex[right=1 of c] (d);

    \vertex[below=2 of a] (e);
    \vertex[right=1 of e] (f);
    \vertex[right=2 of f] (g);
    \vertex[right=1 of g] (h);
    \vertex[right=1 of f] (ff);

    \vertex[below=2 of e] (i);
    \vertex[right=1 of i] (j);
    \vertex[right=2 of j] (k);
    \vertex[right=1 of k] (l);

    \vertex[below=1.2 of b] (m);

    \vertex[below=0.4 of c] (n);
    \vertex[below=1.6 of c] (o);

    \vertex[right=1 of b] (p);
    \vertex[below=2 of p] (q);
    \vertex[below=2 of q] (r);

    \vertex[right=1 of n] (s);
    \vertex[right=1 of o] (t);

    \vertex[below=1 of c] (cc);
    \vertex[above=.5 of h] (gg);
    \vertex[below=.5 of d] (dd);
    
    \diagram*[]{
        (a) -- [] (b);
        (b) -- [] (c);
        (c) -- [] (d);
        
        (e) -- [] (f);
        (f) -- [] (g);
        (g) -- [] (h);

         (c) -- [anti majorana] (ff);
         (b) -- [dashed] (ff);
         (ff) -- [fermion] (gg);
         (c) -- [fermion] (dd);
        };
    \vertex[left=.43 of cc, square dot,rotate=0] (dot2) {};
    \vertex[right=-.07 of c, dot] (dot2) {};
    \vertex[right=-.07 of b, dot] (dot2) {};

        \vertex[above=-.12 of ff, dot, minimum height=7] (dot) {};
        \vertex[above=-.1 of ff, dot, white, minimum height=6] (dot) {};
        \vertex[above=-.07 of ff, dot] (dot) {};    
    \end{feynman}
\end{tikzpicture}}
\end{minipage}
\end{center}

\setcounter{figure}{0} 
\captionsetup{labelformat=default,justification=RaggedRight}
\caption{Examples of pion-neutrino loops contributing to the two-body neutrino potential $V_{\nu,\, 2}$. 
Plain and double lines denotes nucleons and $\Delta$ baryons, respectively. Dashes denote pions. Plain lines with arrows denote electron and neutrinos, and a square denotes an insertion of the neutrino Majorana mass. 
Dots and circled dots indicate LO and NLO interactions from the pion and nucleon Lagrangians, respectively. The diagrams in the first row 
contribute at N$^2$LO. The diagrams in the second row indicate  N$^2$LO corrections arising in a theory with explicit $\Delta$ degrees of freedom.
Diagrams in the third row would contribute at N$^3$LO.
}
\captionsetup{labelformat=empty,justification=centering}
\label{fig:n2lo}
\end{figure}
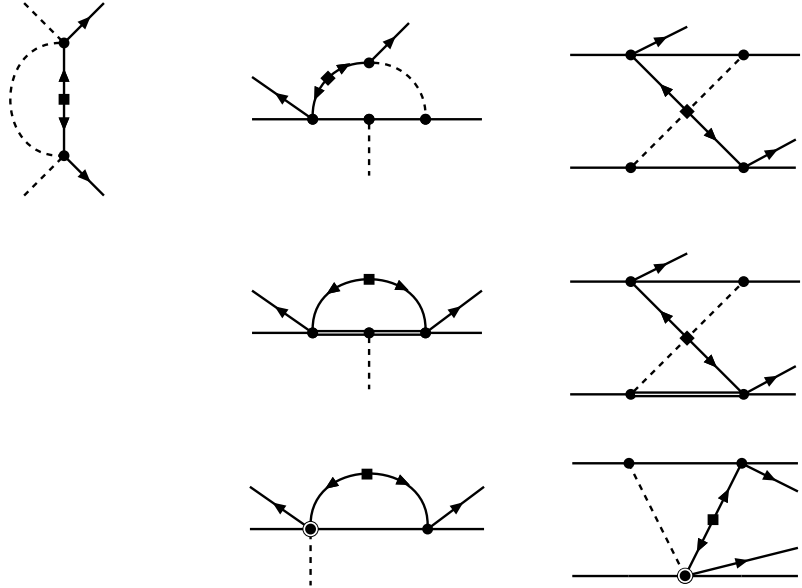
In Eq. \eqref{pot1}, the first term comes from the vector-vector component of the neutrino exchange and is clearly identical to the Coulomb potential, up to a rescaling.  The terms proportional to $g_A^2$ in Eq. \eqref{pot1}
come from the axial-axial ($AA$) piece, which does not appear in the electromagnetic potential. On the other hand, $\delta m_\pi^2$ only appears in the electromagnetic case.
In Ref. \cite{Cirigliano:2019vdj}, it was also shown that no NLO LNV or CIB counterterm is necessary. 

The next corrections arise  at N$^2$LO, where there are three
classes of contributions.
First, one has corrections from the momentum dependence of the nucleon axial and vector form factors, which can be accounted for by rewriting the neutrino potential as 
\begin{eqnarray}\label{pot2}
V^{(0+2)}_{\nu,2} &=&    \sum_{i \neq j}\tau_i^+ \tau_j^+  \left\{ \frac{1}{\spacevec{q}^2} 
\left[
g^2_V(\spacevec q^2)- \frac{2g^2_A(\spacevec q^2)}{3} \boldsigma_{i} \cdot \boldsigma_{j} 
\left(1 + \frac{m_\pi^4}{2(\spacevec q^2 + m_\pi^2)^2} + \frac{\spacevec q^2}{4 m_N^2 } \frac{g^2_M(\spacevec q^2)}{g^2_A(\spacevec q^2)}\right) \nonumber \right. \right. \\
& &  \left. \left.
- \frac{g_A^2(\spacevec q^2)}{3} S_{ij} \left(1 - \frac{m_\pi^4}{(\spacevec q^2 + m_\pi^2)^2} + \frac{\spacevec q^2}{4 m_N^2 } \frac{g^2_M(\spacevec q^2)}{g^2_A(\spacevec q^2)}
\right)
\right]  + 2 g_\nu^{\rm NN}  \right\}.
\end{eqnarray}
Expanding in $\spacevec q^2$,
in the $^1S_0$ channel, the weak magnetic moment 
and the nucleon vector and axial radii
provide a shift to $g_\nu^{\rm NN}$, which should be accounted for when extracting $g_\nu^{\rm NN}$ from lattice QCD or from phenomenological calculations of the $nn \rightarrow pp$ scattering amplitude (see, for example the  discussion in Ref. \cite{Wirth:2021pij}). 
In addition, the nucleon axial radius and magnetic moment induce a short-range tensor component, which contributes in $S=1$ waves, such as the $^3P_J$
 waves.
A study of the effects of such component and whether it requires short-range LNV $P$-wave operators at N$^2$LO remains to be done.

The second class of  two-nucleon N$^2$LO contributions 
contains $\pi$-$\nu$ loops, first considered  in Ref.
\cite{Cirigliano:2017tvr} in the case of a $\Delta$-less theory. 
Examples of the relevant diagrams are shown in Fig. \ref{fig:n2lo}.
The diagrams are closely related to the
$\pi$-$\gamma$ loops 
calculated in Ref.  \cite{vanKolck:1997fu}
for the CIB potential.
The momentum-space expressions for the 
$\Delta$-less potentials are given in Ref. \cite{Cirigliano:2017tvr}.
The loops are ultraviolet divergent and the divergences are absorbed by the $g_\nu^{\pi\pi}$, $g_{\nu}^{\pi N}$ and $g_{\nu}^{\rm NN}$ couplings introduced in Section \ref{Sec:Lag2}. 
The diagrams in the second row of Fig. \ref{fig:n2lo} show 
N$^2$LO corrections that would arise in a $\Delta$-full theory. These diagrams have not yet been computed. They will induce  calculable corrections to the loop functions derived in Ref. \cite{Cirigliano:2017tvr}, with an explicit dependence on $m_\pi/\Delta$. In addition, as the diagrams are naively logarithmically divergent, they will affect the renormalization of
$g_{\nu}^{\pi N}$ and $g_{\nu}^{\rm NN}$.
As we go to N$^3$LO in the three-body sector, in the third row of Fig. \ref{fig:n2lo} we  show  N$^3$LO corrections to the two-body potential induced by the same pion-nucleon couplings that contribute to the three-nucleon potential. Similar two-body diagrams induced by $c_D$ are not shown. 
These diagrams were first considered in Ref. \cite{Wang:2018htk}, but a full calculation is still missing. 
Ref. \cite{Wang:2018htk} noticed that the N$^3$LO diagrams are linearly divergent if computed with a cut-off. 
These divergences will be absorbed by the renormalization of $g_{\nu}^{\pi N}$ and $g_{\nu}^{\rm NN}$, which will then depend on the regulator adopted in the calculation of the $\pi$-$\nu$ loops. Schematically, we can think of the LECs has having a chiral expansion, e.g. for $g_{\nu}^{\rm NN}$ we can write 
\begin{equation}
    g_{\nu}^{\rm NN} = g^{(0)}_{\nu}(\Lambda_{\rm LS}) + 
    g^{(1)}_{\nu}(\Lambda_{\rm LS})+ g^{(2)}_{\nu}(\Lambda_{\rm LS}, \Lambda_{\pi\nu}) + g^{(3)}_{\nu}(\Lambda_{\rm LS}, \Lambda_{\pi\nu}) +\ldots.
\end{equation} 
At LO and NLO, $g_\nu^{\rm NN}$ only depends on the regulator $\Lambda_{\rm LS}$ used in the solution of the Lippmann-Schwinger or Schr\"odinger equation \cite{Cirigliano:2019vdj}.
Starting at N$^{2}$LO, $g_\nu^{\rm NN}$ also depends on the regularization scheme and scale chosen to regulate the loops in Fig. \ref{fig:n2lo}, denoted by $\Lambda_{\pi \nu}$. In principle, then, the value of $g_\nu^{\rm NN}$ needs to be extracted consistently with the corrections included in the long- and medium range neutrino potential, and, especially when using cut-off schemes, the numerical value of the LEC can show large variations with the chiral order.  
After $g_\nu^{\rm NN}$ is fixed to renormalize the two-body LNV amplitudes, the contribution of the two-body potential $V_{\nu, 2}$ to $0\nu\beta\beta$ NMEs should become independent on 
the renormalization scales $\Lambda_{\rm LS}$
and $\Lambda_{\pi \nu}$, up to power corrections. The NMEs of the three-nucleon potential $V_{\nu, 3}$ can then be studied in isolation. Any dependence of three-body NMEs on $\Lambda_{\rm LS}$ 
and $\Lambda_{\pi\nu}$ would be absorbed by three-body counterterms, and any large cancellation between the two- and three-body NMEs would be most likely accidental. 
Therefore, while it is desirable
to calculate both $V_{\nu,\, 2}$ and
$V_{\nu,\, 3}$ at N$^3$LO,  we focus in what follows on the three-nucleon sector. 

The third class of N$^2$LO corrections 
can arise from short-range LNV operators that are enhanced with respect to the expectation of WPC, for example because of renormalization arguments.  The analysis  of Ref. \cite{Cirigliano:2019vdj} showed that no new divergeces arise at NLO, but also that the derivative $^1S_0$ coupling
$g_{2\, \nu}^{\rm NN}$  contributes at N$^2$LO, rather than N$^4$LO as expected in WPC.  
For other contact interactions, there is not at the moment a consistent renormalization analysis.

\section{Three-nucleon sector}\label{Sec:4}

\subsection{Three-nucleon neutrino potential at N$^2$LO}\label{Sec:4a}

\begin{figure}[t]
\begin{minipage}{0.25\textwidth}
\centering
    \scalebox{.75}{
    \begin{tikzpicture}[line width=1.2 pt]
    \begin{feynman}
    \vertex (a);
    \vertex[right=1 of a] (b);
    \vertex[right=-0.075 of b, dot] (dot) {};
    \vertex[right=2 of b] (c);
    \vertex[right=1 of c] (d);

    \vertex[below=2 of a] (e);
    \vertex[right=1 of e] (f);
    \vertex[right=2 of f] (g);
    \vertex[right=1 of g] (h);

    \vertex[below=2 of e] (i);
    \vertex[right=1 of i] (j);
    \vertex[right=2 of j] (k);
    \vertex[right=1 of k] (l);

    \vertex[below=1.2 of b] (m);

    \vertex[below=0.4 of c] (n);
    \vertex[below=1.6 of c] (o);

    \vertex[below=0.925 of b, square dot] (dot) {};
    
    \diagram*{
        (a) -- [] (b);
        (b) -- [] (c);
        (c) -- [] (d);
        
        (e) -- [] (f);
        (f) -- [double] (g);
        (g) -- [] (h);

        (i) -- [] (j);
        (j) -- [] (k);
        (k) -- [] (l);

        (b) -- [anti majorana] (f);

        (b) -- [fermion] (n);
        (f) -- [fermion] (o);

        (k) -- [dashed] (g);
        };

        \vertex[right=-0.075 of f, dot] (dot) {};
        \vertex[right=-0.075 of g, dot] (dot) {};
        \vertex[right=-0.075 of k, dot] (dot) {};
    \end{feynman}
\end{tikzpicture}}
\caption{(2a)}
\end{minipage}\hfill
\begin{minipage}{0.25\textwidth}
\centering
    \scalebox{.75}{
    \begin{tikzpicture}[line width=1.2 pt]
    \begin{feynman}
    \vertex (a);
    \vertex[right=1 of a] (b);
    \vertex[right=-0.075 of b, dot] (dot) {};
    \vertex[right=2 of b] (c);
    \vertex[right=1 of c] (d);

    \vertex[below=2 of a] (e);
    \vertex[right=1 of e] (f);
    \vertex[right=-0.075 of f, dot] (dot) {};
    \vertex[right=2 of f] (g);
    \vertex[right=1 of g] (h);

    \vertex[below=2 of e] (i);
    \vertex[right=1 of i] (j);
    \vertex[right=2 of j] (k);
    \vertex[right=1 of k] (l);

    \vertex[below=0.8 of b] (m);
    \vertex[right=-0.075 of m, dot] (dot) {};
    \vertex[below=1.15 of c] (n);
    \vertex[below=1.55 of c] (o);

    \vertex[below=1.31 of b, square dot] (dot) {};
    
    \diagram*{
        (a) -- [] (b);
        (b) -- [] (c);
        (c) -- [] (d);
        
        (e) -- [] (f);
        (f) -- [double] (g);
        (g) -- [] (h);

        (i) -- [] (j);
        (j) -- [] (k);
        (k) -- [] (l);

        (b) -- [dashed] (m);
        (m) -- [anti majorana] (f);

        (m) -- [fermion] (n);
        (f) -- [fermion] (o);

        (k) -- [dashed] (g);
        };

        \vertex[right=-0.075 of f, dot] (dot) {};
        \vertex[right=-0.075 of g, dot] (dot) {};
        \vertex[right=-0.075 of k, dot] (dot) {};
    \end{feynman}
\end{tikzpicture}}
\caption{(2b)}
\end{minipage}\hfill
\begin{minipage}{0.25\textwidth}
\centering
    \scalebox{.75}{
    \begin{tikzpicture}[line width=1.2 pt]
    \begin{feynman}
    \vertex (a);
    \vertex[right=1 of a] (b);
    \vertex[right=-0.075 of b, dot] (dot) {};
    \vertex[right=2 of b] (c);
    \vertex[right=1 of c] (d);

    \vertex[below=2 of a] (e);
    \vertex[right=1 of e] (f);
    \vertex[right=-0.075 of f, dot] (dot) {};
    \vertex[right=2 of f] (g);
    \vertex[right=1 of g] (h);

    \vertex[below=2 of e] (i);
    \vertex[right=1 of i] (j);
    \vertex[right=2 of j] (k);
    \vertex[right=1 of k] (l);

    \vertex[below=1.2 of b] (m);
    \vertex[right=-0.075 of m, dot] (dot) {};
    \vertex[below=0.35 of c] (n);
    \vertex[below=0.85 of c] (o);

    \vertex[below=0.51 of b, square dot] (dot) {};
    
    \diagram*{
        (a) -- [] (b);
        (b) -- [] (c);
        (c) -- [] (d);
        
        (e) -- [] (f);
        (f) -- [double] (g);
        (g) -- [] (h);

        (i) -- [] (j);
        (j) -- [] (k);
        (k) -- [] (l);

        (b) -- [anti majorana] (m);
        (m) -- [dashed] (f);

        (b) -- [fermion] (n);
        (m) -- [fermion] (o);

        (k) -- [dashed] (g);
        };

        \vertex[right=-0.075 of f, dot] (dot) {};
        \vertex[right=-0.075 of g, dot] (dot) {};
        \vertex[right=-0.075 of k, dot] (dot) {};
    \end{feynman}
\end{tikzpicture}}
\caption{(2c)}
\end{minipage}\hfill
\begin{minipage}{0.25\textwidth}
\centering
    \scalebox{.75}{
    \begin{tikzpicture}[line width=1.2 pt]
    \begin{feynman}
    \vertex (a);
    \vertex[right=1 of a] (b);
    \vertex[right=2 of b] (c);
    \vertex[right=1 of c] (d);

    \vertex[below=2 of a] (e);
    \vertex[right=1 of e] (f);
    \vertex[right=2 of f] (g);
    \vertex[right=1 of g] (h);

    \vertex[below=2 of e] (i);
    \vertex[right=1 of i] (j);
    \vertex[right=2 of j] (k);
    \vertex[right=1 of k] (l);

    \vertex[below=1.2 of b] (m);
    \vertex[below=1.6 of b] (m);
    \vertex[above=1.2 of m] (p);
    
    \vertex[below=0.75 of c] (n);
    \vertex[below=1.25 of c] (o);

    \vertex[below=.91 of b, square dot] (dot) {};
    
    \diagram*{
        (a) -- [] (b);
        (b) -- [] (c);
        (c) -- [] (d);
        
        (e) -- [] (f);
        (f) [square dot]-- [double] (g) [square dot];
        (g) -- [] (h);

        (i) -- [] (j);
        (j) -- [] (k);
        (k) -- [] (l);

        (p) [crossed dot] -- [anti majorana] (m);
        (m) -- [dashed] (f);
        (p) -- [dashed] (b);

        (p) -- [fermion] (n);
        (m) -- [fermion] (o);

        (k) -- [dashed] (g);
        };

        \vertex[right=-0.075 of m, dot] (dot) {};
        \vertex[right=-0.075 of f, dot] (dot) {};
        \vertex[right=-0.075 of p, dot] (dot) {};
        \vertex[right=-0.075 of b, dot] (dot) {};
        \vertex[right=-0.075 of k, dot] (dot) {};
        \vertex[right=-0.075 of g, dot] (dot) {};
    \end{feynman}
\end{tikzpicture}}
\caption{(2d)}
\end{minipage}

\setcounter{figure}{1} 
\captionsetup{labelformat=default,justification=RaggedRight}
\caption{ Diagrams contributing to the neutrino potential at N$^2$LO.
Plain and double lines denotes nucleons and $\Delta$ baryons, respectively. Dashed lines denote pions. Plain lines with arrows denote electrons and neutrinos, and a square denotes an insertion of the neutrino Majorana mass. Dots indicate LO interactions from the pion and nucleon Lagrangians.}
\captionsetup{labelformat=empty,justification=centering}\label{fig:3body_n2lo}
\end{figure}

The N$^2$LO contributions to the neutrino potential are induced by the diagrams in Fig. \ref{fig:3body_n2lo}, with the exchange of an intermediate $\Delta$ baryon, mediated 
by the nucleon-$\Delta$-pion and nucleon-$\Delta$-axial current couplings. In momentum space, the potential assumes the form 
\begin{equation}
    V_{\nu,\, 3}^{(2)} = V_{(2a)} + V_{(2b)} + V_{(2c)} + V_{(2d)} \, ,
\end{equation}
with 
\begin{align}
\begin{split}
    V_{(2a)}&=-\frac{2h_A^2g_A^2}{9F_\pi^2\Delta}\sum_{i\neq j\neq k}\frac{(\boldsigma_k\cdot\boldsymbol{q}_k)}{\boldsymbol{q}_i^2(\boldsymbol{q}_k^2+m_\pi^2)}\left[4(\boldsigma_i\cdot\boldsymbol{q}_k)\tau_i^+\tau_k^++\boldsigma_j\cdot(\boldsymbol{q}_k\times\boldsigma_i)\tau_i^+(\boldsymbol{\tau}_k\times\boldsymbol{\tau}_j)^+\right], \label{eq:n2loa}
\end{split}
\\
\begin{split}
    V_{(2b)}&= V_{(2c)} \\&=\frac{2h_A^2g_A^2}{9F_\pi^2\Delta}\sum_{i\neq j\neq k}\frac{(\boldsigma_i\cdot\boldsymbol{q}_i)(\boldsigma_k\cdot\boldsymbol{q}_k)}{\boldsymbol{q}_i^2(\boldsymbol{q}_i^2+m_\pi^2)(\boldsymbol{q}_k^2+m_\pi^2)}\left[4(\boldsymbol{q}_i\cdot\boldsymbol{q}_k)\tau_i^+\tau_k^++\boldsigma_j\cdot(\boldsymbol{q}_k\times\boldsymbol{q}_i)\tau_i^+(\boldsymbol{\tau}_k\times\boldsymbol{\tau}_j)^+\right], \label{eq:n2lob}
\end{split}
\\
\begin{split}
    V_{(2d)}&=-\frac{2h_A^2g_A^2}{9F_\pi^2\Delta}\sum_{i\neq j\neq k}\frac{(\boldsigma_i\cdot\boldsymbol{q}_i)(\boldsigma_k\cdot\boldsymbol{q}_k)}{(\boldsymbol{q}_i^2+m_\pi^2)^2(\boldsymbol{q}_k^2+m_\pi^2)}\left[4(\boldsymbol{q}_i\cdot\boldsymbol{q}_k)\tau_i^+\tau_k^++\boldsigma_j\cdot(\boldsymbol{q}_k\times\boldsymbol{q}_i)\tau_i^+(\boldsymbol{\tau}_k\times\boldsymbol{\tau}_j)^+\right], \label{eq:n2loc}
\end{split}
\end{align}
where we define
\begin{align}
    i(\boldsymbol{\tau}_i\times\boldsymbol{\tau}_j)^+=\tau_i^+\tau_j^3-\tau_i^3\tau_j^+ \, .
\end{align}

\subsection{Three-nucleon neutrino potential at N$^3$LO}\label{Sec:4b}

\begin{figure}[t]
\begin{minipage}{0.31\textwidth}
\centering
\scalebox{.75}{
    \begin{tikzpicture}[line width=1.2 pt]
    \begin{feynman}
    \vertex (a);
    \vertex[right=1 of a] (b);
    \vertex[right=2 of b] (c);
    \vertex[right=1 of c] (d);

    \vertex[below=2 of a] (e);
    \vertex[right=1 of e] (f);
    \vertex[right=2 of f] (g);
    \vertex[right=1 of g] (h);

    \vertex[below=2 of e] (i);
    \vertex[right=1 of i] (j);
    \vertex[right=2 of j] (k);
    \vertex[right=1 of k] (l);

    \vertex[below=1.2 of b] (m);

    \vertex[below=0.4 of c] (n);
    \vertex[below=1.6 of c] (o);

    \vertex[right=1 of b] (p);
    \vertex[below=2 of p] (q);
    \vertex[below=2 of q] (r);

    \vertex[right=1 of n] (s);
    \vertex[right=1 of o] (t);
    
    \diagram*{
        (a) -- [] (b);
        (b) -- [] (c);
        (c) -- [] (d);
        
        (e) -- [] (f);
        (f) -- [] (g);
        (g) -- [] (h);

        (i) -- [] (j);
        (j) -- [] (k);
        (k) -- [] (l);

        (p) -- [anti majorana] (q);
        (q) -- [dashed] (r);
        (p) -- [fermion] (s);
        (q) -- [fermion] (t);
        };

    \vertex[below=0.925 of p, square dot] (dot1) {};
    \vertex[right=-0.075 of p, dot] (dot) {};
    \vertex[right=-0.075 of q, dot] (dot) {};
    \vertex[right=-0.075 of r, dot] (dot) {};

        \vertex[above=1 of dot1, dot, black, minimum height=7] (dot) {};
        \vertex[above=1 of dot1, dot, white, minimum height=6] (dot) {};
        \vertex[above=1 of dot1, dot, black] (dot) {};
    
    \end{feynman}
\end{tikzpicture}
}
\caption{(3a)}
\end{minipage}\hspace{10pt}
\begin{minipage}{0.31\textwidth}
\centering
\scalebox{.75}{
    \begin{tikzpicture}[line width=1.2 pt]
    \begin{feynman}
    \vertex (a);
    \vertex[right=1 of a] (b);
    \vertex[right=2 of b] (c);
    \vertex[right=1 of c] (d);

    \vertex[below=2 of a] (e);
    \vertex[right=1 of e] (f);
    \vertex[right=2 of f] (g);
    \vertex[right=1 of g] (h);

    \vertex[below=2 of e] (i);
    \vertex[right=1 of i] (j);
    \vertex[right=2 of j] (k);
    \vertex[right=1 of k] (l);

    \vertex[below=1.2 of b] (m);

    \vertex[below=0.4 of c] (n);
    \vertex[below=1.6 of c] (o);

    \vertex[right=1 of b] (p);
    \vertex[below=2 of p] (q);
    \vertex[below=2 of q] (r);

    \vertex[right=1 of n] (s);
    \vertex[right=1 of o] (t);

    \vertex[below=1 of p] (u);
    \vertex[below=2 of u] (v);
    \vertex[below=1 of t] (w);
    
    \diagram*{
        (a) -- [] (b);
        (b) -- [] (c);
        (c) -- [] (d);
        
        (e) -- [] (f);
        (f) -- [] (g);
        (g) -- [] (h);

        (i) -- [] (j);
        (j) -- [] (k);
        (k) -- [] (l);

        (q) -- [dashed] (r);
        (p) -- [fermion] (s);
        (p) -- [anti majorana, quarter left] (v);
        (v) -- [fermion] (w);
        };

    \vertex[right=0.5 of q] (dot) {};
    \vertex[above=0.5 of dot] (dot);
    \vertex[right=-2.5pt of dot, square dot] (sdot) {};
    \vertex[right=-0.075 of p, dot] (dot) {};
    \vertex[right=-0.075 of q, dot] (dot1) {};
    \vertex[right=-0.075 of r, dot] (dot) {};
    \vertex[right=-0.075 of v, dot] (dot) {};

        \vertex[above=2 of dot1, dot, black, minimum height=7] (dot) {};
        \vertex[above=2 of dot1, dot, white, minimum height=6] (dot) {};
        \vertex[above=2 of dot1, dot, black] (dot) {};
    
    \end{feynman}
\end{tikzpicture}}
\caption{(3b)}
\end{minipage}\hspace{10pt}
\begin{minipage}{.31\textwidth}
    \centering
    \scalebox{.75}{
    \begin{tikzpicture}[line width=1.2 pt]
    \begin{feynman}
    \vertex (a);
    \vertex[right=1 of a] (b);
    \vertex[right=-0.075 of b, dot] (dot) {};
    \vertex[right=2 of b] (c);
    \vertex[right=1 of c] (d);

    \vertex[below=2 of a] (e);
    \vertex[right=1 of e] (f);
    \vertex[right=2 of f] (g);
    \vertex[right=1 of g] (h);

    \vertex[below=2 of e] (i);
    \vertex[right=1 of i] (j);
    \vertex[right=2 of j] (k);
    \vertex[right=1 of k] (l);

    \vertex[below=1.2 of b] (m);

    \vertex[below=0.4 of c] (n);
    \vertex[below=1.6 of c] (o);

    \vertex[above right=of g] (p);
    \vertex[below=0.6 of p] (q);
    \vertex[left=1 of g] (r);
    
    \diagram*{
        (a) -- [] (b);
        (b) -- [] (c);
        (c) -- [] (d);
        
        (e) -- [] (f);
        (f) -- [double] (g);
        (g) -- [] (h);

        (i) -- [] (j);
        (j) -- [] (k);
        (k) -- [] (l);

        (b) -- [dashed] (f);

        (k) -- [dashed] (g);
        (r) -- [fermion] (p);
        (r) -- [fermion] (q);
        };

        \vertex[right=-0.075 of f, dot] (dot) {};
        \vertex[right=-0.075 of g, dot] (dot) {};
        \vertex[right=-0.075 of k, dot] (dot) {};
        \node[diamond,draw,fill=white,inner sep=0pt,minimum size=7pt] (d) at (2,-2) {};
    \end{feynman}
\end{tikzpicture}}
\caption{(3c)}
\end{minipage}
\\
\vspace{10pt}
\begin{minipage}{0.22\textwidth}
    \scalebox{.75}{
    \begin{tikzpicture}[line width=1.2 pt]
    \begin{feynman}
    \vertex (a);
    \vertex[right=1 of a] (b);
    \vertex[right=2 of b] (c);
    \vertex[right=1 of c] (d);

    \vertex[below=2 of a] (e);
    \vertex[right=1 of e] (f);
    \vertex[right=2 of f] (g);
    \vertex[right=1 of g] (h);

    \vertex[below=2 of e] (i);
    \vertex[right=1 of i] (j);
    \vertex[right=2 of j] (k);
    \vertex[right=1 of k] (l);

    \vertex[below=1.2 of b] (m);

    \vertex[below=0.4 of c] (n);
    \vertex[below=1.6 of c] (o);

    \vertex[right=1 of b] (p);
    \vertex[below=2 of p] (q);
    \vertex[below=2 of q] (r);

    \vertex[right=1 of n] (s);
    \vertex[right=1 of o] (t);
    
    \diagram*{
        (a) -- [] (b);
        (b) -- [] (c);
        (c) -- [] (d);
        
        (e) -- [] (f);
        (f) -- [] (g);
        (g) -- [] (h);

        (i) -- [] (j);
        (j) -- [] (k);
        (k) -- [] (l);

        (p) -- [anti majorana] (q);
        (q) -- [dashed] (r);
        (p) -- [fermion] (s);
        (q) -- [fermion] (t);
        };

    \vertex[below=0.925 of p, square dot] (dot1) {};
    \vertex[right=-0.075 of p, dot] (dot) {};
    \vertex[right=-0.075 of q, dot] (dot) {};
    \vertex[right=-0.075 of r, dot] (dot) {};

        \vertex[below=1 of dot1, dot, black, minimum height=7] (dot) {};
        \vertex[below=1 of dot1, dot, white, minimum height=6] (dot) {};
        \vertex[below=1 of dot1, dot, black] (dot) {};
    
    \end{feynman}
\end{tikzpicture}}
\centering
\caption{(3d)}
\end{minipage}\hspace{10pt}
\begin{minipage}{0.22\textwidth}
    \scalebox{.75}{
    \begin{tikzpicture}[line width=1.2 pt]
    \begin{feynman}
    \vertex (a);
    \vertex[right=1 of a] (b);
    \vertex[right=2 of b] (c);
    \vertex[right=1 of c] (d);

    \vertex[below=2 of a] (e);
    \vertex[right=1 of e] (f);
    \vertex[right=2 of f] (g);
    \vertex[right=1 of g] (h);

    \vertex[below=2 of e] (i);
    \vertex[right=1 of i] (j);
    \vertex[right=2 of j] (k);
    \vertex[right=1 of k] (l);

    \vertex[below=1.2 of b] (m);

    \vertex[below=0.3 of c] (n);
    \vertex[below=1.7 of c] (o);

    \vertex[right=1 of b] (p);
    \vertex[below=2 of p] (q);
    \vertex[below=2 of q] (r);

    \vertex[right=1 of n] (s);
    \vertex[right=1 of o] (t);

    \vertex[below=1 of p] (u);
    \vertex[below=1 of s] (v);
    
    \diagram*{
        (a) -- [] (b);
        (b) -- [] (c);
        (c) -- [] (d);
        
        (e) -- [] (f);
        (f) -- [] (g);
        (g) -- [] (h);

        (i) -- [] (j);
        (j) -- [] (k);
        (k) -- [] (l);

        (u) -- [anti majorana] (q);
        (u) -- [dashed] (p);
        (q) -- [dashed] (r);
        (u) -- [fermion] (v);
        (q) -- [fermion] (t);
        };

    \vertex[below=1.415 of p, square dot] (dot1) {};
    \vertex[right=-0.075 of p, dot] (dot) {};
    \vertex[right=-0.075 of q, dot] (dot) {};
    \vertex[right=-0.075 of r, dot] (dot) {};
    \vertex[right=-0.075 of u, dot] (dot) {};

        \vertex[below=.51 of dot1, dot, black, minimum height=7] (dot) {};
        \vertex[below=.51 of dot1, dot, white, minimum height=6] (dot) {};
        \vertex[below=.51 of dot1, dot, black] (dot) {};
    
    \end{feynman}
\end{tikzpicture}}
\centering
\caption{(3e)}
\end{minipage}\hspace{10pt}
\begin{minipage}{0.22\textwidth}
    \scalebox{.75}{
    \begin{tikzpicture}[line width=1.2 pt]
    \begin{feynman}
    \vertex (a);
    \vertex[right=1 of a] (b);
    \vertex[right=2 of b] (c);
    \vertex[right=1 of c] (d);

    \vertex[below=2 of a] (e);
    \vertex[right=1 of e] (f);
    \vertex[right=2 of f] (g);
    \vertex[right=1 of g] (h);

    \vertex[below=2 of e] (i);
    \vertex[right=1 of i] (j);
    \vertex[right=2 of j] (k);
    \vertex[right=1 of k] (l);

    \vertex[below=1.2 of b] (m);

    \vertex[below=0.3 of c] (n);
    \vertex[below=1.7 of c] (o);

    \vertex[right=1 of b] (p);
    \vertex[below=2 of p] (q);
    \vertex[below=2 of q] (r);

    \vertex[right=1 of n] (s);
    \vertex[right=1 of o] (t);

    \vertex[below=1 of p] (u);
    \vertex[below=1 of s] (v);

    \vertex[above=1 of v] (w);
    \vertex[above=1 of t] (x);
    
    \diagram*{
        (a) -- [] (b);
        (b) -- [] (c);
        (c) -- [] (d);
        
        (e) -- [] (f);
        (f) -- [] (g);
        (g) -- [] (h);

        (i) -- [] (j);
        (j) -- [] (k);
        (k) -- [] (l);

        (u) -- [dashed] (q);
        (u) -- [anti majorana] (p);
        (q) -- [dashed] (r);
        (p) -- [fermion] (w);
        (u) -- [fermion] (x);
        };

    \vertex[below=0.445 of p, square dot] (dot1) {};
    \vertex[right=-0.075 of p, dot] (dot) {};
    \vertex[right=-0.075 of q, dot] (dot) {};
    \vertex[right=-0.075 of r, dot] (dot) {};
    \vertex[right=-0.075 of u, dot] (dot) {};

        \vertex[below=1.47 of dot1, dot, black, minimum height=7] (dot) {};
        \vertex[below=1.47 of dot1, dot, white, minimum height=6] (dot) {};
        \vertex[below=1.47 of dot1, dot, black] (dot) {};
    
    \end{feynman}
\end{tikzpicture}}
\centering
\caption{(3f)}
\end{minipage}\hspace{10pt}
\begin{minipage}{0.22\textwidth}
    \scalebox{.75}{
    \begin{tikzpicture}[line width=1.2 pt]
    \begin{feynman}
    \vertex (a);
    \vertex[right=1 of a] (b);
    \vertex[right=2 of b] (c);
    \vertex[right=1 of c] (d);

    \vertex[below=2 of a] (e);
    \vertex[right=1 of e] (f);
    \vertex[right=2 of f] (g);
    \vertex[right=1 of g] (h);

    \vertex[below=2 of e] (i);
    \vertex[right=1 of i] (j);
    \vertex[right=2 of j] (k);
    \vertex[right=1 of k] (l);

    \vertex[below=1.2 of b] (m);

    \vertex[below=0.3 of c] (n);
    \vertex[below=1.7 of c] (o);

    \vertex[right=1 of b] (p);
    \vertex[below=2 of p] (q);
    \vertex[below=2 of q] (r);

    \vertex[right=1 of n] (s);
    \vertex[right=1 of o] (t);

    \vertex[below=1 of p] (u);
    \vertex[below=1 of s] (v);

    \vertex[below=0.4 of p] (w);
    \vertex[below=1.2 of w] (x);

    \vertex[above=0.6 of u] (y);
    \vertex[below=1.2 of y] (z);
    \vertex[above=0.6 of v] (aa);
    \vertex[below=0.6 of aa] (bb);
    
    \diagram*{
        (a) -- [] (b);
        (b) -- [] (c);
        (c) -- [] (d);
        
        (e) -- [] (f);
        (f) -- [] (g);
        (g) -- [] (h);

        (i) -- [] (j);
        (j) -- [] (k);
        (k) -- [] (l);

        (w) -- [anti majorana] (x);
        (w) -- [dashed] (p);
        (q) -- [dashed] (x);
        (y) -- [fermion] (aa);
        (z) -- [fermion] (bb);
        (q) -- [dashed] (r);
        };

    \vertex[below=0.915 of p, square dot] (dot1) {};
    \vertex[right=-0.075 of p, dot] (dot) {};
    \vertex[right=-0.075 of q, dot] (dot) {};
    \vertex[right=-0.075 of r, dot] (dot) {};
    \vertex[right=-0.075 of w, dot] (dot) {};
    \vertex[right=-0.075 of x, dot] (dot) {};
    
        \vertex[below=1 of dot1, dot, black, minimum height=7] (dot) {};
        \vertex[below=1 of dot1, dot, white, minimum height=6] (dot) {};
        \vertex[below=1 of dot1, dot, black] (dot) {};
    
    \end{feynman}
\end{tikzpicture}}
\centering
\caption{(3g)}
\end{minipage}
\\
\vspace{10pt}
\begin{minipage}{0.22\textwidth}
    \scalebox{.75}{
    \begin{tikzpicture}[line width=1.2 pt]
    \begin{feynman}
    \vertex (a);
    \vertex[right=1 of a] (b);
    \vertex[right=2 of b] (c);
    \vertex[right=1 of c] (d);

    \vertex[below=1 of a] (e);
    \vertex[right=1 of e] (f);
    \vertex[right=2 of f] (g);
    \vertex[right=1 of g] (h);

    \vertex[below=2 of e] (i);
    \vertex[right=1 of i] (j);
    \vertex[right=2 of j] (k);
    \vertex[right=1 of k] (l);

    \vertex[below=1.2 of b] (m);

    \vertex[below=0.4 of c] (n);
    \vertex[below=1.6 of c] (o);

    \vertex[right=1 of b] (p);
    \vertex[below=2 of p] (q);
    \vertex[below=2 of q] (r);

    \vertex[right=1 of n] (s);
    \vertex[right=1 of o] (t);
    
    \diagram*{
        (a) -- [] (b);
        (b) -- [] (c);
        (c) -- [] (d);

        (e) -- [] (l);
        (i) -- [] (h);

        (p) -- [anti majorana] (q);
        (p) -- [fermion] (s);
        (q) -- [fermion] (t);
        };

    \vertex[below=0.925 of p, square dot] (dot1) {};
    \vertex[right=-0.075 of p, dot] (dot) {};
    \vertex[right=-0.075 of q, dot] (dot) {};

        \vertex[below=1 of dot1, dot, black, minimum height=7] (dot) {};
        \vertex[below=1 of dot1, dot, white, minimum height=6] (dot) {};
        \vertex[below=1 of dot1, dot, black] (dot) {};
    
    \end{feynman}
\end{tikzpicture}}
\centering
\caption{(3h)}
\end{minipage}\hspace{10pt}
\begin{minipage}{0.22\textwidth}
    \scalebox{.75}{
    \begin{tikzpicture}[line width=1.2 pt]
    \begin{feynman}
    \vertex (a);
    \vertex[right=1 of a] (b);
    \vertex[right=2 of b] (c);
    \vertex[right=1 of c] (d);

    \vertex[below=1 of a] (e);
    \vertex[right=1 of e] (f);
    \vertex[right=2 of f] (g);
    \vertex[right=1 of g] (h);

    \vertex[below=2 of e] (i);
    \vertex[right=1 of i] (j);
    \vertex[right=2 of j] (k);
    \vertex[right=1 of k] (l);

    \vertex[below=1.2 of b] (m);

    \vertex[below=0.4 of c] (n);
    \vertex[below=1.6 of c] (o);

    \vertex[right=1 of b] (p);
    \vertex[below=2 of p] (q);
    \vertex[below=2 of q] (r);

    \vertex[right=1 of n] (s);
    \vertex[right=1 of o] (t);

    \vertex[below=.8 of p] (u);
    
    \diagram*{
        (a) -- [] (b);
        (b) -- [] (c);
        (c) -- [] (d);

        (e) -- [] (l);
        (i) -- [] (h);

        (u) -- [anti majorana] (q);
        (u) -- [fermion] (s);
        (q) -- [fermion] (t);

        (p) -- [dashed] (u);
        };

    \vertex[below=1.305 of p, square dot] (dot1) {};
    \vertex[right=-0.075 of p, dot] (dot) {};
    \vertex[right=-0.075 of q, dot] (dot) {};
    \vertex[right=-0.075 of u, dot] (dot) {};

        \vertex[below=.6 of dot1, dot, black, minimum height=7] (dot) {};
        \vertex[below=.6 of dot1, dot, white, minimum height=6] (dot) {};
        \vertex[below=.6 of dot1, dot, black] (dot) {};
    
    \end{feynman}
\end{tikzpicture}}
\centering
\caption{(3i)}
\end{minipage}\hspace{10pt}
\begin{minipage}{0.22\textwidth}
    \scalebox{.75}{
    \begin{tikzpicture}[line width=1.2 pt]
    \begin{feynman}
    \vertex (a);
    \vertex[right=1 of a] (b);
    \vertex[right=2 of b] (c);
    \vertex[right=1 of c] (d);

    \vertex[below=1 of a] (e);
    \vertex[right=1 of e] (f);
    \vertex[right=2 of f] (g);
    \vertex[right=1 of g] (h);

    \vertex[below=2 of e] (i);
    \vertex[right=1 of i] (j);
    \vertex[right=2 of j] (k);
    \vertex[right=1 of k] (l);

    \vertex[below=1.2 of b] (m);

    \vertex[below=0.4 of c] (n);
    \vertex[below=1.6 of c] (o);

    \vertex[right=1 of b] (p);
    \vertex[below=2 of p] (q);
    \vertex[below=2 of q] (r);

    \vertex[right=1 of n] (s);
    \vertex[right=1 of o] (t);

    \vertex[below=1.2 of p] (u);

    \vertex[below=.3 of s] (tt);
    
    \diagram*{
        (a) -- [] (b);
        (b) -- [] (c);
        (c) -- [] (d);

        (e) -- [] (l);
        (i) -- [] (h);

        (u) -- [dashed] (q);
        (p) -- [fermion] (s);
        (u) -- [fermion] (tt);

        (p) -- [anti majorana] (u);
        };

    \vertex[below=0.51 of p, square dot] (dot1) {};
    \vertex[right=-0.075 of p, dot] (dot) {};
    \vertex[right=-0.075 of q, dot] (dot) {};
    \vertex[right=-0.075 of u, dot] (dot) {};

        \vertex[below=1.4 of dot1, dot, black, minimum height=7] (dot) {};
        \vertex[below=1.4 of dot1, dot, white, minimum height=6] (dot) {};
        \vertex[below=1.4 of dot1, dot, black] (dot) {};
    
    \end{feynman}
\end{tikzpicture}}
\centering
\caption{(3j)}
\end{minipage}\hspace{10pt}
\begin{minipage}{0.22\textwidth}
    \scalebox{.75}{
    \begin{tikzpicture}[line width=1.2 pt]
    \begin{feynman}
    \vertex (a);
    \vertex[right=1 of a] (b);
    \vertex[right=2 of b] (c);
    \vertex[right=1 of c] (d);

    \vertex[below=1 of a] (e);
    \vertex[right=1 of e] (f);
    \vertex[right=2 of f] (g);
    \vertex[right=1 of g] (h);

    \vertex[below=2 of e] (i);
    \vertex[right=1 of i] (j);
    \vertex[right=2 of j] (k);
    \vertex[right=1 of k] (l);

    \vertex[below=1.2 of b] (m);

    \vertex[below=0.4 of c] (n);
    \vertex[below=1.6 of c] (o);

    \vertex[right=1 of b] (p);
    \vertex[below=2 of p] (q);
    \vertex[below=2 of q] (r);

    \vertex[right=1 of n] (s);
    \vertex[right=1 of o] (t);

    \vertex[below=.45 of p] (u);
    \vertex[below=1.15 of u] (v);

    \vertex[above=.9 of t] (tt);
    \vertex[right=2 of u] (ss);
    
    \diagram*{
        (a) -- [] (b);
        (b) -- [] (c);
        (c) -- [] (d);

        (e) -- [] (l);
        (i) -- [] (h);

        (u) -- [anti majorana] (v);
        (u) -- [fermion] (ss);
        (v) -- [fermion] (tt);

        (p) -- [dashed] (u);
        (v) -- [dashed] (q);
        };

    \vertex[below=0.93 of p, square dot] (dot1) {};
    \vertex[right=-0.075 of p, dot] (dot) {};
    \vertex[right=-0.075 of q, dot] (dot) {};
    \vertex[right=-0.075 of u, dot] (dot) {};
    \vertex[right=-0.075 of v, dot] (dot) {};

        \vertex[below=1 of dot1, dot, black, minimum height=7] (dot) {};
        \vertex[below=1 of dot1, dot, white, minimum height=6] (dot) {};
        \vertex[below=1 of dot1, dot, black] (dot) {};
    
    \end{feynman}
\end{tikzpicture}}
\centering
\caption{(3k)}
\end{minipage}
\setcounter{figure}{2} 
\captionsetup{labelformat=default,justification=RaggedRight}
\caption{Diagrams contributing to the neutrino potential at N$^3$LO. Circled dots denote NLO interactions in the strong chiral Lagrangians, Eqs. \eqref{Eq:LagPiN_nlo} and \eqref{eq:LagCD}, and the empty diamond denotes an NLO interaction in the electromagnetic Lagrangian, Eq. \eqref{eq:f1}. 
}
\captionsetup{labelformat=empty,justification=centering}\label{fig:n3lo}
\end{figure}

The N$^3$LO three-body neutrino potential is given by
\begin{equation}
    V_{\nu,\, 3}^{(3)} = \sum_{\rho = a}^k V_{(3\rho)} \, .
\end{equation}
Diagrams $(3a)$ and $(3b)$ contain NLO corrections to the nucleon vector and axial current induced by the Lagrangian \eqref{Eq:LagPiN_nlo}.
As we remarked earlier, we neglect pure recoil corrections coming from operators with coefficients fixed by reparameterization invariance. The only correction to the current then arises from the nucleon isovector magnetic moment term, proportional to $1+\kappa_1$. This term contributes in conjunction with the coupling of the weak vector current to a nucleon and a pion (diagram $(3a)$) and to two pions (diagram $(3b)$). Finally, $\kappa_1$ induces a coupling of the axial current to a nucleon and a pion, which contributes together with the LO nucleon axial current (diagram $(3d)$).
In all, the contributions proportional to the magnetic moment give  
\begin{align}
    V_{(3a)}&=\frac{g_A^2(1+\kappa_1)}{4m_NF_\pi^2}\sum_{i\neq j\neq k}\frac{\boldsigma_k\cdot\boldsymbol{q}_k}{\boldsymbol{q}_i^2(\boldsymbol{q}_k^2+m_\pi^2)}\boldsymbol{q}_i\cdot(\boldsigma_i\times\boldsigma_j)\tau_i^+(\boldsymbol{\tau}_k\times\boldsymbol{\tau}_j)^+ \, ,
    \\
    V_{(3b)}&=\frac{g_A^2(1+\kappa_1)}{4m_NF_\pi^2}\sum_{i\neq j\neq k}\frac{(\boldsigma_j\cdot\boldsymbol{q}_j)(\boldsigma_k\cdot\boldsymbol{q}_k)}{\boldsymbol{q}_i^2(\boldsymbol{q}_j^2+m_\pi^2)(\boldsymbol{q}_k^2+m_\pi^2)}\boldsigma_i\cdot(\boldsymbol{q}_i\times\boldsymbol{q}_j)\tau_i^+(\boldsymbol{\tau}_k\times\boldsymbol{\tau}_j)^+  \, ,\\
        V_{(3d)\kappa_1}&=-\frac{g_A^2(1+\kappa_1)}{4m_NF_\pi^2}\sum_{i\neq j\neq k}\frac{\boldsigma_k\cdot\boldsymbol{q}_k}{\boldsymbol{q}_i^2(\boldsymbol{q}_k^2+m_\pi^2)}\boldsymbol{q}_i\cdot(\boldsigma_i\times\boldsigma_j)\tau_i^+(\boldsymbol{\tau}_k\times\boldsymbol{\tau}_j)^+ \, .
\end{align}
We therefore see that the contributions in diagrams $(3a)$ and $(3d)$ cancel, leaving behind only the ``pion-in-flight'' term.

Diagram $(3c)$ is induced by the LNV coupling of two $\Delta$ to two electrons in  Eq. \eqref{eq:f1} and is given by:
\begin{align}\label{eq:potf45}
    \begin{split}
        V_{(3c)}&=\frac{2g_A^2h_A^2(f^\Delta_4+f^\Delta_5)}{9F_\pi^2\Delta^2}\sum_{i\neq j\neq k}\frac{(\boldsigma_i\cdot\boldsymbol{q}_i)(\boldsigma_k\cdot\boldsymbol{q}_k)}{(\boldsymbol{q}^2_i+m_\pi^2)(\boldsymbol{q}_k^2+m_\pi^2)}\left(\boldsymbol{q}_i\cdot\boldsymbol{q}_k\tau_i^+\tau_k^++\frac{1}{2}\boldsigma_j\cdot(\boldsymbol{q}_i\times\boldsymbol{q}_k)\tau_j^+(\boldsymbol{\tau}_i\times\boldsymbol{\tau}_k)^+\right) \, .
    \end{split}
\end{align}
The spin-isospin structure in Eq. \eqref{eq:potf45} are the same as for $V_{(2b)}$ and $V_{(2c)}$, but the operator has shorter-range, with $\spacevec{q}_i^{-2} \rightarrow (\Delta \Lambda_\chi)^{-1}$.

Diagrams $(3d)$ and $(3e)$ receive contributions from the coupling of the nucleon to the axial current and a pion, induced by the operators $c_3$ and $c_4$ in Eq. \eqref{Eq:LagPiN_nlo},
while $(3f)$ and $(3g)$ from two-pion two-nucleon couplings, proportional to $c_{1,3,4}$, followed by a coupling of at least one axial current to the pion. These diagrams give
\begin{align}
    V_{(3d-3g)c_1}&=\frac{4g_A^2m_\pi^4c_1}{F_\pi^2}\sum_{i\neq j\neq k}\frac{(\boldsigma_k\cdot\boldsymbol{q}_k)(\boldsigma_i\cdot\boldsymbol{q}_i)}{\boldsymbol{q}_i^2(\boldsymbol{q}_k^2+m_\pi^2)(\boldsymbol{q}_i^2+m_\pi^2)^2}\tau_i^+\tau_k^+ \, ,
\\
    \begin{split}
        V_{(3d-3g)c_3}&=\frac{2g_A^2c_3}{F_\pi^2}\sum_{i\neq j\neq k}\frac{\boldsigma_k\cdot\boldsymbol{q}_k}{\boldsymbol{q}_i^2(\boldsymbol{q}_k^2+m_\pi^2)}\left(\boldsigma_i\cdot\boldsymbol{q}_k-(\boldsigma_i\cdot\boldsymbol{q}_i)(\boldsymbol{q}_i\cdot\boldsymbol{q}_k)\frac{\boldsymbol{q}_i^2+2m_\pi^2}{(\boldsymbol{q}_i^2+m_\pi^2)^2}\right)\tau_i^+\tau_k^+ \, , 
    \end{split}
    \\
    \begin{split}V_{(3d-3g)c_4}&=\frac{g_A^2c_4}{F_\pi^2}\sum_{i\neq j\neq k}\frac{\boldsigma_k\cdot\boldsymbol{q}_k}{\boldsymbol{q}_i^2(\boldsymbol{q}_k^2+m_\pi^2)}\left(-\boldsymbol{q}_k\cdot(\boldsigma_i\times\boldsigma_j)\phantom{\frac{1}{1}}\right.\\&\left.-(\boldsigma_i\cdot\boldsymbol{q}_i)\boldsigma_j\cdot(\boldsymbol{q}_i\times\boldsymbol{q}_k)\frac{\boldsymbol{q}_i^2+2m_\pi^2}{(\boldsymbol{q}_i^2+m_\pi^2)^2}\right)\tau_i^+(\boldsymbol{\tau}_k\times\boldsymbol{\tau}_j)^+ \, .
    \end{split}
\end{align}
We notice that the N$^2$LO potentials $V_{(2a)}$ to $V_{(2d)}$ could be obtained by replacing 
\begin{equation}
    c_3 \rightarrow c_3 - \frac{4}{9} \frac{h_A^2}{\Delta}\, , \qquad  c_4 \rightarrow c_4 + \frac{2}{9} \frac{h_A^2}{\Delta}\, , 
\end{equation}
corresponding to the $\Delta$ saturation hypothesis for these two LECs.

Finally, diagrams  $(3h)$ to $(3k)$ are proportional to $c_D$
\begin{align}
    \begin{split}
        V_{(3h-3k)c_D}&=-\frac{g_Ac_D}{2\Lambda_\chi F_\pi^2}\sum_{i\neq j\neq k}\frac{1}{\boldsymbol{q}_i^2}\left(\boldsigma_i\cdot\boldsigma_k-(\boldsigma_i\cdot\boldsymbol{q}_i)(\boldsigma_k\cdot\boldsymbol{q}_i)\frac{\boldsymbol{q}_i^2+2m_\pi^2}{(\boldsymbol{q}_i^2+m_\pi^2)^2}\right)\tau_i^+\tau_k^+  \, .
    \end{split}
\label{eq:cD}
\end{align}

Neglecting recoil terms, the NLO $\Delta$-$N$
Lagrangian contains a nucleon-$\Delta$ transition magnetic moment, $b_1$, and two operators, $b_{4,5}$,
which induce couplings of a nucleon and a $\Delta$ baryon to two pions, or to one pion and one axial current
\cite{Hemmert:1996rw,Hemmert:1997ye,Siemens:2020vop}.
$b_{4,5}$ do not induce three-nucleon diagrams at tree level. After a $b_{4,5}$ interaction, the $\Delta$  needs to be converted back into a nucleon, resulting in a one-loop configuration. $b_1$ could induce diagrams as those in Fig. \ref{fig:3body_n2lo}, with the nucleon-$\Delta$-axial current vertex replaced by $b_1$. When this vertex is contracted with the neutrino propagator and the LO nucleon vector current, the diagram vanishes. If we contract with the LO nucleon axial current, the resulting neutrino potential is parity odd, and thus vanishes in $0^+$ nuclei. The NLO $\Delta$-$N$ Lagrangian therefore does not contribute to the three-nucleon potential at N$^3$LO.

\subsection{Three-nucleon neutrino potential with short-ranged contributions from $g_\nu^{\mathrm{NN}}$}\label{Sec:4c}

The $\abs{\Delta L}=2$ contact interaction of Eq. (\ref{C12def}) can be expanded in pion fields to construct three-nucleon neutrino potentials. The first contribution is the tree-level diagram of Fig. \ref{gnuNNdiagrams}, which appears at N$^2$LO in WPC.  This potential has odd parity and therefore does not contribute to $0^+\rightarrow0^+$ transitions. 

\begin{figure}[t]
    \centering
    \begin{minipage}{0.22\textwidth}
    \scalebox{.75}{
    \begin{tikzpicture}[line width=1.2 pt]
    \begin{feynman}
    \vertex (a);
    \vertex[right=1 of a] (b);
    \vertex[right=2 of b] (c);
    \vertex[right=1 of c] (d);

    \vertex[below=1 of a] (e);
    \vertex[right=1 of e] (f);
    \vertex[right=2 of f] (g);
    \vertex[right=1 of g] (h);

    \vertex[below=2 of e] (i);
    \vertex[right=1 of i] (j);
    \vertex[right=2 of j] (k);
    \vertex[right=1 of k] (l);

    \vertex[below=1.2 of b] (m);

    \vertex[below=0.4 of c] (n);
    \vertex[below=1.6 of c] (o);

    \vertex[right=1 of b] (p);
    \vertex[below=2 of p] (q);
    \vertex[below=2 of q] (r);

    \vertex[right=1 of n] (s);
    \vertex[right=1 of o] (t);

    \vertex[right=1 of p] (pion1) {};
    \vertex[left=1 of p] (pion2) {};

    \vertex[right=1.925 of a, dot] (pion2) {};
    \vertex[right=2 of a] (pion1);

    \vertex[above=.6 of l] (e1);
    \vertex[below=.6 of h] (e2);
    
    \diagram*{
        (a) -- [] (b);
        (b) -- [] (c);
        (c) -- [] (d);

        (e) -- [] (l);
        (i) -- [] (h);


        (q) -- [fermion] (e1);
        (q) -- [fermion] (e2);
        (pion1) -- [dashed] (q);
        };

    \vertex[below=0.925 of p] (dot1) {};
    \vertex[right=-0.075 of p, dot] (dot);
    \vertex[right=-0.075 of q, dot] (dot);
    \vertex[right=-0.075 of q, dot] (dot) {};

    \end{feynman}
\end{tikzpicture}}
\centering
\caption{(4a)}
\label{fig:3a}
\end{minipage}\hspace{10pt}
\begin{minipage}{0.22\textwidth}
    \scalebox{.75}{
    \begin{tikzpicture}[line width=1.2 pt]
    \begin{feynman}
    \vertex (a);
    \vertex[right=1 of a] (b);
    \vertex[right=2 of b] (c);
    \vertex[right=1 of c] (d);

    \vertex[below=1 of a] (e);
    \vertex[right=1 of e] (f);
    \vertex[right=2 of f] (g);
    \vertex[right=1 of g] (h);

    \vertex[below=2 of e] (i);
    \vertex[right=1 of i] (j);
    \vertex[right=2 of j] (k);
    \vertex[right=1 of k] (l);

    \vertex[below=1.2 of b] (m);

    \vertex[below=0.4 of c] (n);
    \vertex[below=1.6 of c] (o);

    \vertex[right=1 of b] (p);
    \vertex[below=2 of p] (q);
    \vertex[below=2 of q] (r);

    \vertex[right=1 of n] (s);
    \vertex[right=1 of o] (t);

    \vertex[right=1 of p, dot] (pion1) {};
    \vertex[left=1 of p, dot] (pion2) {};

    \vertex[above=.6 of l] (e1);
    \vertex[below=.6 of h] (e2);
    
    \diagram*{
        (a) -- [] (b);
        (b) -- [] (c);
        (c) -- [] (d);

        (e) -- [] (l);
        (i) -- [] (h);

        (pion1) -- [dashed] (q);
        (pion2) -- [dashed] (q);

        (q) -- [fermion] (e1);
        (q) -- [fermion] (e2);
        };

    \vertex[right=-0.075 of p, dot] (dot);
    \vertex[right=-0.075 of q, dot] (dot);
    \vertex[right=-0.075 of q, dot] (dot) {};

    \end{feynman}
\end{tikzpicture}}
\centering
\caption{(4b)}
\end{minipage}\hspace{10pt}
\begin{minipage}{0.22\textwidth}
    \scalebox{.75}{
    \begin{tikzpicture}[line width=1.2 pt]
    \begin{feynman}
    \vertex (a);
    \vertex[right=1 of a] (b);
    \vertex[right=2 of b] (c);
    \vertex[right=1 of c] (d);

    \vertex[below=1 of a] (e);
    \vertex[right=1 of e] (f);
    \vertex[right=2 of f] (g);
    \vertex[right=1 of g] (h);

    \vertex[below=2 of e] (i);
    \vertex[right=1 of i] (j);
    \vertex[right=2 of j] (k);
    \vertex[right=1 of k] (l);

    \vertex[below=1.2 of b] (m);

    \vertex[below=0.4 of c] (n);
    \vertex[below=1.6 of c] (o);

    \vertex[right=1 of b] (p);
    \vertex[below=2 of p] (q);
    \vertex[below=2 of q] (r);

    \vertex[right=1 of n] (s);
    \vertex[right=1 of o] (t);

    \vertex[right=1 of p] (pion1) {};
    \vertex[left=1 of p] (pion2) {};

    \vertex[right=1.925 of a, dot] (pion2) {};
    \vertex[right=2 of a] (pion1);

    \vertex[above=.6 of l] (e1);
    \vertex[below=.6 of h] (e2);
    
    \diagram*{
        (a) -- [] (b);
        (b) -- [] (c);
        (c) -- [] (d);

        (e) -- [] (l);
        (i) -- [] (h);


        (q) -- [fermion] (e1);
        (q) -- [fermion] (e2);
        (q) -- [dashed, quarter left] (pion1);
        (q) -- [dashed, quarter right] (pion2);
        };

    \vertex[below=0.925 of p] (dot1) {};
    \vertex[right=-0.075 of p, dot] (dot);
    \vertex[right=-0.075 of q, dot] (dot);
    \vertex[right=-0.075 of q, dot] (dot) {};

    \end{feynman}
\end{tikzpicture}}
\centering
\caption{(4c)}
\label{fig:3a}
\end{minipage}\hspace{10pt}
    \begin{minipage}{0.22\textwidth}
    \scalebox{.75}{
    \begin{tikzpicture}[line width=1.2 pt]
    \begin{feynman}
    \vertex (a);
    \vertex[right=1 of a] (b);
    \vertex[right=2 of b] (c);
    \vertex[right=1 of c] (d);

    \vertex[below=1 of a] (e);
    \vertex[right=1 of e] (f);
    \vertex[right=2 of f] (g);
    \vertex[right=1 of g] (h);

    \vertex[below=2 of e] (i);
    \vertex[right=1 of i] (j);
    \vertex[right=2 of j] (k);
    \vertex[right=1 of k] (l);

    \vertex[below=1.2 of b] (m);

    \vertex[below=0.4 of c] (n);
    \vertex[below=1.6 of c] (o);

    \vertex[right=1 of b] (p);
    \vertex[below=2 of p] (q);
    \vertex[below=2 of q] (r);

    \vertex[right=1 of n] (s);
    \vertex[right=1 of o] (t);

    \vertex[right=1 of p] (pion1) {};
    \vertex[left=1 of p] (pion2) {};

    \vertex[right=1.925 of a, dot] (pion2) {};
    \vertex[right=2 of a] (pion1);

    \vertex[above=.6 of l] (e1);
    \vertex[below=.6 of h] (e2);
    \vertex[below=2.45 of b, dot] (pion3) {};
    
    \diagram*{
        (a) -- [] (b);
        (b) -- [] (c);
        (c) -- [] (d);

        (e) -- [] (l);
        (i) -- [] (h);


        (q) -- [fermion] (e1);
        (q) -- [fermion] (e2);
        (pion1) -- [dashed] (q);
        (pion3) -- [dashed, half right] (q);
        };

    \vertex[below=0.925 of p] (dot1) {};
    \vertex[right=-0.075 of p, dot] (dot);
    \vertex[right=-0.075 of q, dot] (dot);
    \vertex[right=-0.075 of q, dot] (dot) {};

    \end{feynman}
\end{tikzpicture}}
\centering
\caption{(4d)}
\label{fig:3a}
\end{minipage}\hspace{10pt}
\setcounter{figure}{3} 
\captionsetup{labelformat=default}
    \caption{Diagrams proportional to the LEC $g_\nu^{\mathrm{NN}}$.}
\captionsetup{labelformat=empty}
\label{gnuNNdiagrams}
\end{figure}

Recently, Ref. \cite{Cirigliano:2024ocg} identified a new set of three-nucleon forces which appear at N$^5$LO in WPC but are enhanced by renormalization and by a factor of $\pi$ due to their loop structure, promoting them by two/three orders of magnitude. After expanding Eq. \eqref{C12def} keeping terms with up to two pions, diagram $(4b)$ can be constructed in analogy with the $D_2$ operator of Ref. \cite{Cirigliano:2024ocg}, and its potential is provided in Eq. \eqref{eq:3b}. The third diagram $(4c)$ vanishes due to the isospin antisymmetry of the LO $\Bar{N}N\pi\pi$ vertex, and the fourth diagram $(4d)$ vanishes due to an odd power of loop momentum. There are other contributions up to this order that we are not considering here. We only evaluate these set of loop diagrams due to their expected importance relative to their position in WPC. We obtain
\begin{align}
    V_{(4b)g_\nu^{\mathrm{NN}}}&=-\frac{3m_\pi g_A^2g_\nu^{\mathrm{NN}}}{8\pi F_\pi^4}\sum_{i\neq j\neq k}\left(1+\left(\frac{2m_\pi^2+\boldsymbol{q}_j^2}{2m_\pi\abs{\boldsymbol{q}_j}}\right)\arctan\left(\sqrt{\frac{\boldsymbol{q}_j^2}{4m_\pi^2}}\right)\right)\tau_i^+\tau_k^+\label{eq:3b},
\end{align}
where the loop function is the same as in Ref. \cite{Cirigliano:2024ocg}. 
Counting $g_{\nu}^{\rm NN} \sim F_\pi^{-2}$, and $c_{3,4} \sim \Lambda_\chi^{-1}$,  this contribution is suppressed by 
\begin{equation}
    \frac{V_{(4b) g_{\nu}^{\rm NN}} }{V_{(3d-3g)c_{3,4}} } = \mathcal O\left(\pi\frac{Q}{\Lambda_\chi}\right),
\end{equation}
and thus, depending on how seriously we want to consider the $\pi$ enhancement, it could be numerically comparable with the contributions of $c_{3,4}$.   

\subsection{Coordinate-space potentials}\label{coordspace}

For use in QMC and other many-body methods, we provide the coordinate-space expressions of the three-nucleon neutrino potentials. The full list of these potentials, which we denote by $\tilde{V}$, can be found in Appendix \ref{sec.A}. These expressions can naturally be grouped by their spin-isospin operator structure, analogous to the decomposition of the two-nucleon potentials into Fermi, Gamow-Teller, and Tensor components. In the case of the three-nucleon potentials of this work, there are additional operators built from Pauli matrices and the unit vectors between pairs of nucleons. Each of these three-nucleon operators can be rewritten as a commutator or anticommutator of Gamow-Teller and Tensor operators acting on only a pair of nucleons, as done in, for example, Ref. \cite{Carlson:1982} for the standard three-nucleon forces. For example, one such decomposition is
\begin{align}
    (\boldsigma_2\cdot\hat{\boldsymbol{r}}_{12})(\boldsigma_3\cdot\boldsigma_1\times\hat{\boldsymbol{r}}_{12})=\frac{1}{6i}[(S_{12}+\boldsigma_1\cdot\boldsigma_2),\boldsigma_1\cdot\boldsigma_3] \, ,
\end{align}
where the Tensor operator in coordinate-space is defined as 
\begin{align}
    S_{12}=3(\boldsigma_1\cdot\hat{\boldsymbol{r}}_{12})(\boldsigma_2\cdot\hat{\boldsymbol{r}}_{12})-\boldsigma_1\cdot\boldsigma_2 \, .
\end{align}

In addition to being a potentially easier operator structure to implement in a many-body code, this organization will be important in interpreting the numerical result in the sections that follow. 
In Sec. \ref{Sec:5}, we will present results for the 8 spin-isospin structures
\begin{align}
    &\left\{ (S_{ij} + \boldsigma_i \cdot \boldsigma_j), (S_{jk} + \boldsigma_j \cdot \boldsigma_k)  \right\} \tau^+_i \tau^+_k  ,& \qquad
 &    \left[ (S_{ij} + \boldsigma_i \cdot \boldsigma_j), (S_{jk} + \boldsigma_j \cdot \boldsigma_k)  \right] \left[\isovec{\tau}_j \times \isovec{\tau}_k\right]^+  \tau^+_i \, ,\\
    &\left\{ (S_{ij} + \boldsigma_i \cdot \boldsigma_j),  \boldsigma_j \cdot \boldsigma_k  \right\} \tau^+_i \tau^+_k \, ,& \qquad
 &    \left[ (S_{ij} + \boldsigma_i \cdot \boldsigma_j),  \boldsigma_j \cdot \boldsigma_k  \right] \left[\isovec{\tau}_j \times \isovec{\tau}_k\right]^+  \tau^+_i  \, ,\\
     &\left\{ \boldsigma_i \cdot \boldsigma_j, (S_{jk} + \boldsigma_j \cdot \boldsigma_k)  \right\} \tau^+_i \tau^+_k \, ,& \qquad
 &    \left[ \boldsigma_i \cdot \boldsigma_j, (S_{jk} + \boldsigma_j \cdot \boldsigma_k)  \right] \left[\isovec{\tau}_j \times \isovec{\tau}_k\right]^+  \tau^+_i \, ,\\    &\left\{ \boldsigma_i \cdot \boldsigma_j,  \boldsigma_j \cdot \boldsigma_k  \right\} \tau^+_i \tau^+_k \, ,& \qquad
 &    \left[ \boldsigma_i \cdot \boldsigma_j, \boldsigma_j \cdot \boldsigma_k  \right] \left[\isovec{\tau}_j \times \isovec{\tau}_k\right]^+  \tau^+_i \, ,
\end{align}
which can be read off the momentum space expressions in Sections \ref{Sec:4a}
and \ref{Sec:4b} by using the identities listed in Appendix \ref{Sec:B}.

Lastly, there are a small number of radial functions, which depend only on the magnitude of the pair distances within a triplet, that arise in these coordinate-space potentials. We define these functions as
\begin{align}
    Y(r)&=\frac{e^{-m_\pi r}}{m_\pi r}, &&
    \;\;T(r)=Y(r)\left(1+\frac{3}{m_\pi r}+\frac{3}{m_\pi^2 r^2}\right),
    \\
    W_1(r)&=-T(r)+Y(r)-\frac{4\pi}{m_\pi^3}\delta^{(3)}(r), &&
    W_2(r)=\frac{3}{m_\pi^3 r^3}-T(r),
\end{align}
where $Y(r)$ is the well-known Yukawa function from which $T(r)$, $W_1(r)$, and $W_2(r)$ are built. The functions $T(r)$ and $W_1(r)$ are approximately equal in magnitude and opposite in sign for large $r$, which causes cancellations between certain operators in many of the potentials. This is discussed in greater detail in Sec. \ref{reduced}.

\subsection{Comparison with the literature}\label{Sec:4d}

Two-body currents in $0\nu\beta\beta$ were first considered in Ref. \cite{Menendez:2011qq} by normal ordering chiral two-body operators with respect to a spin and isospin symmetric reference state to obtain an effective one-body density. Constructing this effective density using the $c_3$, $c_4$, and $c_D$ contributions to the two-body currents, the authors found a $10\%$ enhancement to $35\%$ reduction of the NMEs in medium and heavy nuclei.

Ref. \cite{Wang:2018htk} performed a more complete analysis by directly constructing products of one- and two-body chiral currents rather than resorting to effective one-body densities, finding a smaller ($\sim10\%$) quenching of the NMEs. The expected leading component of these three-body operators, as considered in Ref. \cite{Wang:2018htk}, are products of the one-body current with short- or pion-ranged two-body currents, i.e. diagrams $(2d)$ and $(2h)$ in this work. 
The main differences of our derivation of the three-nucleon neutrino potential  with these works are:
\begin{itemize}
    \item We include the $\Delta$-resonance as an explicit degree of freedom, which shifts contributions from the pion-ranged two-body current by one order. The effect of the $\Delta$ is implicitly included in Ref. \cite{Menendez:2008jp,Walzl:2000cx} in the saturation of LECs $c_3$ and $c_4$.  With the N$^2$LO potentials having a similar operator structure to the $c_3$ and $c_4$ potentials, one would expect the inclusion of these $\Delta$s to simply improve the order-by-order chiral expansion rather than change the physics;
\item Refs. \cite{Menendez:2011qq,Wang:2018htk} explicitly include 
the contributions of the diagrams 
$(3d)$ and $(3h)$ in Fig. \ref{fig:n3lo}. However, they do not include the iteration of these diagrams with additional pseudoscalar insertions of pions coupled to the neutrino, as shown in diagrams $(3e-3g)$ and $(3i-3k)$, which appear at the same order in WPC. 
As we will see in the QMC results for light nuclei, diagrams $(3e-3g)$ and $(3i-3k)$ play an important role in determine the behavior of transitions densities at $r\lesssim 2$ fm, where $r$ denotes the distance between the two decaying neutrons.
We address this with the inclusion of these additional contributions that have not been previously considered.
\item The pion diagrams $(3f)$ and $(3g)$ induce a dependence of the three-nucleon neutrino potential on $c_1$, in addition to $c_3$ and $c_4$. For the nuclei we study in Sec. \ref{Sec:5}, the contribution proportional to $c_1$ turns out to be small;
\item  At N$^3$LO,  $\Delta$ exchanges induce a new qualitative effect, via the LNV $\Delta \Delta e e$ coupling $f_{4}^\Delta + f_{5}^\Delta$. This coupling induces an operator with similar spin-isospin structure as  
$V_{(2b)}$ or $V_{(2d)}$, but shorter range. The LEC $f_{4}^\Delta + f_{5}^\Delta$ is not well known, but assuming $f_{4}^\Delta \sim f_{5}^\Delta$,
this contribution also appears to be small for the nuclei that we study;
\item We include three-nucleon diagrams with contribution from the nucleon isovector anomalous magnetic moment, which were not considered in Refs. \cite{Menendez:2011qq,Wang:2018htk}.
\end{itemize}
In closing, as a check of the consistency of the 3N $0\nu\beta\beta$ potentials and their Fourier transforms with other electroweak operators, we can examine the form of the short-ranged contact contribution in these potentials. The axial current has a two-body contact as N$^3$LO whose size, when written in coordinate-space, is determined by the adimensional LEC \cite{Baroni:2018fdn}
\begin{align}
    z_0\propto -\frac{m_\pi}{4g_A\Lambda_\chi}c_D+\frac{m_\pi}{3}(c_3+2c_4)+\frac{m_\pi}{6m_N}.
\end{align}
Upon Fierz transforming the coordinate-space potential $\tilde{V}_{(3d-3g)c_3}$ given in Appendix \ref{sec.A}, this same combination of $c_D$, $c_3$, and $c_4$ appears in the LNV operators of $\tilde{V}_{(3d-3g)c_3}$, $\tilde{V}_{(3d-3g)c_4}$, $\tilde{V}_{(3d-3g)c_D}$, which are proportional to $\delta^{(3)}(r_{jk})$.

\section{Calculations in light nuclei}\label{Sec:5}

\subsection{Variational Monte Carlo Method}\label{Sec:5a}

The neutrinoless double beta decay matrix elements are evaluated using Variational Monte Carlo (VMC) \cite{Carlson:2014vla} based on the many-body Hamiltonian

\vspace{-10pt}
\begin{align}
    H=\sum_i K_i+\sum_{i<j}v_{ij}+\sum_{i<j<k}V_{ijk} \, ,
\end{align}
where $K_i$ is the non-relativistic nucleon kinetic energy, and $v_{ij}$ and $V_{ijk}$ are the two- and three-body Norfolk (NV2+3) potentials
~\cite{Piarulli:2014bda,Piarulli:2016vel,Piarulli:2020,Piarulli:2017dwd,Baroni:2018fdn,Baroni:2016,Piarulli:2019cqu,Piarulli:2022hml}.
The variational ansatz for the many-body wave function is

\vspace{-15pt}
\begin{align}    |\Psi_T\rangle=\mathcal{S}\prod_{i<j}\left[1+U_{ij}+\sum_{i<j\neq k}\tilde{U}_{ijk}^{\mathrm{TNI}}\right]|\Psi_J\rangle \, ,
\end{align}

\noindent where $\mathcal{S}$ is the symmetrization operator, $U_{ij}$ is the two-body correlation operator, $\tilde{U}_{ijk}^{\mathrm{TNI}}$ is the three-body correlation operator, and $\Psi_J$ is a Jastrow-like wave function. The antisymmetric state $\Psi_J$ is built from correlation functions, which encode the long-ranged cluster structure of the nucleus, acting on a slater determinant that places nucleons in $s$- and $p$-shell orbitals. The correlation operators contain the variational parameters which are varied to optimize the wave function, $\Psi_T$, by minimizing the expectation value

\vspace{-10pt}
\begin{align}    E_T=\frac{\langle\Psi_T|H|\Psi_T\rangle}{\langle\Psi_T|\Psi_T\rangle}\geq E_0 \, .
\end{align}

The $^6\mathrm{He}$ wave function has spatial symmetries $^1\mathrm{S}_0$[2] and $^3\mathrm{P}_0$[11], where $[n]$ denotes the Young pattern~\cite{Wiringa:2006ih}. For $^6\mathrm{Be}$, the charge-symmetric wave function of $^6\mathrm{He}$ is used, which results in large transition densities with no nodal structure. The amplitudes of the two (five) spatial symmetry components of $^8\mathrm{He}$ and $^{12}\mathrm{Be}$ ($^8\mathrm{Be}$ and $^{12}\mathrm{C}$) are given in Tables \ref{tab:wfs8} and \ref{tab:wfs12}. The small overlap between the dominant spatial symmetry components of the initial and final nuclei in the $^8\mathrm{He}\rightarrow {^8}\mathrm{Be}$ and $^{12}\mathrm{Be}\rightarrow {^{12}}\mathrm{C}$ transitions results in small transition densities. The matrix elements of these transitions are additionally suppressed due to the nodal structure of their densities, as required by the isospin orthogonality of the wave functions.

\begin{table}[H]
    \centering
    \begin{tabular}{c | c c  c c c}
    \hline
    \hline
      & $^1\mathrm{S}_0$[44] & $^3\mathrm{P}_0$[431] & $^1\mathrm{S}_0$[422] & $^5\mathrm{D}_0$[422] & $^3\mathrm{P}_0$[4211] \\
     \hline
     $^{8}\mathrm{He}$ & \ding{56} & \ding{56} & 0.8749 & \ding{56} & -0.4843\\
     $^{8}\mathrm{Be}$& 0.9959 & -0.0815 & 0.0124 & -0.0313 & 0.0219 \\
     \hline
     \hline
    \end{tabular}
    \caption{Amplitudes of the spatial symmetry components of $A=8$ wave functions for model Ia$^\star$.}
    \label{tab:wfs8}
\end{table}

\begin{table}[H]
    \centering
    \begin{tabular}{c | c c c c c}
    \hline
    \hline
      & $^1\mathrm{S}_0$[444] & $^3\mathrm{P}_0$[4431] & $^1\mathrm{S}_0$[4422] & $^5\mathrm{D}_0$[4422] & $^3\mathrm{P}_0$[4332] \\
     \hline
     $^{12}\mathrm{Be}$ & \ding{56} & \ding{56} & 0.9684 & \ding{56} & 0.2496\\
     $^{12}\mathrm{C}$& 0.9496 & 0.3047 & 0.0602 & 0.0249 & 0.0328\\
     \hline
     \hline
    \end{tabular}
    \caption{Amplitudes of the spatial symmetry components of $A=12$ wave functions for model Ia$^\star$.}
    \label{tab:wfs12}
\end{table}

The VMC wave function could be further refined with an imaginary time propagation, which eliminates contaminations from excited states. The Green's Function Monte Carlo (GFMC) algorithm is one such implementation of this method \cite{Carlson:2014vla}. Previous QMC calculations of electroweak matrix elements indicate that the GFMC method provides a small adjustment ($\lesssim 3\%$) to the VMC solution \cite{Pastore:2017uwc,King:2020wmp} as well as utilizing more refined VMC wave functions having updated spatial symmetry amplitudes~\cite{Weiss:2021rig}, which are beyond the precision needed for this work. 

The Norfolk two-body (NV2) potential~\cite{Piarulli:2014bda,Piarulli:2016vel,Piarulli:2020,Piarulli:2017dwd,Baroni:2018fdn,Baroni:2016,Piarulli:2019cqu,Piarulli:2022hml} is built from interactions up to N$^3$LO, which includes one- and two-pion exchange terms as well as contact interactions parameterized in terms of 26 unknown LECs. These LECs are determined by fitting to $NN$ scattering data \cite{Navarro:2013,Navarro:2014,Navarro:2014_stat}, and the deuteron binding energy. The radial functions in these potentials diverge at the origin with $1/r^n$ behavior for $n\leq 6$. Ensuing divergences are removed by a long-ranged regulator, 

\vspace{-10pt}
\begin{align}
    F(r)=1-\frac{1}{(r/R_L)^6e^{2(r-R_L)/R_L}+1} \, ,\label{eq:RL}
\end{align}

\noindent with long-range cutoff $R_L$. Delta-functions entering the potentials are represented with a Gaussian,

\vspace{-10pt}
\begin{align}
    \delta(r)=\frac{e^{-(r/R_s)^2}}{\pi^{3/2}R_S^3} \, ,
\end{align}

\noindent with short-range cutoff $R_S$, which determines the width of the smearing.

The four classes of NV2 models used in this work are denoted by I (II) for fits to $NN$
scattering data up to 125 (200) MeV and denoted by a (b) for the choice of regulator cutoff [$R_L,R_S$]=[1.2 fm, 0.8 fm] ([$R_L,R_S$]=[1.0 fm, 0.7 fm]). The Norfolk three-body (NV3) potential includes two additional LECs: $c_D$ and $c_E$. Models denoted without a star ($\star$) are fit to triton ground state energy and $nd$ doublet scattering length \cite{Piarulli:2018}, and models with a star are fit to the triton ground state energy and Gamow-Teller $\beta$-decay matrix element \cite{Baroni:2018fdn}. The values of these LECs are provided in Table \ref{tab:table_cD}.

\begin{table}[H]
    \centering
    \begin{tabular}{c | c c c c c}
    \hline
    \hline
      & Ia & Ia$^\star$ & IIb & IIb$^\star$\\
     \hline
     $c_D$ & 3.666 & -0.635 & -4.480 & -5.25 \\
     $c_E$& -1.638 & -0.090 & -0.412 & 0.05 \\
     $z_0$ & 0.090 & 1.035 & 2.806 & 3.059 \\
     \hline
     \hline
    \end{tabular}
    \caption{Adimensional values of LECs $c_D$, $c_E$, and $z_0$ given in the NV3 interactions used in this work.}
    \label{tab:table_cD}
\end{table}

The matrix element between the initial and final nuclear states, $|\Psi_i\rangle$ and $|\Psi_f\rangle$, is defined as

\begin{align}
    M=\langle \Psi_f|O_n|\Psi_i\rangle \, , 
\end{align}
where the two- and three-nucleon operators are given by
\begin{align}
    O_\mathrm{2N}=(4\pi R_A)\sum_{i\neq j}\mathcal{O}_{ij} \, , &&
    O_\mathrm{3N}=(4\pi R_A)\sum_{i\neq j\neq k}\mathcal{O}_{ijk} \, ,
\end{align}
with nuclear radius $R_A=1.2 A^{1/3}$ fm, which makes the matrix element dimensionless. The two- and three-nucleon operators $\mathcal{O}_{ij}$ and $\mathcal{O}_{ijk}$ depend on the spin and isospin of the nucleons as well as the radial distance between nucleons in the pair or triplet. We can study these matrix elements in terms of their density in coordinate space,
\begin{align}
    C(r_{12},r_{13},r_{23})&=\langle \Psi_f|\sum_{i\neq j\neq k}\mathcal{O}_{ijk}\delta(r_{ij}-r_{12})\delta(r_{ik}-r_{13})\delta(r_{jk}-r_{23})|\Psi_i\rangle\, , 
    \\
    M&=(4\pi R_A)\int dr_{12} \;dr_{13}\; dr_{23}\; C(r_{12},r_{13},r_{23}) \, .
\end{align}
In principle, the three-nucleon densities can be constructed with any combination of the three pair distances and three angles within a triplet. However, in analogy with the two-nucleon densities, we show densities in terms of only the pair distances between nucleons, such that integrating the density over these distances recovers the $0\nu\beta\beta$ matrix element. 

For operators $\mathcal{O}_{ijk}$ with the first isospin structure, $\tau_i^+\tau_k^+$, the pair distance $r_{13}$ corresponds to the distance between the two decaying neutrons. For operators of the second isospin structure, $i\tau_i^+(\boldsymbol{\tau}_k\times\boldsymbol{\tau}_j)^+=\tau_i^+(\tau_k^+\tau_j^3-\tau_k^3\tau_j^+)$, the nucleon indices on the second term, along with the indices on the corresponding spin- and radial-component of $\mathcal{O}_{ijk}$, are relabeled such that $r_{13}$ still corresponds to the distance between decaying neutrons. All of the densities shown in this work are binned such that $r_{13}$ corresponds to this pair distance.

\subsection{Results and Discussion}\label{Sec:5b}

In this section, we report calculations of the matrix elements of the three-nucleon $0\nu\beta\beta$ potentials and their densities in coordinate space. Table \ref{tab:MatrixElements} lists two- and three-nucleon matrix elements for $^6\mathrm{He}$, $^8\mathrm{He}$, and $^{12}\mathrm{Be}$ decays. We investigate two kind of transitions based on the change in total isospin between initial and final nuclear states, \textit{i.e.}, $\Delta T=\abs{T_i-T_f}=0,2$. While the $\Delta T=0$ transitions are not realized in the nuclei of experimental interest, we study these decays to benchmark with other many-body methods that can reach heavier nuclei.

Table \ref{tab:MatrixElements} lists VMC matrix elements using four classes of the NV2+3 models for the $A=6,8$ transitions and only two classes of the $A=12$ transition, due to the computational cost of these calculations. The column $M_{\mathrm{\nu,2}}$ denotes the light-neutrino exchange two-nucleon matrix elements which include the Fermi, Gamow-Teller, and Tensor operators with single nucleon form factors \cite{Pastore:2017ofx}. 
 As we use it mostly as a reference point to assess the importance of three-body operators, we
only consider contributions to $M_{\mathrm{\nu,2}}$ arising from the  long-range part of the neutrino potential, effectively setting
 $g_{\nu}^{\rm NN} =0$ in this part of the calculation. The theoretical uncertainties in
 $M_{\mathrm{\nu,2}}$, coming from the incomplete knowledge of $g_\nu^{\rm NN}$ and from missing N$^2$LO two-body potential, will affect the exact ratio of three- and two-body matrix elements, but they will not significantly change the discussion below. 
The matrix elements of the two-nucleon potentials were not regulated with the long-ranged regulator of Eq.~(\ref{eq:RL}). Including this regulator in only the three-nucleon matrix elements thus allows us to isolate the model dependence of the three-nucleon contributions in relation to the leading-order two-nucleon matrix element, as demonstrated in Fig. \ref{fig:ratio}.

\begin{table}[]
    \centering
    \resizebox{\textwidth}{!}{%
    \begin{tabular}{c|c|c|c|cccccc}
    \hline
    \hline
    \multirow{2}{4em}{$T_i\rightarrow T_f$} & \multirow{2}{4em}{\centering Model} & \multicolumn{1}{c}{$M_{\nu,2}^{\mathrm{}}$} & \multicolumn{1}{|c|}{$M_{\nu,3}^{\mathrm{N^2LO}}\;(10^{-2})$} & \multicolumn{6}{c}{$M_{\nu,3}^{\mathrm{N^3LO}}\;(10^{-2})$}\\
    \cline{3-10}
         &  & $M_{\mathrm{F+GT+T}}$ & $M_{\mathrm{h_A^2}}$ & $M_{\kappa_1}^{J=0}$ & $M_{f_4^\Delta+f_5^\Delta}$ & $M_{\mathrm{c_1}}$ & $M_{\mathrm{c_3}}$ & $M_{\mathrm{c_4}}$ & $M_{\mathrm{c_D}}$ \\
        \hline\hline
        \multirow{4}{8em}{\centering$^6$He(1)$\rightarrow$$^6$Be(1)}
         & Ia$\phantom{^\star}$ & 7.629 & 2.316 & 0.334 & -0.082 & 0.062 & 0.798 & -0.614 & 3.157 \\
        & Ia$^\star$ & 7.567 & 1.881 & 0.328 & -0.064 & 0.059 & 0.697 & -0.663 & -0.499 \\
        & IIb$\phantom{^\star}$ & 7.712 & 1.301 & 0.338 & -0.059 & 0.050 & 0.470 & -0.418 & -3.035 \\
        & IIb$^\star$ & 7.769 & 1.395 & 0.349 & -0.062 & 0.052 & 0.518 & -0.496 & -3.828 \\
        \hline
        \multirow{4}{8em}{\centering$^8$He(2)$\rightarrow$$^8$Be(0)}
         &
    Ia$\phantom{^\star}$ & 0.327 & -3.900 & -0.052(1) & -0.065(3) & 0.021 & -1.656 & 2.085 & -2.750 \\
        & Ia$^\star$ & 0.293 & -3.684 & -0.050(1) & -0.059(2) & 0.016 & -1.576 & 2.009 & 0.456 \\
        & IIb$\phantom{^\star}$ & 0.263 & -3.929 & -0.052(1) & -0.031(2) & 0.030 & -1.561 & 1.740 & 3.125 \\
        & IIb$^\star$ & 0.285 & -4.163 & -0.053(1) & -0.030(3) & 0.029 & -1.657 & 1.853 & 3.867 \\
        \hline
        \multirow{2}{8em}{\centering$^{12}$Be(2)$\rightarrow$$^{12}$C(0)}
        & Ia$\phantom{^\star}$ & 1.591(37) & -9.16(62) & 0.054(9) & -0.983(51) & 0.113(7) & -3.61(15) & 3.96(23) & -2.56(19) \\
        & Ia$^\star$ & 1.418(30) & -9.49(63) & 0.064(1)& -0.980(55) & 0.108(6) & -3.67(15) & 3.86(23) & 0.44(4) \\

     \hline
     \hline
    \end{tabular}}
\captionsetup{justification=RaggedRight}
    \caption{Two- and three-nucleon $0\nu\beta\beta$ VMC matrix elements in light nuclei. Statistical uncertainty is shown in parentheses if more than $2\%$, otherwise it is omitted. These results are based on LECs provided in Tables 
    \ref{tab:lecs} and \ref{tab:table_cD}.}
    \captionsetup{justification=centering}
    \label{tab:MatrixElements}
\end{table}

\begin{figure}[t]
\centering
\captionsetup{labelformat=default}
  \centering
  \includegraphics[width=\linewidth]{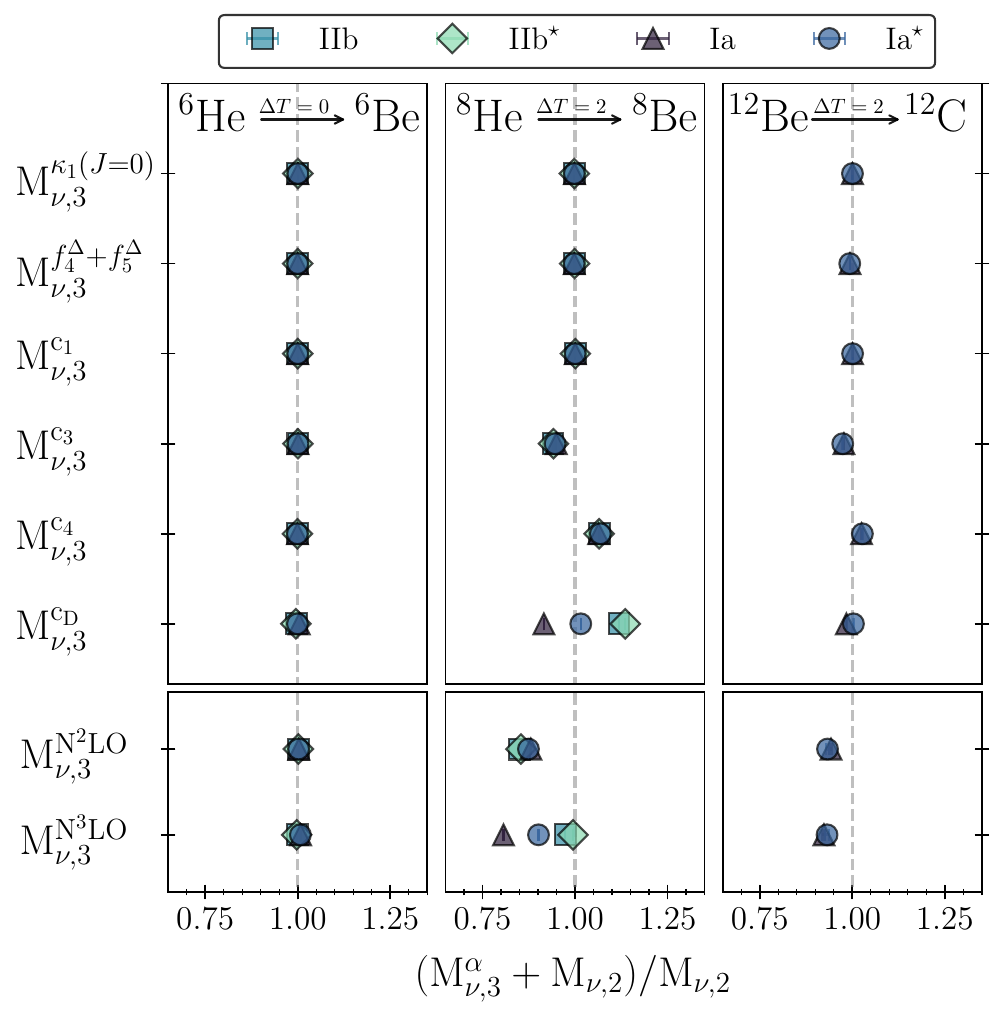}
  \label{fig:ratios}
  \captionsetup{justification=raggedright,singlelinecheck=false}
\caption{Ratios of three- and two-nucleon matrix elements with respect to the two-nucleon matrix elements. $M_{\nu, 2}$ includes LO long-range neutrino exchange diagrams and single nucleon form-factor corrections. The top panels denote the N$^3$LO contributions grouped by LEC. The bottom two panels are the full three-nucleon matrix element contributions up to either N$^2$LO or N$^3$LO.}
\label{fig:ratio}
\captionsetup{justification=centering}
\end{figure}

We report three-nucleon matrix elements separated by the contributions of the different LECs. The N$^2$LO diagrams, each having an intermediate $\Delta$, are all proportional to $h_A^2$. These matrix elements comprise the largest contribution in the three-nucleon sector with only the $c_D$ contribution being comparable in size for the $A=6,8$ transitions. The N$^3$LO diagrams have six LEC contributions, given in the last columns of Table \ref{tab:MatrixElements}. The largest contributions at this order come from $c_3$, $c_4$, and $c_D$. 
There are two approximations in these NMEs. For $M_{f_4^\Delta+f_{5}^\Delta}$, we set $f_5^\Delta=f_4^\Delta=1.6$ GeV since the value of $f_5^\Delta$ is unknown but of natural size $f_i^\Delta=M_\rho F_\pi^{-2}(4\pi)^{-2}$. The Fourier transform of $V_{(3b)}$ is nontrivial. In this work, we consider only the $J=0$ projection of this operator, as detailed in Appendix \ref{Aprojection}.

There is little model variation throughout both N$^2$LO and N$^3$LO three-nucleon matrix elements, except for the $c_D$ contribution. This is expected in the Norfolk model classes because the value of $c_D$ varies both in sign and magnitude, as provided in Table \ref{tab:table_cD}. 
The contributions proportional to $c_3$ and $c_4$ are rather stable across models, but are affected by the uncertainty in the determination of the LECs, which  is about 30\%.
Because of the partial cancellation of the $c_{3,4}$ contributions, this uncertainty is amplified. For example, the correction from $c_3$ and $c_4$ to the $^8$He matrix elements range from $[0.2,0.4] \cdot 10^{-2}$ using the LECs in Table \ref{tab:lecs}, but it is reduced to a value in the  $[0.15,-0.03] \cdot 10^{-2}$ range if one 
uses the extraction of Ref. \cite{Siemens:2016jwj}, reported in
Eq. \eqref{eq:c34M}. Similarly, in the case of $^{12}$Be, the contribution of $c_{3,4}$ would go from $0.35 \cdot 10^{-2}$ ($0.19 \cdot 10^{-2}$) 
to $-0.11 \cdot 10^{-2}$ ($-0.23 \cdot 10^{-2}$)
in model Ia (Ia$^\star$). The same cancellation that amplifies the effect of theory uncertainties, however, also reduces the overall importance of the $c_{3,4}$ contributions, making them subleading compared to the corrections proportional to $h^2_A$ and $c_D$. We notice that our estimate of the uncertainties due to $c_3$ and $c_4$ is only qualitative, as we varied them only in the neutrino potential and not in the two- and three-body forces that determine the nuclear wavefunctions. A similar model dependence has also been observed in single-beta decay in light nuclei~\cite{King:2020wmp,Baroni:2018fdn,King:2024zbv}, where, {\it e.g.}, the two-nucleon contact contribution proportional to $z_0$ undergoes up to a  $\sim 50$ \% variation depending on the adopted Norfolk model.  The model dependence of these predictions may point to the need for improved
three-nucleon interactions. In particular, sub-leading components of the three-nucleon force, having currently only two unconstrained parameters, $c_D$ and $c_E$, may improve the description of low-energy nuclear observables. For example, studies including additional contact LECs in the three-nucleon sector~\cite{Girlanda:2019} found a better agreement with the data, ${\it e.g.}$, for polarization observables of  $N-d$ elastic scattering~\cite{Sekiguchi:2019xvh}.

To understand the sizes of the matrix elements given in Table \ref{tab:MatrixElements} across the different transition types, we explore the densities of these matrix elements in the following sections.

\subsubsection{Three-nucleon densities}

To understand the structure of the three-nucleon transition densities we first compute the simplest operator, having no spin- or radial-dependence, which is contained in each neutrino potential: $\tau^+\otimes\tau^+$. Figure \ref{fig:6Hetautau}\textcolor{blue}{a} shows the three-nucleon density of this operator in all isosceles configurations with $r_{12}=r_{23}$ for the transition $^6\mathrm{He}$ $\rightarrow$ $^6 \mathrm{Be}$. There are two distinct regions in the density for this configuration. Most of the contribution of the density is when $r_{13}\lesssim$ 2 fm and 2 $\lesssim r_{12} \lesssim 6$ fm. In this region, the two valence neutrons of $^6$He are close to each other and the third nucleon in the $\alpha$ core can be far or close to the decaying pair. The second contribution of this density is along the line $r_{13}=2r$. In this configuration, the two valence neutrons are on opposite sides of the $\alpha$ core. Integrating over this three-nucleon density gives 8, as expected by simple counting: the unconstrained sum over $i\neq j\neq k$ adds one for each nucleon in the $\alpha$ core with an overall factor of two for the valence neutrons which can undergo the decay. 

\begin{figure}[t]
\centering
\captionsetup{labelformat=default,labelformat=empty}
\begin{minipage}{.47\textwidth}
  \centering
  \includegraphics[width=\linewidth]{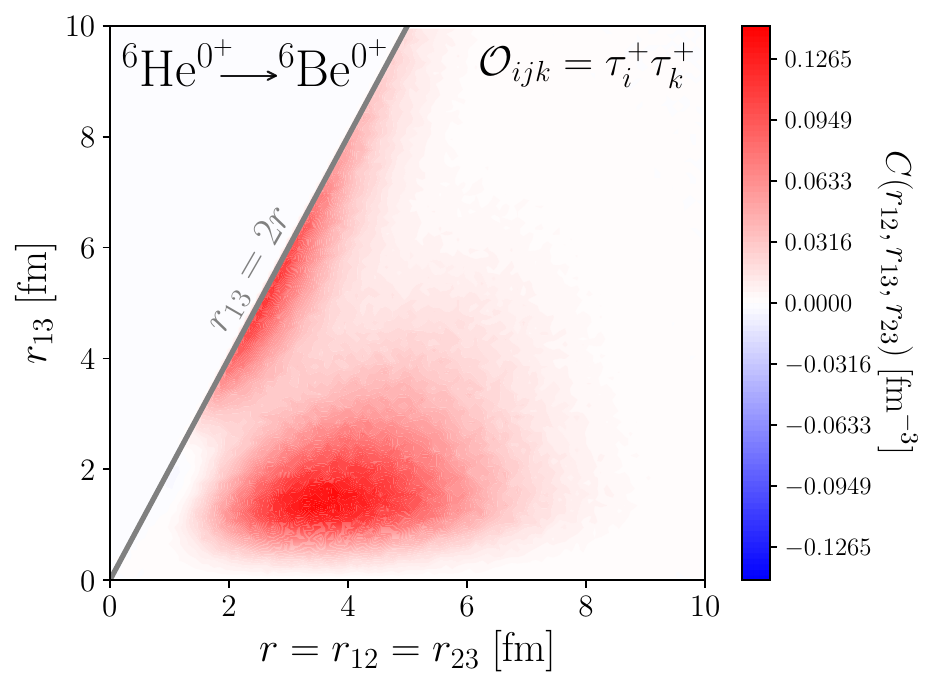}
  \caption{(a)}
  \label{fig:6Hetautau}
\end{minipage}%
\hfill
\begin{minipage}{.47\textwidth}
  \centering
  \includegraphics[width=\linewidth]{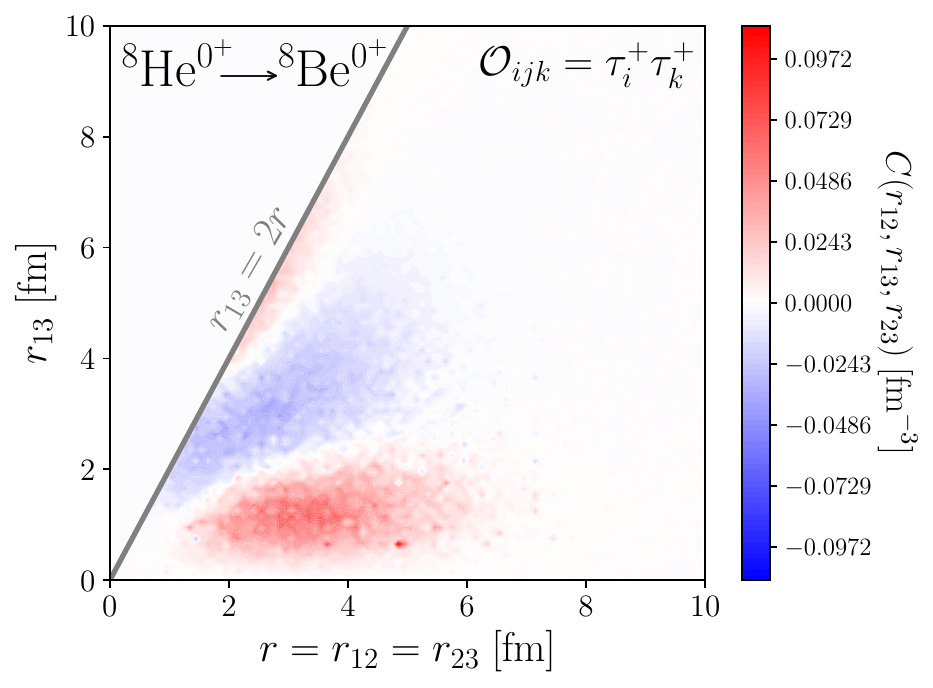}
  \caption{(b)}
  \label{fig:8He_tautau}
\end{minipage}
\setcounter{figure}{5} 
\captionsetup{justification=default,labelformat=default}
\captionsetup{justification=raggedright,singlelinecheck=false}
\caption{Three-nucleon density of $\tau_i^+\tau_k^+$ for the $^6\mathrm{He}$ $\rightarrow$ $^6 \mathrm{Be}$ (a) and $^8\mathrm{He}$ $\rightarrow$ $^8 \mathrm{Be}$ (b) transition for the isosceles configuration $r_{12}=r_{23}$ using model Ia$^\star$. The pair distance $r_{13}$ corresponds to the distance between the two neutrons which decay. The grey line at $r_{13}=2r$ denotes the boundary of the density. All possible configurations of the triplet exist in the parameter space below this line.}
\label{fig:8Hetautau}
\captionsetup{justification=centering}
\end{figure}

Figure \ref{fig:6Hetautau}\textcolor{blue}{b} shows the same isospin density but for $^8\mathrm{He}$ $\rightarrow$ $^8\mathrm{Be}$. The parameter space is also dominated by $2\lesssim r_{12}\lesssim 6$ fm. However, there are now positive and negative regions, denoted by red and blue, due to the nodal structure of the transition density. Integrating over this density gives 0, as expected due to the orthogonality of the spatial components of the intial and final wave function. Introducing additional spin-, isospin-,  and radial-dependence to this operator will break this exact cancellation.

\subsubsection{Reduced three-nucleon densities}\label{reduced}

To compare to the two-nucleon densities, we can integrate over two of the pair distances of the three-nucleon density. This allows us to study the three-nucleon matrix elements as a function of only the distance between the decaying neutrons,

\begin{align}
    C(r)\equiv C(r_{13})=\int dr_{12}\;dr_{23}\; C(r_{12},r_{13},r_{23}) \, .
\end{align}

\begin{figure}[t]
\centering
\captionsetup{labelformat=empty}
\begin{minipage}{.5\textwidth}
  \centering
  \includegraphics[width=\linewidth]{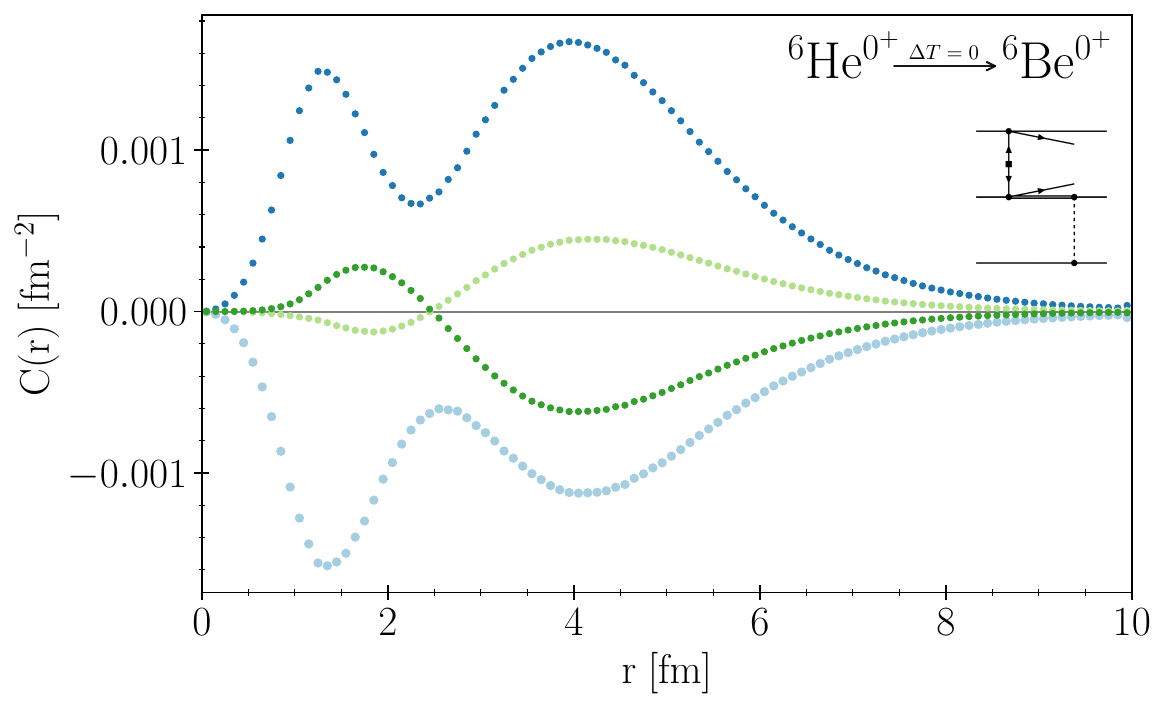}
  \caption{(a)}
  \label{fig:6He_V1a}
\end{minipage}%
\hfill
\begin{minipage}{.5\textwidth}
  \centering
  \includegraphics[width=\linewidth]{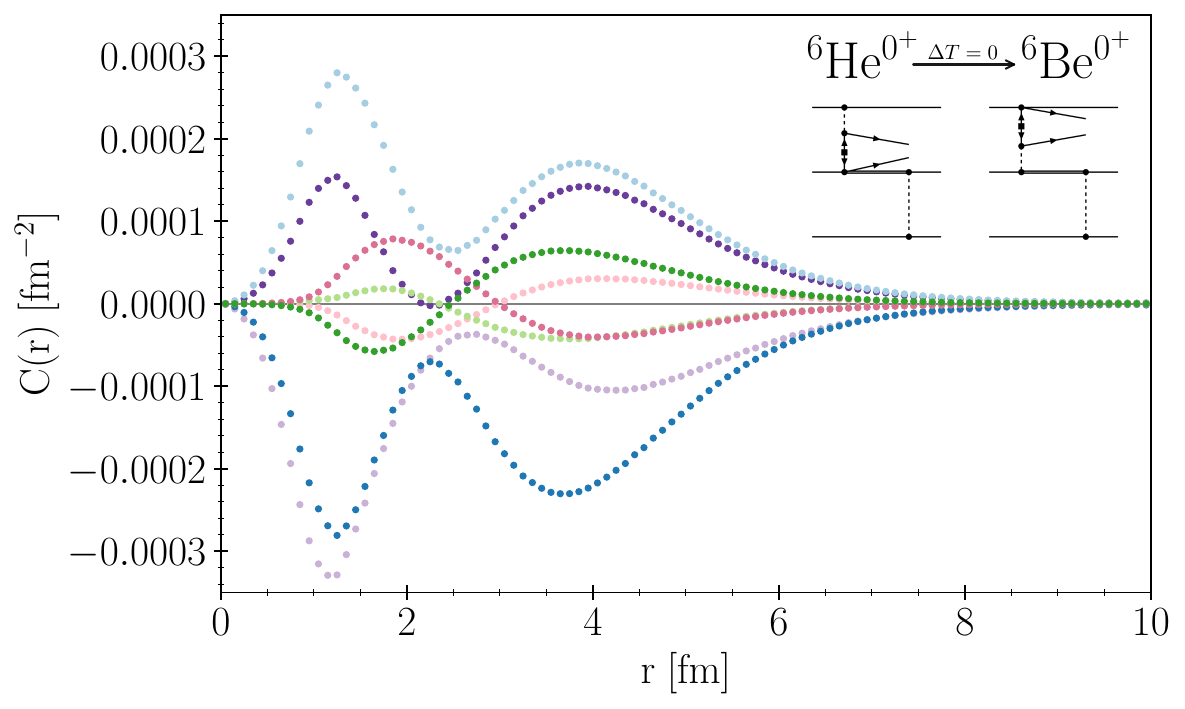}
  \caption{(b)}
  \label{fig:6He_V1b}
\end{minipage}
\\
\begin{minipage}{.9\textwidth}
  \centering
  \includegraphics[width=\linewidth]{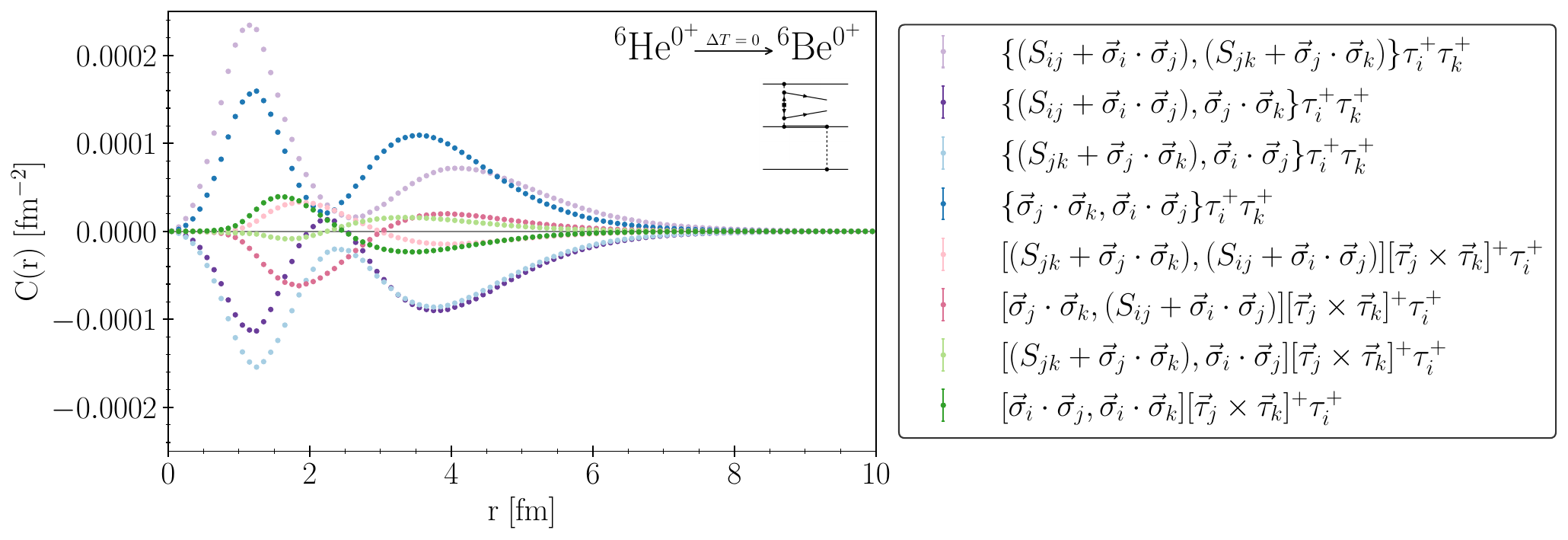}
  \caption{(c)}
  \label{fig:6He_V1d}
\end{minipage}
\setcounter{figure}{6} 
\captionsetup{justification=default,labelformat=default}
\captionsetup{justification=raggedright,singlelinecheck=false}
\caption{Three-nucleon densities of the N$^2$LO potentials decomposed into operator type calculated in $^6\mathrm{He}$ $\rightarrow$ $^6\mathrm{Be}$ using model Ia$^\star$ as a function of the distance between the two decaying neutrons $r$. Plots $a$, $b$, and $c$ show the contribution from diagrams $(2a)$, $(2b)$ and $(2c)$, and $(2d)$. Potential $\tilde{V}_{2a}$ has four spin-isospin operators, which are shown in plot $a$. Potentials $\tilde{V}_{2b}=\tilde{V}_{2c}$ and $\tilde{V}_{2d}$ have these same four operators types and an additional four operators.}
\label{fig:N2LO_he6}
\captionsetup{justification=centering}
\end{figure}

One might expect the leading contribution of the three-nucleon potentials to be from diagram (2a), which has a long-ranged, Coulomb-like interaction between one of the three nucleon pairs due to the light-neutrino exchange. The coordinate-space expression for this potential, which can be found in Appendix \ref{sec.A}, shows four distinct spin-isospin operators multiplied by different radial functions. There exists a large cancellation between these operators, as demonstrated in Fig. \ref{fig:6He_V1a}. This can be also appreciated directly from the form of the potential, whereby the first two and last two terms of \eqref{eq:A6} have similar spin-isospin operators, only differing by a Tensor contribution, and approximately opposite radial functions at large distances.

Fig. \ref{fig:N2LO_he6} shows the densities of different operator structures in the coordinate-space potentials $\tilde{V}_{(2a)}$, $\tilde{V}_{(2b)}$, and $\tilde{V}_{(2d)}$, which correspond to diagrams $(2a)$, $(2b)$, and $(2d)$, respectively. The operators are grouped in pairs, denoted by light and dark colors, based on the operator they cancel with. 

As shown in Fig. \ref{fig:8Hetautau}, there is a large portion of the parameter space available for decays in which the two decaying neutrons are close to each other ($\simle 2$ fm) and far less available space where that decaying pair is within $2$ fm of the third nucleon in the triplet. For diagram $\tilde{V}_{2a}$, whose entire contribution is shown in Fig. \ref{fig:N2LO}, the density between similar operators almost exactly cancels in the region where the two decaying neutrons are within $2$ fm from each other. For a distance of larger than $2$ fm, the cancellation is not as exact, resulting in a large contribution at long range. A similar cancellation between different spin-isospin operators also occurs in the other N$^2$LO diagrams as well as certain N$^3$LO potentials, namely, $\tilde{V}_{(3c)}$, $\tilde{V}_{(3d-3g)c_3}$, and $\tilde{V}_{(3d-3g)c_4}$.


The different radial functions in these operators, $T(x)$ and $W_1(x)$, are functions of the distance between one of the decaying neutrons and the nucleon which does not decay, as indicated in Sec. \ref{coordspace}. For $^6$He $\rightarrow$ $^6$Be, most of these pair distances fall within $2\lesssim r_{23}\lesssim 6$ fm, as demonstrated in Fig. \ref{fig:8Hetautau}. These two functions differ greatly in their short-ranged behavior but are approximately equal in size and opposite in sign within this region, as shown in Figure \ref{fig:N2LO}\textcolor{blue}{b}. These two functions arise from a laplacian acting on the Yukawa function, $i.e.$, a diagram containing a one-pion exchange between two nucleons. For the long-ranged interaction of diagram $(2a)$, the potential allows for contribution within the region where $T(x)\approx-W_1(x)$ as well as at short distance where the two functions differ. On the other hand, the shorter-ranged interactions of diagram ($2b-2d$) biases towards the region of parameter space where $T(x)$ and $-W_1(x)$ differ. The cancellation between operator types is a combination of this difference in radial functions, which grows larger when the decaying neutron pair is close to the third nucleon in the triplet, and Tensor contributions, which are longer-ranged with respect to each of the pair distances in the triplet. The size of the $\tilde{V}_{(2a)}$ contribution with respect to the other N$^2$LO contribution is consistent across the three transitions investigated in this work. However, whether this delicate cancellation, which depends largely on the structure of the transition, will also hold for the systems of experimental interest should be investigated. In principle, the contribution of the diagrams containing intermediate pions could be enhanced with respect to the neutrino-only-exchange diagrams due to structure effects.

\begin{figure}[t]
\centering
\captionsetup{labelformat=empty}
\begin{minipage}{.53\textwidth}
  \centering
  \includegraphics[width=\linewidth]{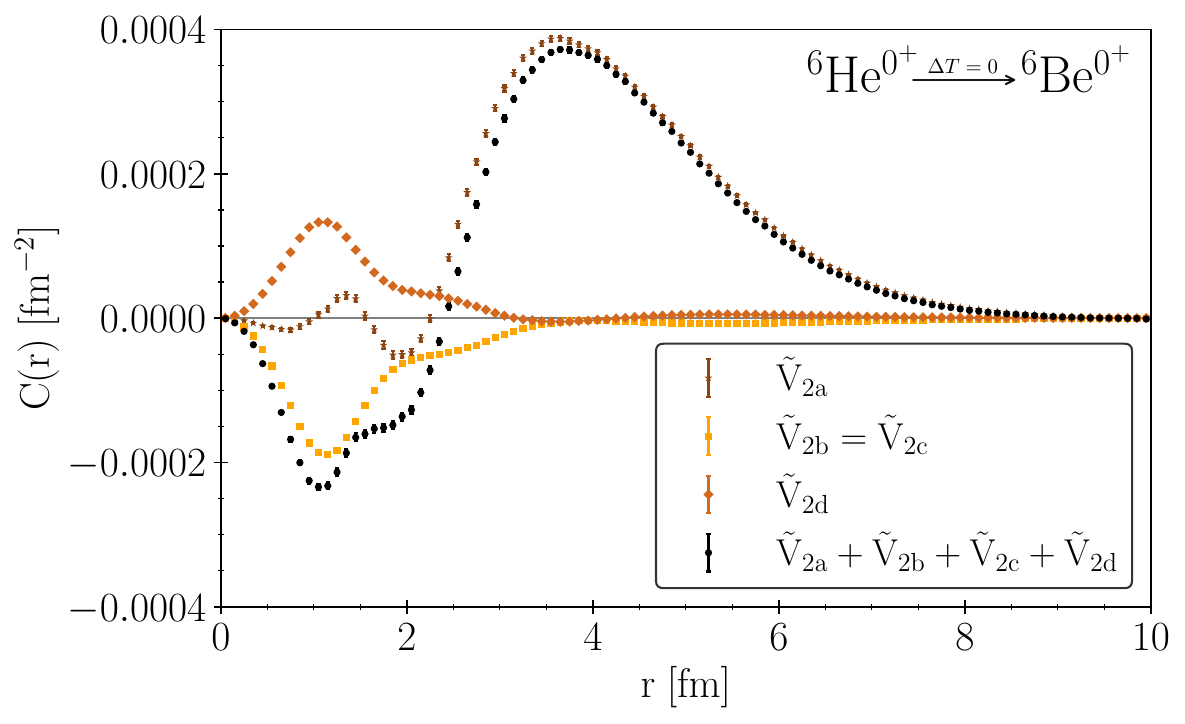}
  \caption{(a)}
  \label{fig:N2LO}
\end{minipage}%
\hfill
\begin{minipage}{.47\textwidth}
  \centering
  \includegraphics[width=\linewidth]{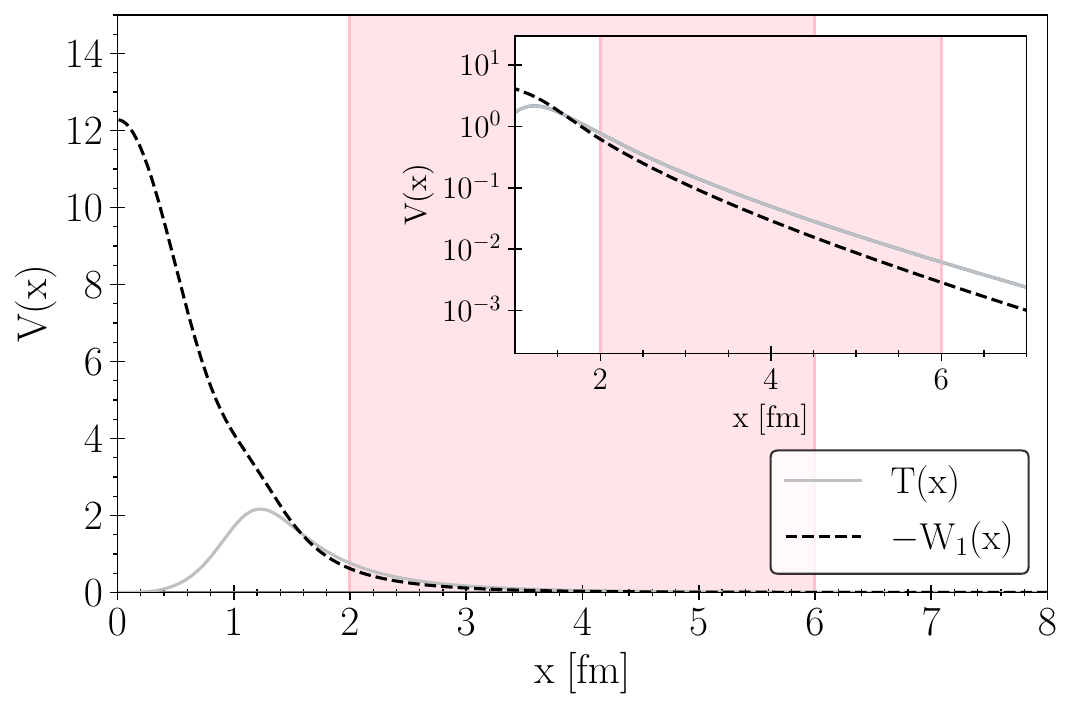}
  \caption{(b)}  \label{fig:radial_comparison}
\end{minipage}
\setcounter{figure}{7} 
\captionsetup{justification=default,labelformat=default}
\captionsetup{justification=raggedright,singlelinecheck=false}
\caption{Left: plot of the four potentials at N$^2$LO and their sum as a function of the distance between the decaying neutrons. Right: comparison between adimensional radial functions $T(x)$ and $W_1(x)$, as defined in Appendix \ref{sec.A}. The pink region indicates the typical range of the distance between one of the decaying neutrons and the third nucleon which does not decay in the $^6$He $\rightarrow$ $^6$Be transition. The regulators used in this plot have cutoffs [$R_L$, $R_S$] = [1.2 fm, 0.8 fm].}
\captionsetup{justification=centering}
\end{figure}

The size of the two- and three-nucleon matrix elements can be appreciated from their densities, as shown in Fig. \ref{fig:full}, for different transitions. The height of the two-nucleon densities, shown in orange, are about two orders of magnitude larger than the three-nucleon densities, which are shown in light (dark) blue at N$^2$LO (N$^3$LO). The form of the two-nucleon density varies between the $\Delta T=0$ and $\Delta T=2$ transitions, with the characteristic lack of node in the former case. The latter case, where the total isospin changes, is more interesting as the decays of experimental interest are of this kind. In this case, the two-nucleon density has a node, which suppresses the matrix element even further. 

\begin{figure}[t]
\centering
\captionsetup{labelformat=empty}
\begin{minipage}{.5\textwidth}
  \centering
  \includegraphics[width=\linewidth]{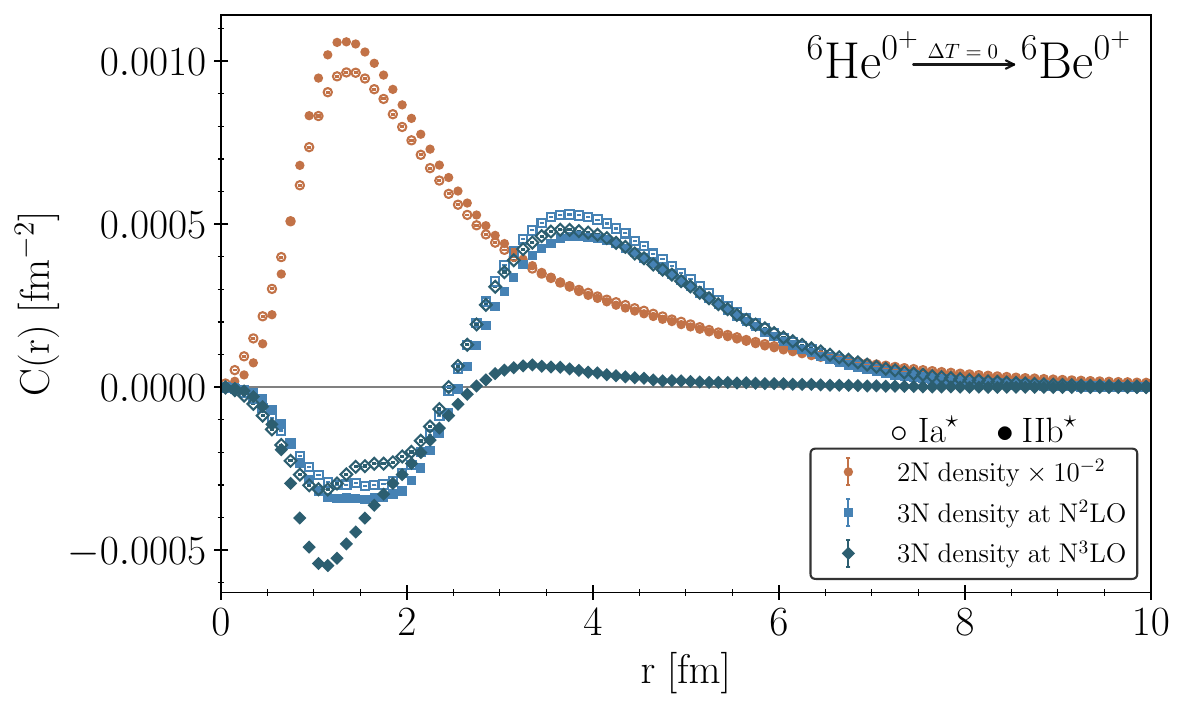}
    \caption{(a)}
  \label{fig:6He_full}
\end{minipage}%
\hfill
\begin{minipage}{.5\textwidth}
  \centering
  \includegraphics[width=\linewidth]{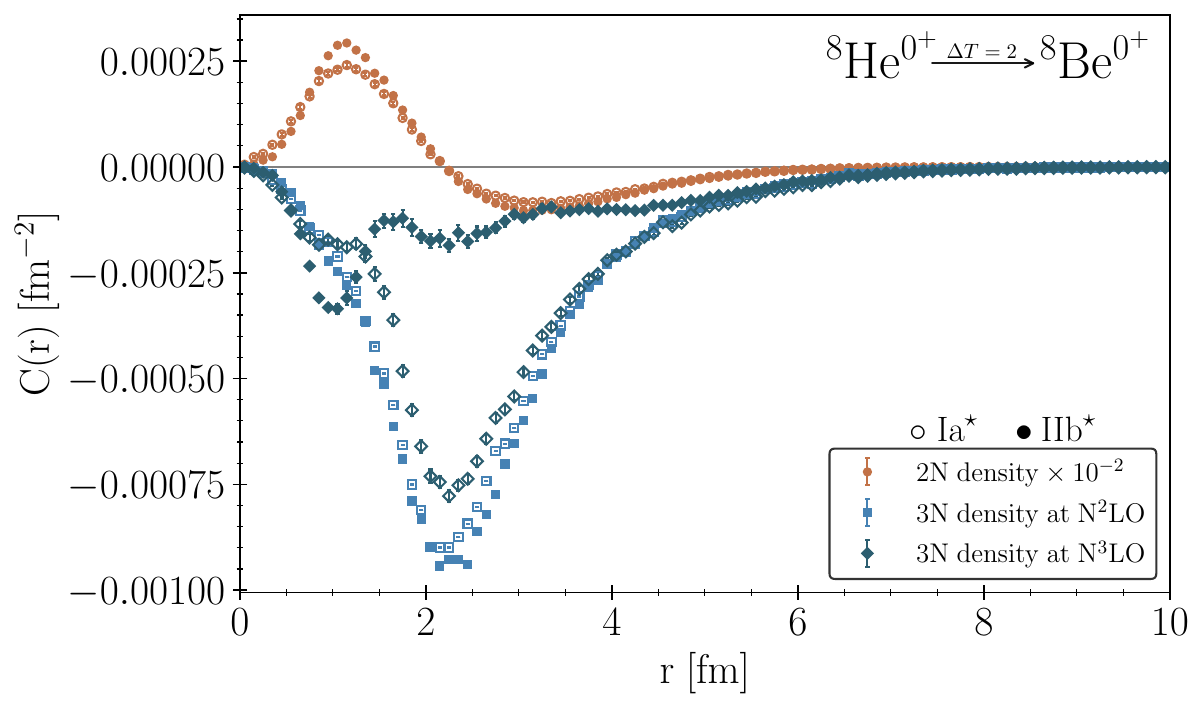}
    \caption{(b)}
  \label{fig:8He_full}
\end{minipage}
\\
\begin{minipage}{.5\textwidth}
  \centering
  \includegraphics[width=\linewidth]{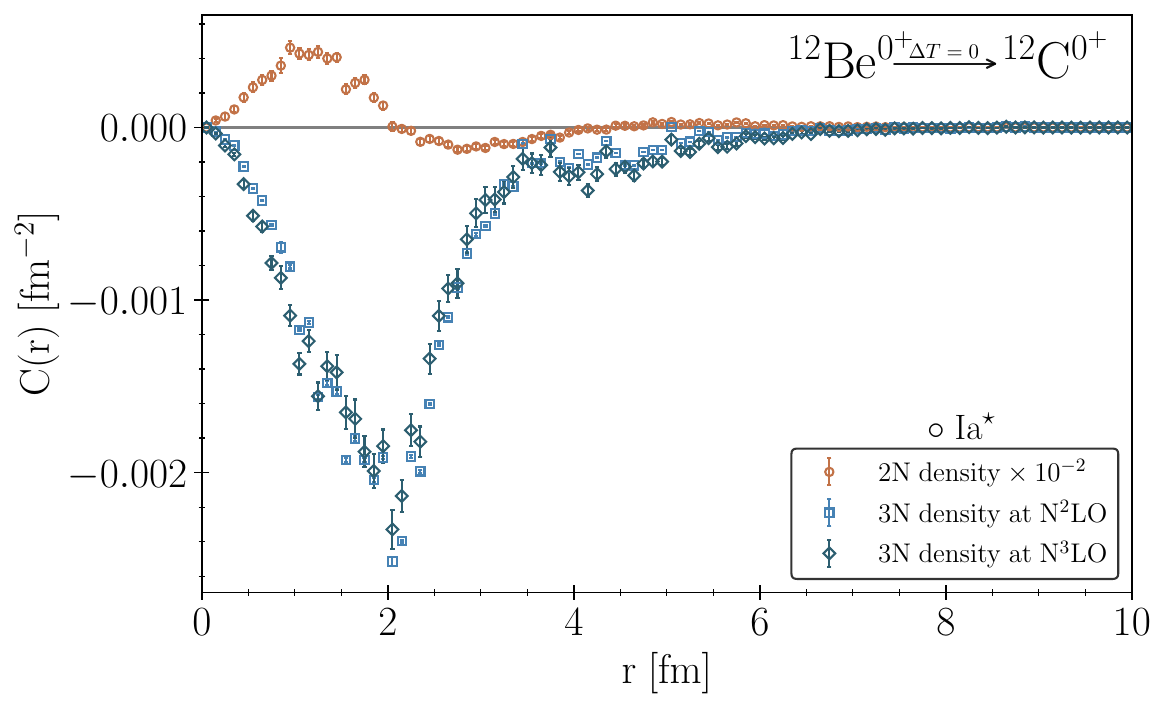}
  \caption{(c)}
  \label{fig:12Be_full}
\end{minipage}
\setcounter{figure}{8} 
\captionsetup{justification=default,labelformat=default}
\captionsetup{justification=raggedright,singlelinecheck=false}
\caption{2N and integrated 3N densities for $^6\mathrm{He}$ $\rightarrow$ $^6\mathrm{Be}$ and $^8\mathrm{He}$ $\rightarrow$ $^8\mathrm{Be}$. Models Ia$^\star$ (IIb$^\star$) is denoted by open (closed) symbols. The 2N density in orange is multiplied by $10^{-2}$ for plotting purposes to compare to the N$^2$LO (light blue) and N$^3$LO (dark blue) 3N densities. }
\label{fig:full}
\captionsetup{justification=centering}
\end{figure}

Depending on the transition, the three-nucleon densities exhibit a nodal structure, which is due to a combination of the antisymmetry of the spatial wave function and competing terms in the potential which have different signs. The amount of quenching of the two- and three-nucleon matrix elements due to this nodal structure can be appreciated in Fig. \ref{fig:quenching_ratios}, which shows the ratio of the integral of these densities with respect to the integral of the absolute value of the densities. A shift away from $1$ indicates a quenching of the matrix element due to this nodal structure. For example, the two-nucleon density ratio for $^6\mathrm{He}\rightarrow\,^6\mathrm{Be}$ is exactly $1$ because this density does not change signs as a function of $r$. This results in the two-nucleon matrix elements being overwhelmingly dominant over the three-nucleon matrix elements for this transition. The three-nucleon matrix elements provide a $\sim1\%$ correction in this case.

In comparison, the two-nucleon densities in the $\Delta T=2$ transitions feature this quenching. As shown in Fig. \ref{fig:quenching_ratios}, the cancellation in the two-nucleon case is greater than the cancellation in the three-nucleon case. The effect is a promotion of the importance of the three-nucleon potentials for these transitions, providing a much larger correction.

\subsubsection{Model dependence}

Figs. \ref{fig:6He_full} and \ref{fig:8He_full} show the model dependence of these matrix elements with open (closed) symbols denoting Norfolk model Ia$^\star$ (IIb$^\star$). For both the two-nucleon and N$^2$LO three-nucleon densities, there is little variation between these models as well as with models Ia and IIb, which are not shown in these plots. However, there exists large differences in the behavior of the three-nucleon density at N$^3$LO.

\begin{figure}[t]
\centering
\captionsetup{labelformat=default}
\begin{minipage}{\textwidth}
  \centering
  \includegraphics[width=\linewidth]{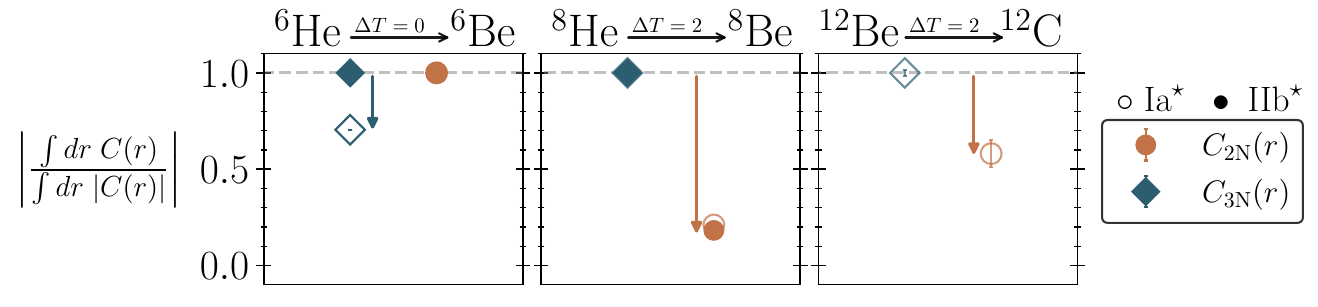}
  \label{fig:ratios}
\end{minipage}%
\captionsetup{justification=default}
\captionsetup{justification=raggedright,singlelinecheck=false}
\caption{Ratios of the integrated densities shown in Fig. \ref{fig:full} with respect to the integrated absolute value of the density for the two-nucleon and three-nucleon neutrino potentials. A shift from 1 indicates a quenching of the matrix element due to the nodal structure of the density. For $\Delta T=2$ transitions, the two-nucleon densities have a larger quenching with respect to the three-nucleon contributions. This causes the relative contribution of the three-nucleon potentials to be larger in the $\Delta T=2$ transition compared to $\Delta T=0$.}
\label{fig:quenching_ratios}
\captionsetup{justification=centering}
\end{figure}

This model dependence can be completely understood by examining the densities at N$^3$LO decomposed into the different contributions from each LEC. Fig. \ref{fig:densityN3LO} shows these densities for the three transitions. There are negligible differences between the densities from $\kappa_1$, $f_4^\Delta+f_5^\Delta$, and $c_1$ and slightly larger differences between the short-ranged behavior of $c_3$ and $c_4$ for each of the three transitions. The large model dependence of Figs. \ref{fig:full} and \ref{fig:quenching_ratios} is driven entirely by the $c_D$ contribution, which varies both in size and sign in the Norfolk models, as indicated in Table \ref{tab:table_cD}.

The impact of these three-nucleon matrix elements are shown in Fig. \ref{fig:ratio}, which shows the ratio of the sum of the two- and three-nucleon matrix elements to the two-nucleon matrix elements for the four Norfolk models. The top panel displays these ratios for the different LEC contributions at N$^3$LO, and the bottom panel shows the impact up to each order. Across each transition, there is negligible variation between NMEs for different models except for the $c_D$ contribution. This variation then propagates to the final N$^3$LO matrix element.

For the $^6\mathrm{He}\rightarrow\,^6\mathrm{Be}$ transition, the impact of the three-nucleon matrix elements is a percent level correction. However, the impact for the $\Delta T=2$ transition is nonnegligible. In the $^8\mathrm{He}\rightarrow\,^8\mathrm{Be}$ transition, $M_{\nu,3}^{\mathrm{N^2 LO}}$, $M_{\nu,3}^{c_3}$, and $M_{\nu,3}^{c_4}$ provide a $-10$ to $-15$, $-5$, and $5\%$ correction, respectively. The $M_{\nu,3}^{c_D}$ ratio has a spread of $-5$ to $15\%$, which results in an N$^3$LO correction of $5-25\%$.

\begin{figure}[t]
\centering
\captionsetup{labelformat=empty}
\begin{minipage}{.5\textwidth}
  \centering
  \includegraphics[width=\linewidth]{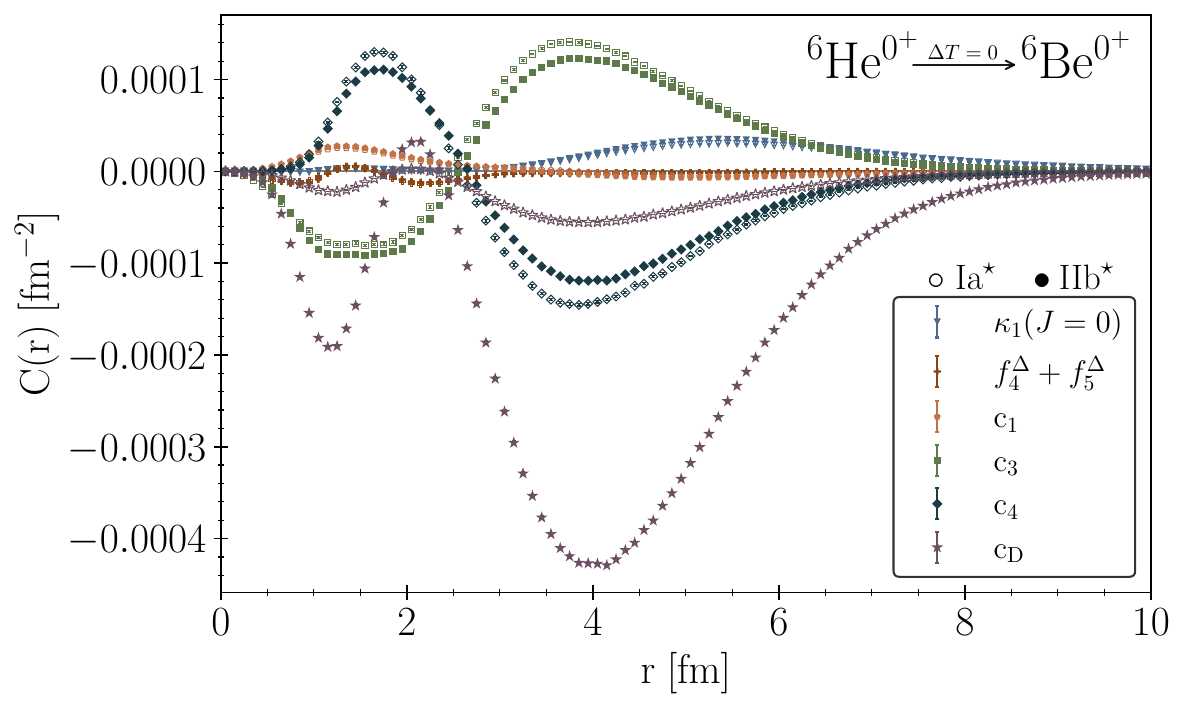}
  \caption{(a)}
  \label{fig:}
\end{minipage}%
\hfill
\begin{minipage}{.5\textwidth}
  \centering
  \includegraphics[width=\linewidth]{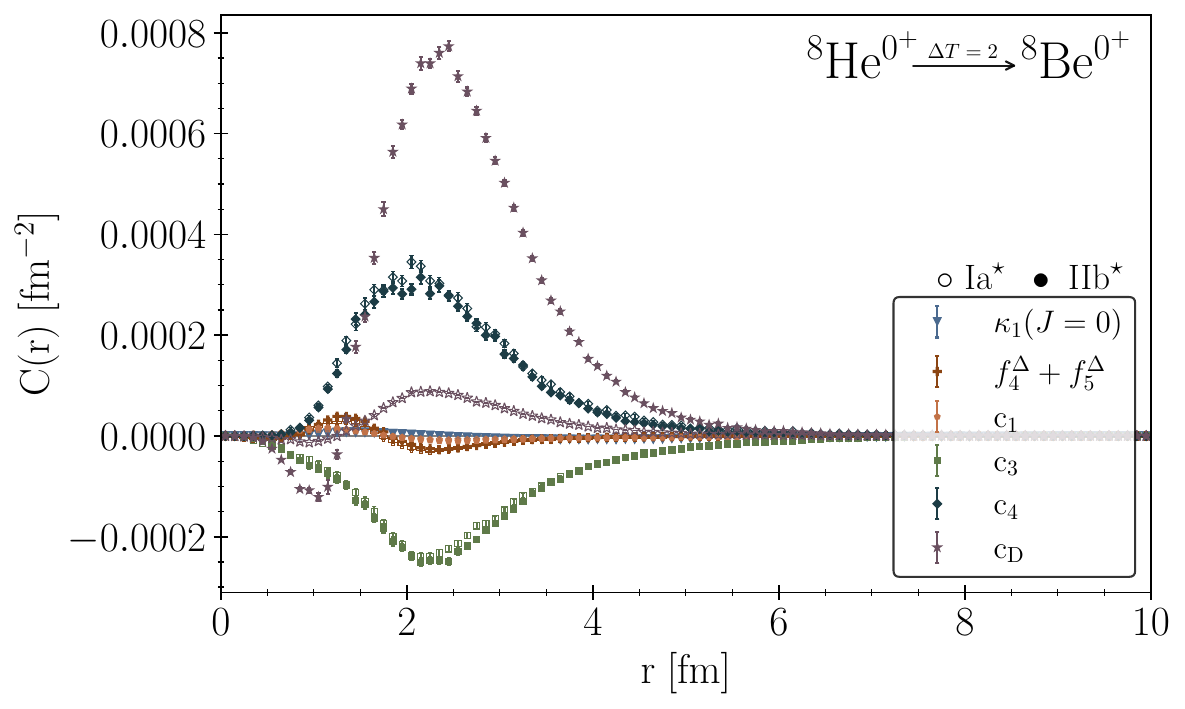}
  \caption{(b)}
  \label{fig:}
\end{minipage}
\\
\begin{minipage}{.5\textwidth}
  \centering
  \includegraphics[width=\linewidth]{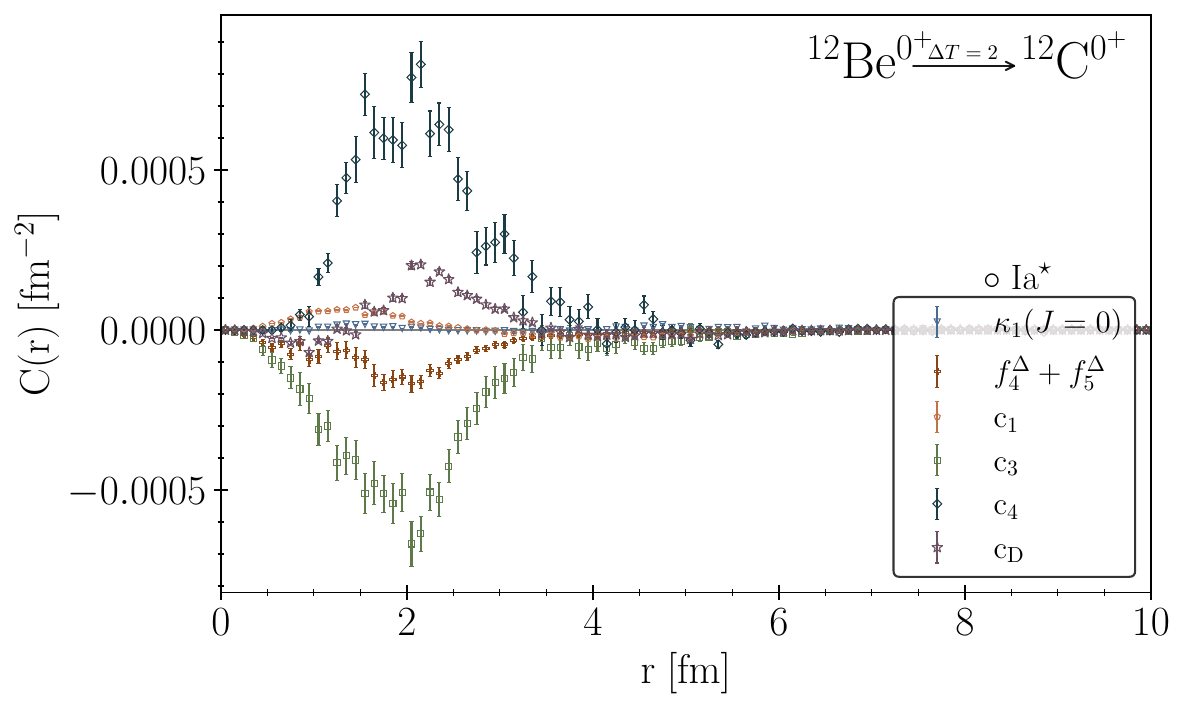}
  \caption{(c)}
  \label{fig:}
\end{minipage}
\setcounter{figure}{10}
\captionsetup{justification=default,labelformat=default}
\captionsetup{justification=raggedright,singlelinecheck=false}
\caption{Reduced three-nucleon densities for the N$^3$LO contributions grouped by LEC for transitions  $^6\mathrm{He}$ $\rightarrow$ $^6\mathrm{Be}$, $^8\mathrm{He}$ $\rightarrow$ $^8\mathrm{Be}$, and $^{12}\mathrm{Be}\rightarrow\,^{12}\mathrm{C}$. Model Ia$^\star$ (IIb$^\star$) is denoted with an open (closed) symbol.}
\label{fig:densityN3LO}
\captionsetup{justification=centering}
\end{figure}

Likewise in $^{12}\mathrm{Be}\rightarrow\,^{12}\mathrm{C}$, the largest contributions come from $c_3$, $c_4$, $c_D$, and the N$^2$LO potentials. However, the contribution from $c_3$ and $c_4$ is larger relative to $c_D$ in the $A=12$ transition compared to the $A=8$. Although the variation in $M_{\nu,3}^{c_D}$ is just as large across the two models used for $^{12}\mathrm{Be}\rightarrow\,^{12}\mathrm{C}$, the relative size of $M_{\nu,3}^{c_3}$ and $M_{\nu,3}^{c_4}$ is greater compared to $M_{\nu,3}^{c_D}$. Therefore, the impact of $c_D$ is smaller in the $A=12$ transition, and the final N$^{3}$LO matrix element has far less model dependence.

\subsubsection{$g_\nu^{\mathrm{NN}}$ in the three-nucleon sector}

Finally, we compute the contribution of the short-ranged LEC $g_\nu^{\mathrm{NN}}$ from diagram $(4b)$, whose coordinate-space expression is provided in Appendix \ref{sec.A}. This potential has two components: a short-ranged term where all three nucleons are close and a longer-ranged term where the decaying neutrons are close and the third nucleon can be further separated. As discussed in Sec. \ref{Sec:Lag2}, a $0\nu\beta\beta$ transition operator with two delta functions should vanish in the infinite cutoff limit. However, when this operator is implemented using a short-ranged regulator, the contribution does not exactly vanish.

The LECs $g_\nu^{\mathrm{NN}}$ and $\mathcal{C}_1$, appearing in Eq. (\ref{C12def}), are related as $g_\nu^{\mathrm{NN}}=\mathcal{C}_1$. However, $\mathcal{C}_1$ and $\mathcal{C}_2$ have not been separately determined. Instead, we use the value of $g_\nu^{\mathrm{NN}}\sim(\mathcal{C}_1+\mathcal{C}_2)/2$ extracted from CIB processes. In particular, we compute the matrix element of this operator for model Ia$^\star$ with $(\mathcal{C}_1+\mathcal{C}_2)/2=-1.03$ (fm$^2$) \cite{Cirigliano:2019vdj}. 

\begin{figure}[t]
\captionsetup{labelformat=empty}
\begin{minipage}{.5\textwidth}
  \includegraphics[width=\linewidth]{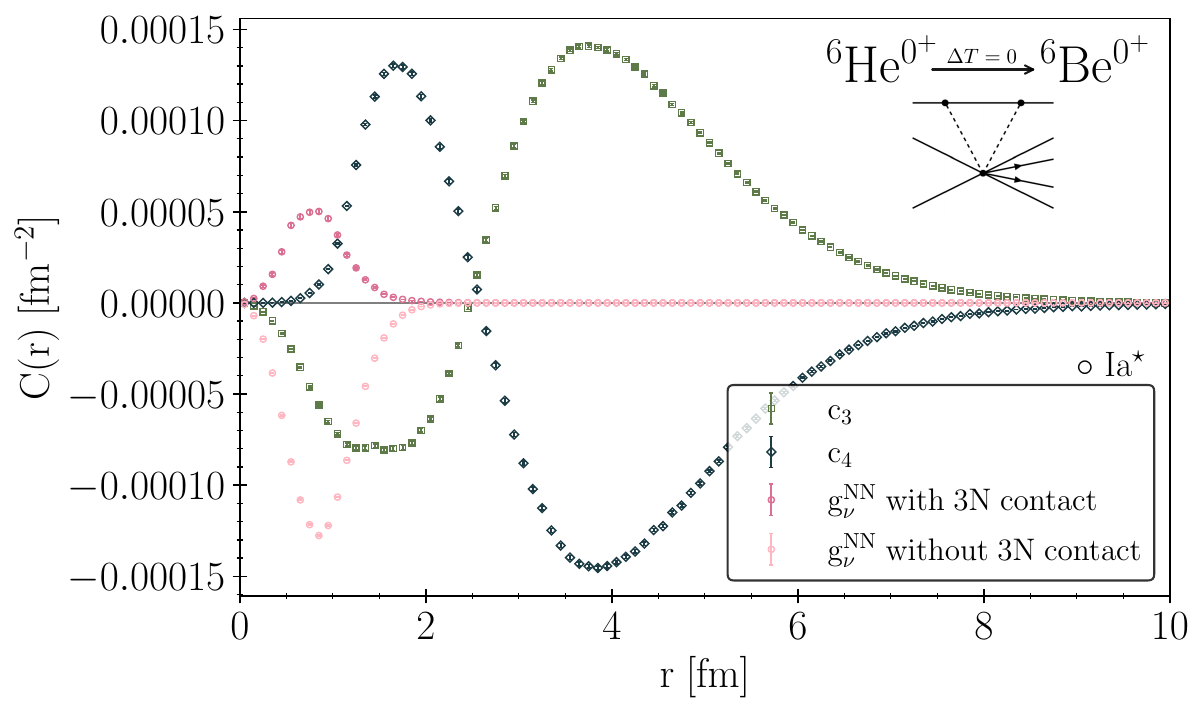}
  \caption{(a)}
  \label{fig:}
\end{minipage}%
\hfill
\begin{minipage}{.5\textwidth}
\includegraphics[width=\linewidth]{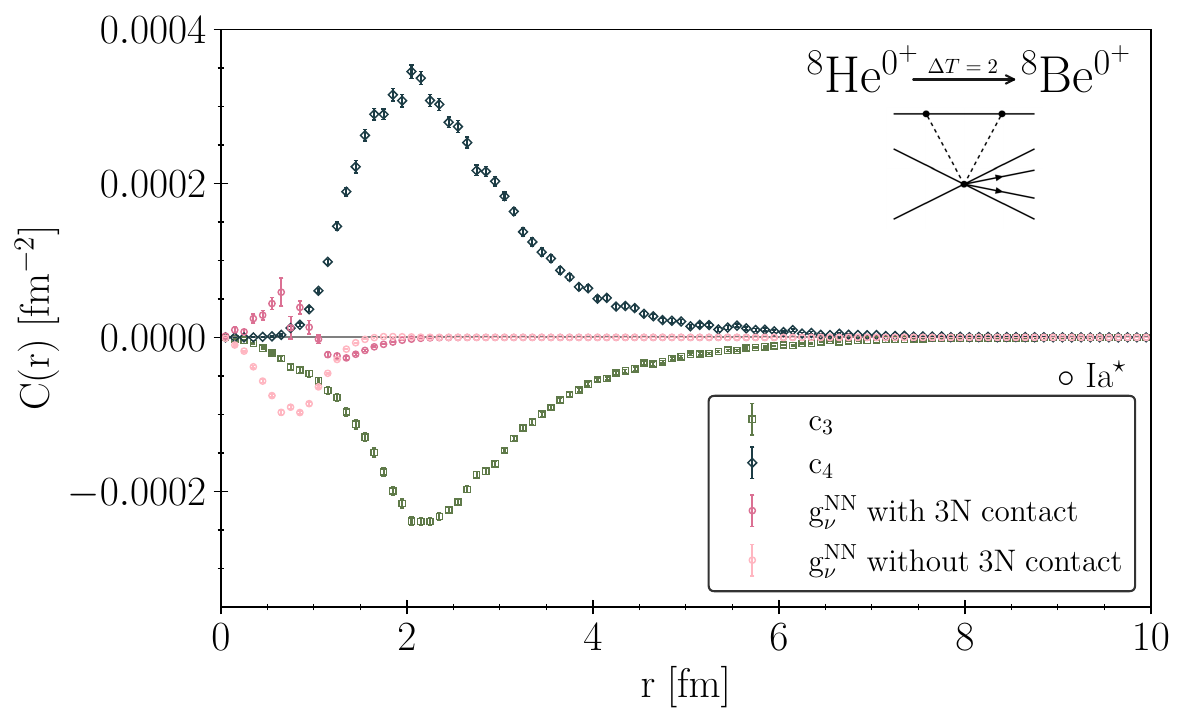}
\caption{(b)}
  \label{fig:}
\end{minipage} 
\setcounter{figure}{11} 
\captionsetup{justification=default,labelformat=default}
\captionsetup{justification=raggedright,singlelinecheck=false}
\caption{Plots of the $g_\nu^{\mathrm{NN}}$ contribution (diagram (3b)), shown in pink, for model Ia$^\star$ for the three transitions. The densities for $c_3$ and $c_4$ are also shown for comparison. The $A=12$ transition was not computed due to the computational cost.}
\label{fig:gnuNN_all}
\captionsetup{justification=centering}
\end{figure}


Fig. \ref{fig:gnuNN_all} shows the densities of this contribution for two of the transitions compared to the $c_3$ and $c_4$ densities. As expected, the density is short-ranged with the distance between decaying neutrons $r \lesssim 2$ fm. The size of this matrix element is small for $^6\mathrm{He}\rightarrow\,^6\mathrm{Be}$ at roughly a quarter the size of either $c_3$ and $c_4$ for this model, providing only a small correction to the total three-nucleon matrix elements. For the $^8\mathrm{He}\rightarrow\,^8\mathrm{Be}$ transition, the magnitude of the density is even smaller relative to $c_3$ and $c_4$. There is an additional suppression of this three-nucleon $g_\nu^{\mathrm{NN}}$ contribution in this transition due to the nodal structure of this density. For larger fit values such as $(\mathcal{C}_1+\mathcal{C}_2)/2= -1.44$ fm$^2$ in other Norfolk models (IIa$^\star$ and Ic) \cite{Cirigliano:2019vdj}, the size of this matrix element may increase by $50\%$. However, this $g_\nu^{\mathrm{NN}}$ contribution will remain a subpercent correction to the total NME.

We show the contributions with and without the short-ranged component of this operator, denoted by dark and light pink, respectively, in Fig. \ref{fig:gnuNN_all}. Although one would expect this component to be vanishingly small, we find a larger contribution than expected. A three-nucleon contact operator, which would appear at N$^5$LO in WPC, would in principle be further suppressed at the operator level. However, due to the unexpectedly large contribution of the short-ranged piece of diagram $(4b)$, the importance of $g_\nu^{\mathrm{3N}}$ may be greater when computed in a many-body method than expected at the operator level.

\subsection{Physics implications}

Converting the bound on the experimental $0\nu\beta\beta$ half-life to a bound on $m_{\beta\beta}$ in the light-neutrino exchange mechanism requires accurate NMEs for various heavy nuclei. 
First principle calculations of NMEs for nuclei of experimental interest appeared in the literature only recently \cite{Belley:2023btr,Belley:2023lec,Novario2020}. These studies
indicate smaller NMEs than those predicted by phenomenological methods and furthermore suggest that $0\nu\beta\beta$ experiments have yet to probe the non-degenerate inverted hierarchy region~\cite{Belley:2023btr}. The next generation of ton-scale experiments hope to completely probe this region but require precise NMEs for accurate physics interpretations. 

\begin{figure}[t]
    \centering
    \captionsetup{labelformat=default}\includegraphics[width=.8\linewidth]{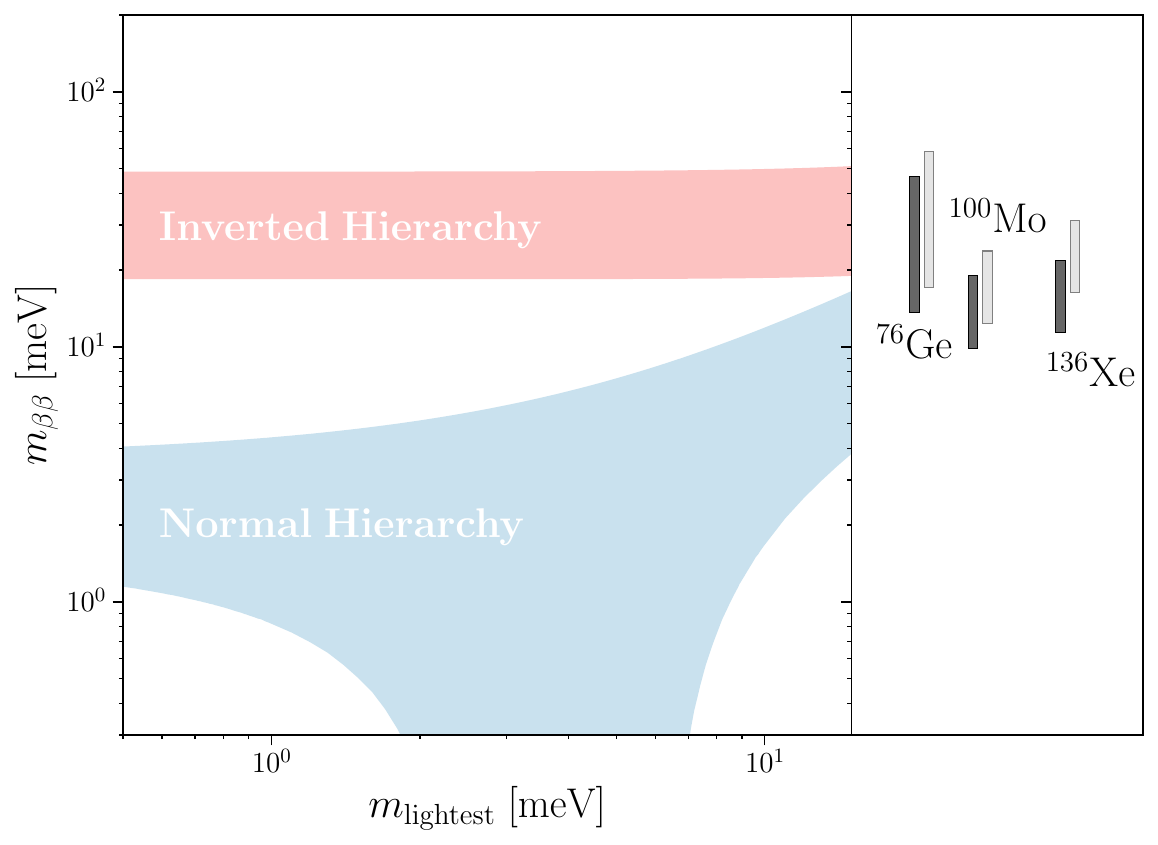}
    \caption{LNV parameter space of $m_{\beta\beta}$ and the lightest neutrino mass with inverted and normal hierarchy bands using best fit oscillation parameters from NuFit-6.0 \cite{Esteban_2024}. The next generation of ton-scale experiments will be able to probe at the scale of the non-degenerate inverted region, as indicated by the black bands. A $20\%$ quenching of the NMEs for these isotopes would shift the exclusion region to larger $m_{\beta\beta}$, as indicated by the gray bands.}\label{fig:lobster}
\end{figure}

In order to explore the effect of three-nucleon potential in nuclei of experimental interest, we perform a qualitative analysis based on the recent IMSRG (for Refs. \cite{Belley:2023lec,Belley:2023btr}) and Shell-Model (for Ref. \cite{Coraggio2022}) results and apply a 20\% quenching on the calculated NMEs, corresponding to the most severe quenching observed in light nuclei.  By using these values, which are smaller compared to phenomenological predictions, we can provide a conservative estimate on the ability of the experimental effort to probe the parameter space featured in Fig. \ref{fig:lobster} and the impact of the three-nucleon contributions. 
Specifically, Fig. \ref{fig:lobster} shows the $m_{\beta\beta}$ and $m_{\mathrm{lightest}}$ parameter space compared to the $m_{\beta\beta}$ bounds that the next generation of $0\nu\beta\beta$ experiments aim to probe.

Black bands indicate the projected reach of the next generation of experiments, and gray bands indicate the reach with the $20\%$ quenching of NMEs using projected half-lives of order $10^{28}$ yr for the future LEGEND-1000, CUPID-1T, and nEXO experiments \cite{Legend1000_2021,Cupid1T2022,nExo2021} and phase space factors of Ref. \cite{Stoica:2013lka}. A $20\%$ reduction in the NME is the most severe quenching we found in light nuclei. Explicit calculations of the three-nucleon operators in heavy nuclei will provide a definite estimate of their impact. 
However, in this crude extrapolation, even in the worst-case quenching scenario, the next generation of experiments should still be able to probe the inverted hierarchy region assuming the light-neutrino-exchange mechanism.

\section{Conclusion}\label{Sec:6}

Experiments searching for neutrinoless double beta decay serve as the most sensitive probes into the Majorana nature of the neutrinos. However, converting limits on the experimental half-lives of various isotopes into constraints on LNV parameters requires precise determinations of $0\nu\beta\beta$ nuclear matrix elements. Performing these calculations for the heavy nuclei of experimental interest, such as $^{76}$Ge, $^{100}$Mo, or $^{136}$Xe, is an insurmountable task if one starts at the fundamental theory of QCD. Instead, connecting theories across different scales, namely, lattice QCD at the hadronic level, chiral effective field theory at the few-nucleon and pion level, and many-body methods at the nuclear level, provides a framework to perform accurate NME calculations. At the level of chiral effective field theory, one must construct transition operators mediated by a LNV interaction up to some order, after which higher-order corrections are expected to be smaller than the desired precision. These operators are then input into a many-body method to calculate the NMEs for different nuclei. Within this approach, two-nucleon transition operators have been derived up to next-to-next-to-leading order within WPC. In this work, we derived three-nucleon transition operators in $\Delta$-full $\chi$EFT up to N$^3$LO and calculated their matrix elements in light nuclei using quantum Monte Carlo methods. Calculations in light nuclei, although not of direct experimental interest, serve as a valuable tool to study the impact of different operators using a computationally accurate many-body method while also carving the path towards  benchmarking in other many-body methods, such as coupled-cluster, shell model, and in-medium similarity renormalization group.

We performed QMC calculations using nuclear wave functions constructed from the Norfolk class of nuclear interactions, and calculate two- and three-nucleon matrix elements of transition operators consistently derived with these nuclear interactions.  We studied two types of transition, those where the total isospin changes by either $\Delta T=0$ or $2$. In the decay of $^6\mathrm{He}\rightarrow\,^6\mathrm{Be}$, where the total isospin does not change, we found the three-nucleon matrix elements to be a small correction, $\mathcal{O}(\sim1\%)$, to the leading-order two-nucleon light neutrino exchange contribution. However, in the case of $\Delta T=2$ transitions, we found a larger three-nucleon contribution. This can be understood, in part, by the structure of the densities of these transitions. In the $\Delta T=2$ transitions, there exists a node in the two- and three-nucleon densities due to the orthogonality between the spatial symmetries of the initial and final wave function. This behavior suppresses the NMEs for these transitions. We find that the two-nucleon matrix elements have a larger quenching in the $^{8}\mathrm{He}\rightarrow\,^8\mathrm{Be}$ and $^{12}\mathrm{Be}\rightarrow\,^{12}\mathrm{C}$ transitions compared to the three-nucleon densities, which promotes the importance of the three-nucleon operators in the final NME. Whether this feature extends to the larger nuclei of experimental interest, which are isospin-changing transitions, should be examined with other many-body methods.

There is virtually no model dependence in the three-nucleon sector using this subset of the Norfolk model classes except for the contribution of $c_D$ at N$^3$LO. This variation propagates to the sum of all three-nucleon contributions up to N$^3$LO, providing a $0$ to $-20\%$ correction to the two-nucleon matrix elements in $^{8}\mathrm{He}\rightarrow\,^8\mathrm{Be}$. For $^{12}\mathrm{Be}\rightarrow\,^{12}\mathrm{C}$, where $c_D$ is a smaller contribution, the three-nucleon operators provide a $-10\%$ correction.

In either case, this is a nonnegligible correction to the predicted inverse half-life, which is proportional to the NMEs squared. This large dependence on $c_D$ suggests the need for a revaluation of the three-nucleon forces, such as including higher-order contributions, which recent work has indicated as important \cite{Cirigliano:2024ocg}, and a revaluation of the fitting procedure.

In the two-nucleon sector, there is a short-ranged contact operator, proportional to the unknown LEC $g_\nu^{\mathrm{NN}}$, which is required at leading-order based on renormalization arguments \cite{Cirigliano:2018hja}. We study the impact of this operator in the three-nucleon sector. The first contribution that participates in $0^+\rightarrow0^+$ transitions is a loop-diagram which appears at N$^5$LO, which has the same structure as the $D_2$ operator of Ref. \cite{Cirigliano:2024ocg}, which
found a numerical promotion of this operator to higher order. We estimate an impact of this $g_\nu^{\mathrm{NN}}$ contribution in light nuclei to be subpercent.

This work can be extended in two ways. First, within the $\chi$EFT approach, NMEs are typically computed in the two-nucleon sector up to N$^2$LO in WPC with higher-order contributions assumed to be small. However, the three-nucleon matrix elements of this work indicate a $10-15\%$ effect at N$^2$LO and a $0-20\%$ effect at N$^3$LO for the two $\Delta T=2$ transitions in this work. At N$^3$LO there are two-nucleon loop contributions, discussed in more detail in Sec. \ref{Sec:3}, induced by the pion-nucleon coupling that appear in the three-nucleon potentials. Whether these two-nucleon contributions have a similar impact on the NMEs at this order has not been tested.

Second, we call for many-body practitioners to compute NMEs of these three-nucleon operators to assess their impact in the heavy nuclei of experimental interest. Whether the same size of quenching due to the complete set of three-nucleon contributions presented in this work also occurs in the heavy nuclei has yet to be investigated. However, a possible $10-20\%$ shift in the NME would have important implications on the experimental extraction of $m_{\beta\beta}$ in a discovery scenario.

\section*{Acknowledgments}

We would like to thank Ronen Weiss and Maria Dawid for useful discussions at various stages of the project;
Wouter Dekens for useful discussions and for pointing out that short-range LNV three-nucleon operators vanish by symmetry; Martin Hoferichter for discussion on the values of the LECs $c_{1,3,4}$; Amy Nicholson for pointing out the need for the $\Delta$-contact diagrams of Appendix \ref{sec.D} in our discussion; and Giovanni De Gregorio, Tokuro Fukui, Luigi Coraggio, and Nunzio Itaco for their hospitality in hosting G.~C. and S.~P. at the Universit\'a degli Studi della Campania “Luigi Vanvitelli” during part of this project.

This project is supported in part by the US Department of Energy under Contracts No. DE-SC0021027 (G.~C. and S.~P.) and DE-AC02-CH11357 (R.~B.~W.), the US Department of Energy 2021 Early Career Award number DE-SC0022002 (M.~P.), the NSF Graduate Research Fellowship Program under Grant No. DGE-213989 (G.~C.), and the Office of Science Graduate Student Research (SCGSR) program under contract number
DE‐SC0014664 (G.~C.). 
Financial support by Los Alamos
National Laboratory's Laboratory Directed Research and Development program under projects
20230047DR  (E.~M.)
and 20240742PRD1 (G.~B.~K.)
is gratefully acknowledged.  Los Alamos National Laboratory is operated by Triad National Security, LLC,
for the National Nuclear Security Administration of U.S.\ Department of Energy (Contract No.\
89233218CNA000001). This project is supported in part by the NSF FRHTP program under award No. PHY-2402275.
We thank the Nuclear Theory for New Physics Topical Collaboration for fostering dynamic collaborations.

The VMC calculations were performed by the National Energy Reserach Scientific Computing Center, a DOE Office of Science User Facility supported by the Office of Science of the U.S. Department of Energy under Contract No. DE-AC02-05CH11231 using NERSC award NP-ERCAP0027147, and the Laboratory Computing Resource Center of Argonne National Laboratory.

\appendix

\section{Three-nucleon neutrino potential in coordinate space}\label{sec.A}

We give in this Appendix the coordinate space expression of the three-nucleon neutrino potential. 
We define the Fourier transform as 
\begin{align}
    \tilde{V}(\boldsymbol{r}_{12},\boldsymbol{r}_{23})=\int \frac{d^3\boldsymbol{q}_1}{(2\pi)^3}\int \frac{d^3 \boldsymbol{q}_3}{(2\pi)^3}V(\boldsymbol{q}_1,\boldsymbol{q}_3)\exp(i\boldsymbol{q}_1\cdot\boldsymbol{r}_{12}-i\boldsymbol{q}_3\cdot\boldsymbol{r}_{23}).
\end{align}
The Fourier transform of the N$^2$LO potentials in Eqs. \eqref{eq:n2loa}, \eqref{eq:n2lob} and \eqref{eq:n2loc} is given by
\begin{align}
\begin{split}
       \tilde{V}_{(2a)}=&-\frac{2h_A^2g_A^2}{9F_\pi^2\Delta}\sum_{i\neq j\neq k}-\frac{m_\pi^4}{96\pi^2}\frac{1}{m_\pi r_{ij}}\left[4T(r_{jk})\left\{(S_{jk}+\boldsigma_j\cdot\boldsigma_k),\boldsigma_i\cdot\boldsigma_j\right\} \tau_i^+\tau_k^+\phantom{\frac{1}{2}}\right.
    \\
    &\left.+4W_1(r_{jk})\{\boldsigma_j\cdot\boldsigma_k,\boldsigma_i\cdot\boldsigma_j\}\tau_i^+\tau_k^+\right.
    \\
    &\left.+T(r_{jk})[(S_{jk}+\boldsigma_j\cdot\boldsigma_k),\boldsigma_i\cdot\boldsigma_j](\tau_k^3\tau_i^+\tau_j^+-\tau_j^3\tau_i^+\tau_k^+)\right.
    \\
    &\left.+W_1(r_{jk})[\boldsigma_j\cdot\boldsigma_k,\boldsigma_i\cdot\boldsigma_j](\tau_k^3\tau_i^+\tau_j^+-\tau_j^3\tau_i^+\tau_k^+)\right],
\end{split}\label{eq:A6}
\end{align}
\begin{align}
\begin{split}
\tilde{V}_{(2b)}=\tilde{V}_{(2c)}=\;&\frac{2h_A^2g_A^2}{9F_\pi^2\Delta}\sum_{i\neq j\neq k}\frac{m_\pi^4}{288\pi^2}\left[4T(r_{jk})W_2(r_{ij})\{(S_{ij}+\boldsigma_i\cdot\boldsigma_j),(S_{jk}+\boldsigma_j\cdot\boldsigma_k)\}\tau_i^+\tau_k^+\right.
    \\
    &\left.+4W_1(r_{jk})W_2(r_{ij})\{(S_{ij}+\boldsigma_i\cdot\boldsigma_j),\boldsigma_j\cdot\boldsigma_k\}\tau_i^+\tau_k^+\right.
    \\
    &\left.-4T(r_{jk})(W_2(r_{ij})+Y(r_{ij}))\{(S_{jk}+\boldsigma_j\cdot\boldsigma_k),\boldsigma_i\cdot\boldsigma_j\}\tau_i^+\tau_k^+\right.
    \\
    &\left.-4W_1(r_{jk})(W_2(r_{ij})+Y(r_{ij})\{\boldsigma_i\cdot\boldsigma_j,\boldsigma_j\cdot\boldsigma_k\}\tau_i^+\tau_k^+\right.
    \\
    &\left.+T(r_{jk})W_2(r_{ij})\left[(S_{jk}+\boldsigma_j\cdot\boldsigma_k),(S_{ij}+\boldsigma_i\cdot\boldsigma_j))\right](\tau_k^3\tau_i^+\tau_j^+-\tau_j^3\tau_i^+\tau_k^+)\right.
    \\
    &\left.-T(r_{jk})(W_2(r_{ij})+Y(r_{ij}))[(S_{jk}+\boldsigma_j\cdot\boldsigma_k),\boldsigma_i\cdot\boldsigma_j](\tau_k^3\tau_i^+\tau_j^+-\tau_j^3\tau_i^+\tau_k^+)\right.
    \\
    &\left.+W_1(r_{jk})W_2(r_{ij})[\boldsigma_j\cdot\boldsigma_k,(S_{ij}+\boldsigma_i\cdot\boldsigma_j)](\tau_k^3\tau_i^+\tau_j^+-\tau_j^3\tau_i^+\tau_k^+)\right.
    \\
    &\left.-W_1(r_{jk})(W_2(r_{ij})+Y(r_{ij}))[\boldsigma_i\cdot\boldsigma_j,\boldsigma_i\cdot\boldsigma_k](\tau_k^3\tau_i^+\tau_j^+-\tau_j^3\tau_i^+\tau_k^+)\right],
 \end{split}
 \end{align}
\begin{align}
\begin{split}
    \tilde{V}_{(2d)}=&-\frac{2h_A^2g_A^2}{9F_\pi^2\Delta}\sum_{i\neq j\neq k}\frac{m_\pi^4}{576\pi^2}\left[4Y(r_{ij})(1+m_\pi r_{ij})T(r_{jk})\{(S_{ij}+\boldsigma_i\cdot\boldsigma_j),(S_{jk}+\boldsigma_j\cdot\boldsigma_k)\}\tau_i^+\tau_k^+\phantom{\frac{1}{1}}\hspace{-8pt}\right.
    \\
    &\left.+4Y(r_{ij})(1+m_\pi r_{ij})W_1(r_{jk})\{(S_{ij}+\boldsigma_i\cdot\boldsigma_j),\boldsigma_j\cdot\boldsigma_k\}\tau_i^+\tau_k^+\right.
    \\
    &\left.-12Y(r_{ij})T(r_{jk})\{(S_{jk}+\boldsigma_j\cdot\boldsigma_k),\boldsigma_i\cdot\boldsigma_j\}\tau_i^+\tau_k^+\right.
    \\
    &\left.-12Y(r_{ij})W_1(r_{jk})\{\boldsigma_i\cdot\boldsigma_j,\boldsigma_j\cdot\boldsigma_k\}\tau_i^+\tau_k^+\right.
    \\
    &\left.+Y(r_{ij})(1+m_\pi r_{ij})T(r_{jk})[(S_{jk}+\boldsigma_j\cdot\boldsigma_k),(S_{ij}+\boldsigma_i\cdot\boldsigma_j)](\tau_k^3\tau_i^+\tau_j^+-\tau_j^3\tau_i^+\tau_k^+)\right.
    \\
    &\left.+Y(r_{ij})(1+m_\pi r_{ij})W_1(r_{jk})[\boldsigma_j\cdot\boldsigma_k,(S_{ij}+\boldsigma_i\cdot\boldsigma_j)](\tau_k^3\tau_i^+\tau_j^+-\tau_j^3\tau_i^+\tau_k^+)\right.
    \\
    &\left.-3Y(r_{ij})T(r_{jk})[(S_{jk}+\boldsigma_j\cdot\boldsigma_k),\boldsigma_i\cdot\boldsigma_j](\tau_k^3\tau_i^+\tau_j^+-\tau_j^3\tau_i^+\tau_k^+)\right.
    \\
    &\left.-3Y(r_{ij})W_1(r_{jk})[\boldsigma_j\cdot\boldsigma_k,\boldsigma_i\cdot\boldsigma_j](\tau_k^3\tau_i^+\tau_j^+-\tau_j^3\tau_i^+\tau_k^+)\right].
\end{split}
\end{align}

In the case of the N$^3$LO potential, the Fourier transform of the contributions proportional to the LECs $f^{\Delta}_{4} + f_5^{\Delta}$, $c_{1,3,4}$ and $c_D$ have simple expressions, given by
\begin{align}
    \begin{split}
        \tilde{V}_{(3c)}=&\;\frac{2g_A^2h_A^2(f_4^\Delta+f_5^\Delta)}{9F_\pi^2\Delta^2}\sum_{i\neq j\neq k}\frac{m_\pi^6}{288\pi^2}\left[\phantom{\frac{1}{1}}\hspace{-7pt}T(r_{ij})T(r_{jk})\{(S_{ij}+\boldsigma_i\cdot\boldsigma_j),(S_{jk}+\boldsigma_j\cdot\boldsigma_k)\}\tau_i^+\tau_k^+\right.
        \\
        &\left.+T(r_{ij})W_1(r_{jk})\{(S_{ij}+\boldsigma_i\cdot\boldsigma_j),\boldsigma_j\cdot\boldsigma_k\}\tau_i^+\tau_k^+\right.
        \\
        &\left.+T(r_{jk})W_1(r_{ij})\{(S_{jk}+\boldsigma_j\cdot\boldsigma_k),\boldsigma_i\cdot\boldsigma_j\}\tau_i^+\tau_k^+\right.
        \\
        &\left.+W_1(r_{ij})W_1(r_{jk})\{\boldsigma_i\cdot\boldsigma_j,\boldsigma_j\cdot\boldsigma_k\}\tau_i^+\tau_k^+\right.
        \\
        &\left.+T(r_{ij})T(r_{jk})[(S_{ij}+\boldsigma_i\cdot\boldsigma_j),(S_{jk}+\boldsigma_j\cdot\boldsigma_k)](\tau_k^3\tau_i^+\tau_j^+-\tau_j^3\tau_i^+\tau_k^+)\right.
        \\
        &\left.+T(r_{ij})W_1(r_{jk})[(S_{ij}+\boldsigma_i\cdot\boldsigma_j),\boldsigma_j\cdot\boldsigma_k](\tau_k^3\tau_i^+\tau_j^+-\tau_j^3\tau_i^+\tau_k^+)\right.
        \\
        &\left.+T(r_{jk})W_1(r_{ij})[\boldsigma_i\cdot\boldsigma_j,(S_{jk}+\boldsigma_j\cdot\boldsigma_k)](\tau_k^3\tau_i^+\tau_j^+-\tau_j^3\tau_i^+\tau_k^+)\right.
        \\
        &\left.+W_1(r_{ij})W_1(r_{jk})[\boldsigma_i\cdot\boldsigma_j,\boldsigma_j\cdot\boldsigma_k](\tau_k^3\tau_i^+\tau_j^+-\tau_j^3\tau_i^+\tau_k^+)\right],
    \end{split}
\end{align}
\begin{align}
    \begin{split}
        \tilde{V}_{(3d-3g)c_1}=&\;\frac{4g_A^2m_\pi^4c_1}{F_\pi^2}\sum_{i\neq j\neq k}\frac{1}{576\pi^2}\left(1+\frac{1}{m_\pi r_{jk}}\right)Y(r_{jk})\frac{1}{m_\pi^2 r_{ij}^2}
        \\
        &\times(2-Y(r_{ij})(2m_\pi r_{ij}+2m_\pi^2 r_{ij}^2+m_\pi^3 r_{ij}^3))
        \\
        &\times\frac{1}{\hat{\boldsymbol{r}}_{ij}\cdot\hat{\boldsymbol{r}}_{jk}} \{(S_{ij}+\boldsigma_i\cdot\boldsigma_j),(S_{jk}+\boldsigma_j\cdot\boldsigma_k)\}\tau_i^+\tau_k^+,
    \end{split}
\end{align}
\begin{align}
\begin{split}
\tilde{V}_{(3d-3g)c_3}=&\;\frac{2g_A^2c_3}{F_\pi^2}\sum_{i\neq j\neq k}-\frac{m_\pi^4}{576\pi^2}\left[\phantom{\frac{1}{1}}\right.
\\
&\left. T(r_{jk})\left(\frac{6}{m_\pi r_{ij}}-4W_2(r_{ij})-Y(r_{ij})\right)\{(S_{jk}+\boldsigma_j\cdot\boldsigma_k),\boldsigma_i\cdot\boldsigma_j\}\tau_i^+\tau_k^+\right.
\\
&\left.+W_1(r_{jk})\left(\frac{6}{m_\pi r_{ij}}-4W_2(r_{ij})-Y(r_{ij})\right)\{\boldsigma_j\cdot\boldsigma_k,\boldsigma_i\cdot\boldsigma_j\}\tau_i^+\tau_k^+\right.
\\
&\left.+T(r_{jk})(4W_2(r_{ij})-Y(r_{ij})(1+m_\pi r_{ij}))\{(S_{jk}+\boldsigma_j\cdot\boldsigma_k),(S_{ij}+\boldsigma_i\cdot\boldsigma_j)\}\tau_i^+\tau_k^+\right.
\\
&\left.+W_1(r_{jk})(4W_2(r_{ij})-Y(r_{ij})(1+m_\pi r_{ij}))\{\boldsigma_j
\cdot\boldsigma_k,(S_{ij}+\boldsigma_i
\cdot\boldsigma_j)\}\tau_i^+\tau_k^+\hspace{-5pt}\phantom{\frac{1}{1}}\right],
\end{split}
\end{align}
\begin{align}
    \begin{split}
        \tilde{V}_{(3d-3g)c_4}=\;&\frac{g_A^2 c_4}{F_\pi^2}\sum_{i\neq j\neq k}\frac{m_\pi^4}{576\pi^2}\left[\phantom{\frac{1}{1}}\right.
        \\
        &\left.T(r_{jk})\left(\frac{6}{m_\pi r_{ij}}-4W_2(r_{ij})-Y(r_{ij})\right)\right.\\&\left.\times[(S_{jk}+\boldsigma_j
        \cdot\boldsigma_k),\boldsigma_i
        \cdot\boldsigma_j](\tau_k^3\tau_i^+\tau_j^+-\tau_j^3\tau_i^+\tau_k^+)\right.
        \\
        &\left.+W_1(r_{jk})\left(\frac{6}{m_\pi r_{ij}}-4W_2(r_{ij})-Y(r_{ij})\right)\right.\\
        &\left.\times[\boldsigma_i\cdot\boldsigma_j,\boldsigma_i\cdot\boldsigma_k](\tau_k^3\tau_i^+\tau_j^+-\tau_j^3\tau_i^+\tau_k^+)\right.
        \\
        &\left.+T(r_{jk})(4W_2(r_{ij})-Y(r_{ij})(1+m_\pi r_{ij}))\right.\\
        &\left.\times[(S_{jk}+\boldsigma_j\cdot\boldsigma_k),(S_{ij}+\boldsigma_i\cdot\boldsigma_j)](\tau_k^3\tau_i^+\tau_j^+-\tau_j^3\tau_i^+\tau_k^+)\right.
        \\
       &\left.+W_1(r_{jk})(4W_2(r_{ij})-Y(r_{ij})(1+m_\pi r_{ij}))\right.\\
       &\left.\times[\boldsigma_j\cdot\boldsigma_k,(S_{ij}+\boldsigma_i\cdot\boldsigma_j)](\tau_k^3\tau_i^+\tau_j^+-\tau_j^3\tau_i^+\tau_k^+)\right],
    \end{split}
\end{align}
\begin{align}
    \begin{split}
        \tilde{V}_{(3h-3k)c_D}=&\frac{g_A c_D}{2\Lambda_\chi F_\pi^2}\sum_{i\neq j\neq k}-\frac{m_\pi}{48\pi}\delta^{(3)}(r_{jk})\left[\phantom{\frac{1}{1}}\right.
        \\
        &\left.(4W_2(r_{ij})-Y(r_{ij})(1+m_\pi r_{ij}))\{(S_{ij}+\boldsigma_i\cdot\boldsigma_j),\boldsigma_j\cdot\boldsigma_k\}\tau_i^+\tau_k^+\right.\\&\left.-\left(4W_2(r_{ij})+Y(r_{ij})-\frac{6}{m_\pi r_{ij}}\right)\{\boldsigma_i\cdot\boldsigma_j,\boldsigma_j\cdot\boldsigma_k\}\tau_i^+\tau_k^+\right].
    \end{split}
\end{align}

The Fourier transform of potential $V_{(3b)}$ is nontrivial due to the three propagators with different momenta. We consider only the $J=0$ component of this potential, as detailed in Appendix \ref{Aprojection}, for our VMC matrix elements. We provide the reduced form of the integral without approximation below, which can in principle be integrated numerically after applying the derivatives for a complete calculation.  After removing the angular dependence of the denominator with a Feynman parameterization, the Fourier transform can be reduced to 

\begin{align}
    \tilde{V}_{(3b)}&=-\frac{g_A^2(1+\kappa_1)}{4m_NF_\pi^2}\sum_{i\neq j\neq k}\boldsymbol{\sigma}_j\cdot(\nabla_{r_{jk}}-\nabla_{r_{ij}})\boldsigma_k\cdot\nabla_{r_{jk}}\boldsigma_i\cdot(\nabla_{r_{ij}}\times\nabla_{r_{jk}})I(\boldsymbol{r}_{ij},\boldsymbol{r}_{jk})\tau_i^+(\boldsymbol{\tau}_k\times\boldsymbol{\tau}_j)^+
\end{align}
where
\begin{align}
    I(\boldsymbol{r}_{ij},\boldsymbol{r}_{jk})=\int_0^1 dx\int_0^\infty dq\;\frac{1}{2\pi^2\abs{\boldsymbol{r}_{jk}+x\boldsymbol{r}_{ij}}}\frac{e^{-r_{ij}\sqrt{q^2(x-x^2)+xm_\pi^2}}}{8\pi\sqrt{q^2(x-x^2)+xm_\pi^2}}\frac{q\sin(q\abs{\boldsymbol{r}_{jk}+x\boldsymbol{r}_{ij}})}{q^2+m_\pi^2}.
\end{align}

The Fourier transform of potential $V_{(4b)g_\nu^{\mathrm{NN}}}$ does not converge, but can be evaluated exploiting the integral form of $\arctan(x)$, namely $\arctan(x)=x\int_0^1 du\;(1+u^2x^2)^{-1}$. By interchanging the order of the momentum integral and integral over $u$, one arrives at 

\begin{align}
    \Tilde{V}_{(4b)g_\nu^{\mathrm{NN}}}&=-\frac{3m_\pi g_A^2g_\nu^{\mathrm{NN}}}{8\pi F_\pi^4}\sum_{i\neq j\neq k}\delta^{(3)}(r_{ik})\left(\delta^{(3)}(r_{jk})-\frac{m_\pi^3}{2\pi}\frac{e^{-2m_\pi r_{jk}}}{2m_\pi^4 r_{jk}^4}(1+m_\pi r_{jk})^2\right)\tau_i^+\tau_k^+
\end{align}

\noindent This has the same form as the Fourier transforms provided in Ref. \cite{Cirigliano:2024ocg} where the authors instead regulated the integral with a small imaginary component of $r$.

\section{Three-body spin operator identities}\label{Sec:B}

The spin component of the three-body operators can be decomposed into two-body operators within (anti)commutators. The identities needed for the $0\nu\beta\beta$ potentials are listed below and can be found in Ref. \cite{Carlson:1982}.

\begin{align}
    \boldsigma_2\cdot\boldsigma_3&=\frac{1}{2}\{\boldsigma_1\cdot\boldsigma_2,\boldsigma_1\cdot\boldsigma_3\}
    \\
    (\boldsigma_2\cdot \hat{\boldsymbol{r}}_{12})(\boldsigma_3\cdot\hat{\boldsymbol{r}}_{12})&=\frac{1}{6}\{(S_{12}+\boldsigma_1\cdot\boldsigma_2),\boldsigma_1\cdot\boldsigma_3\}
    \\
    (\boldsigma_2\cdot\hat{\boldsymbol{r}}_{12})(\boldsigma_3\cdot\hat{\boldsymbol{r}}_{13})(\hat{\boldsymbol{r}}_{12}\cdot\hat{\boldsymbol{r}}_{13})&=\frac{1}{18}\{(S_{12}+\boldsigma_1\cdot\boldsigma_2),(S_{13}+\boldsigma_1\cdot\boldsigma_3)\}
    \\
    \boldsigma_1\cdot\boldsigma_2\times\boldsigma_3&=\frac{1}{2i}[\boldsigma_1\cdot\boldsigma_2,\boldsigma_1\cdot\boldsigma_3]
    \\
    (\boldsigma_2\cdot\hat{\boldsymbol{r}}_{12})(\boldsigma_3\cdot\boldsigma_1\times\hat{\boldsymbol{r}}_{12})&=\frac{1}{6i}[(S_{12}+\boldsigma_1\cdot\boldsigma_2),\boldsigma_1\cdot\boldsigma_3]
    \\
    (\boldsigma_2\cdot\hat{\boldsymbol{r}}_{12})(\boldsigma_3\cdot\hat{\boldsymbol{r}}_{13})(\boldsigma_1\cdot\hat{\boldsymbol{r}}_{12}\times\hat{\boldsymbol{r}}_{13})&=\frac{1}{18i}[(S_{12}+\boldsigma_1\cdot\boldsigma_2),(S_{13}+\boldsigma_1\cdot\boldsigma_3)]
\end{align}

\section{Angular momentum projections}\label{Aprojection}

The potential $V_{(3b)}$ is nontrivial to Fourier transform due to the three propagators with different momenta. We can perform a projection of this operator by decomposing the numerator,

\begin{equation}
    q_j^a \, q_k^b \varepsilon^{c d e} q_i^d q_j^e,
\end{equation}
into spherical tensors.
The product of three vectors can be written as
$\mathbf{3} \otimes \mathbf{3} \otimes  \mathbf{3}  = \mathbf{7} \oplus \mathbf{5} \oplus \mathbf{5} \oplus \mathbf{3} \oplus \mathbf{3} \oplus \mathbf{3} \oplus \mathbf{1}$. 

Rewriting the three vectors in the form
\begin{equation}
    A^{a} B^b C^c = \varepsilon^{a b c} O_0 + 
 + \delta^{a b} O^c_{1, 1}
   + \delta^{b c} O^a_{1, 2} + 
   \delta^{a c} O^b_{1, 3} + \ldots
\end{equation}
the $J=0$ component can be selected by contracting with $\varepsilon^{a b c}$.
We then get
\begin{eqnarray}
    \varepsilon^{a b c} \varepsilon^{d e c} q_j^a q_k^b q_i^d q_j^e = {\bf q}_j \cdot {\bf q}_i \, {\bf q}_k \cdot {\bf q}_j - {\bf q}^2_j \, {\bf q}_k \cdot {\bf q}_i,
\end{eqnarray}
which, using ${\bf q_k} = - {\bf q_i} - {\bf q}_j$ becomes
\begin{eqnarray}
    \varepsilon^{a b c} \varepsilon^{d e c} q_j^a q_k^b q_i^d q_j^e = - {\bf q}_j \cdot {\bf q}_i \,   {\bf q}_i \cdot {\bf q}_j  + {\bf q}^2_j \, {\bf q}^2_i.
\end{eqnarray}
Next we write ${\bf q}_j \cdot {\bf q}_i$ as
\begin{equation}
    {\bf q}_j \cdot {\bf q}_i  = 
\frac{1}{2} ({\bf q}^2_k + m_\pi^2 - {\bf q}^2_j - m_\pi^2 - {\bf q}_i^2)
\end{equation}
and restore the propagators 
\begin{eqnarray}
   & \frac{1}{{\bf q}_i^2}
  \frac{1}{{\bf q}_j^2 + m_\pi^2}
  \frac{1}{{\bf q}_k^2 + m_\pi^2}
   \left( 
   {\bf q}_j \cdot {\bf q}_i 
\frac{1}{2} ({\bf q}^2_k + m_\pi^2 - {\bf q}^2_j - m_\pi^2 - {\bf q}_i^2)
   -  {\bf q}^2_j \, {\bf q}^2_i
   \right) \nonumber \\
   &=
\frac{1}{2}     \frac{  {\bf q}_j \cdot {\bf q}_i}{{\bf q}_i^2}
  \frac{1}{{\bf q}_j^2 + m_\pi^2}
 -  \frac{1}{2}     \frac{  {\bf q}_j \cdot {\bf q}_i}{{\bf q}_i^2}
  \frac{1}{{\bf q}_k^2 + m_\pi^2}
- \frac{1}{2}  \frac{  {\bf q}_j \cdot {\bf q}_i}{{\bf q}_k^2 + m_\pi^2}
  \frac{1}{{\bf q}_j^2 + m_\pi^2}
- \frac{{\bf q}_j^2}{{\bf q}_k^2 + m_\pi^2}
  \frac{1}{{\bf q}_j^2 + m_\pi^2}.
\end{eqnarray}
Each term will now cancel one of the three propagators, allowing the resuling expression to be Fourier transformed. We can also try to isolate the $J=1$ piece, writing
\begin{equation}
    q_j^a q_k^b v^c  = \frac{1}{3} \left( 
\delta^{a b} v^c {\bf q}_j \cdot {\bf q}_k 
    +\delta^{a c} q^a_k\,  {\bf q}_j \cdot {\bf v} + \delta^{b c} q^a_j\, {\bf q}_k \cdot {\bf v}  \right) + \ldots
\end{equation}
with ${\bf v} = {\bf q}_i \times {\bf q}_j$.
Now, ${\bf v} \cdot {\bf q}_{i,j,k} = 0$, so that there is only one vector in the representation.
This leads to 
\begin{eqnarray}
    {\bf q}_j \cdot {\bf q}_k  =
\frac{1}{2} \left[{\bf q}^2_i - [{\bf q}_j^2 + m_\pi^2] - [ {\bf q}_k^2 + m_\pi^2] + 2 m_\pi^2\right].
\end{eqnarray}
However, due to the last term ($2m^2_\pi$), this does not lead immediately to something that can be transformed. The coordinate-space expression for this $J=0$ projection then becomes


\begin{align}
    \begin{split}\tilde{V}_{(3b)}^{(J=0)}&=\frac{g_A^2(1+\kappa_1)}{24m_NF_\pi^2}\sum_{i\neq j\neq k}\left\{\frac{1}{32\pi^2}\left(\nabla_{r_{ij}}^2-\nabla_{r_{ij}}\cdot\nabla_{r_{jk}}\right)\left[\left(\frac{1}{\abs{\boldsymbol{r}_{ij}+\boldsymbol{r}_{jk}}}-\frac{1}{r_{ij}}\right)m_\pi Y(r_{jk})\right.\right.
    \\
    &\left.\left.\phantom{\frac{1}{1}}-m_\pi^2Y(r_{ij})Y(\abs{\boldsymbol{r}_{ij}+\boldsymbol{r}_{jk}})\right]-\frac{m_\pi}{4\pi}Y(\abs{\boldsymbol{r}_{ij}+\boldsymbol{r}_{jk}})\left(\delta^{(3)}(r_{ij})\right.\right.
    \\
    &\left.\left.\phantom{\frac{1}{1}}-\frac{m_\pi^3}{4\pi}Y(r_{ij})\right)\right\}[\boldsigma_i\cdot\boldsigma_j,\boldsigma_i\cdot\boldsigma_k]\tau_i^+(\boldsymbol{\tau}_k\times\boldsymbol{\tau}_j)^+
    \end{split}
\end{align}

\noindent where $\abs{\boldsymbol{r}_{ij}+\boldsymbol{r}_{jk}}=r_{ik}$.

\section{$\Delta$ contribution to $c_D$ potentials}\label{sec.D}

There are additional three-nucleon contributions containing a short-ranged contact interaction, which are produced by replacing the $NN\rightarrow NN$ vertex with $NN\rightarrow N\Delta$ in the $c_D$ diagrams. This vertex is proportional to the not well constrained LEC $D_T$, as given in the interaction below \cite{vanKolck:1994,Krebs:2007rh}. 

\begin{align}
    \mathcal{L} = D_T\bar{T}_i^\mu N \bar{N} S_\mu \tau^i N + \mathrm{h.c.}
\end{align}

\begin{figure}[t]
    \centering
\begin{minipage}{.22\textwidth}
    \begin{center}
    \scalebox{.75}{
    \begin{tikzpicture}[line width=1.2 pt]
    \begin{feynman}
    \vertex (a);
    \vertex[right=1 of a] (b);
    \vertex[right=2 of b] (c);
    \vertex[right=1 of c] (d);

    \vertex[below=1 of a] (e);
    \vertex[right=1 of e] (f);
    \vertex[right=2 of f] (g);
    \vertex[right=1 of g] (h);

    \vertex[below=2 of e] (i);
    \vertex[right=1 of i] (j);
    \vertex[right=2 of j] (k);
    \vertex[right=1 of k] (l);

    \vertex[below=1.2 of b] (m);

    \vertex[below=0.4 of c] (n);
    \vertex[below=1.6 of c] (o);

    \vertex[right=1 of b] (p);
    \vertex[below=2 of p] (q);
    \vertex[below=2 of q] (r);

    \vertex[right=1 of n] (s);
    \vertex[right=1 of o] (t);

    \vertex[below=.45 of p] (u);
    \vertex[below=1.15 of u] (v);

    \vertex[above=.9 of t] (tt);
    \vertex[right=2 of u] (ss);
    \vertex[left=1 of t, dot] (tt);
    \vertex[above=.04 of tt, dot] (ttt) {};
    \vertex[below=2 of p] (pp);
    \vertex[above=1.5 of ttt, dot] (t4) {};
    \vertex[right=1.1 of ttt] (electron1) {};
    \vertex[right=1.1 of t4] (electron2) {};
    \vertex[below=.4 of electron2] (electron3) {};
    
    \diagram*{
        (a) -- [] (b);
        (b) -- [] (c);
        (c) -- [] (d);

        (e) -- [] (l);
        (i) -- [] (pp);
        (pp) -- [double] (ttt);
        (ttt) -- [] (h);
        (ttt) -- [anti majorana] (t4);
        (ttt) -- [fermion] (electron1);
        (t4) -- [fermion] (electron3);
        };

    \vertex[below=0.87 of p] (dot1) {};
    \vertex[right=-0.075 of p] (dot) {};
    \vertex[right=-0.075 of q] (dot) {};
    \vertex[right=-0.075 of u] (dot) {};
    \vertex[right=-0.075 of v] (dot) {};

    \vertex[below=1 of dot1, dot, black] (dot) {};
    \vertex[above=.04 of tt, dot] (t10101) {};
    \vertex[below=.75 of t4, square dot] (dot1) {};

    \end{feynman}
\end{tikzpicture}}
\end{center}
\end{minipage}\hspace{10pt}
\begin{minipage}{.22\textwidth}
    \begin{center}
    \scalebox{.75}{
    \begin{tikzpicture}[line width=1.2 pt]
    \begin{feynman}
    \vertex (a);
    \vertex[right=1 of a] (b);
    \vertex[right=2 of b] (c);
    \vertex[right=1 of c] (d);

    \vertex[below=1 of a] (e);
    \vertex[right=1 of e] (f);
    \vertex[right=2 of f] (g);
    \vertex[right=1 of g] (h);

    \vertex[below=2 of e] (i);
    \vertex[right=1 of i] (j);
    \vertex[right=2 of j] (k);
    \vertex[right=1 of k] (l);

    \vertex[below=1.2 of b] (m);

    \vertex[below=0.4 of c] (n);
    \vertex[below=1.6 of c] (o);

    \vertex[right=1 of b] (p);
    \vertex[below=2 of p] (q);
    \vertex[below=2 of q] (r);

    \vertex[right=1 of n] (s);
    \vertex[right=1 of o] (t);

    \vertex[below=.45 of p] (u);
    \vertex[below=1.15 of u] (v);

    \vertex[above=.9 of t] (tt);
    \vertex[right=2 of u] (ss);
    \vertex[left=1 of t, dot] (tt);
    \vertex[above=.04 of tt, dot] (ttt) {};
    \vertex[below=2 of p] (pp);
    \vertex[above=1.5 of ttt, dot] (t4) {};
    \vertex[right=1.1 of ttt] (electron1) {};
    \vertex[right=1.1 of t4] (electron2) {};
    \vertex[below=.4 of electron2] (electron3) {};
    \vertex[below=1.2 of t4] (t5) {};
    \vertex[right=1.1 of t5] (electron4) {};
    \vertex[below=.2 of electron3] (electron5) {};
    \vertex[left=.2 of t5] (t6) {};
    
    \diagram*{
        (a) -- [] (b);
        (b) -- [] (c);
        (c) -- [] (d);

        (e) -- [] (l);
        (i) -- [] (pp);
        (pp) -- [double] (ttt);
        (ttt) -- [] (h);
        (t5) -- [anti majorana] (t4);
        (t6) -- [fermion] (electron5);
        (t4) -- [fermion] (electron3);
        (t4) -- [dashed] (ttt);
        };

    \vertex[below=0.87 of p] (dot1) {};
    \vertex[right=-0.075 of p] (dot) {};
    \vertex[right=-0.075 of q] (dot) {};
    \vertex[right=-0.075 of u] (dot) {};
    \vertex[right=-0.075 of v] (dot) {};

    \vertex[below=1 of dot1, dot, black] (dot) {};
    \vertex[above=.04 of tt, dot] (t10101) {};
    \vertex[below=1.1 of t4, dot, black] (dot4) {};
    \vertex[below=.59 of t4, square dot] (dot1) {};

    \end{feynman}
\end{tikzpicture}}
\end{center}
\end{minipage}\hspace{10pt}
\begin{minipage}{.22\textwidth}
    \begin{center}
    \scalebox{.75}{
    \begin{tikzpicture}[line width=1.2 pt]
    \begin{feynman}
    \vertex (a);
    \vertex[right=1 of a] (b);
    \vertex[right=2 of b] (c);
    \vertex[right=1 of c] (d);

    \vertex[below=1 of a] (e);
    \vertex[right=1 of e] (f);
    \vertex[right=2 of f] (g);
    \vertex[right=1 of g] (h);

    \vertex[below=2 of e] (i);
    \vertex[right=1 of i] (j);
    \vertex[right=2 of j] (k);
    \vertex[right=1 of k] (l);

    \vertex[below=1.2 of b] (m);

    \vertex[below=0.4 of c] (n);
    \vertex[below=1.6 of c] (o);

    \vertex[right=1 of b] (p);
    \vertex[below=2 of p] (q);
    \vertex[below=2 of q] (r);

    \vertex[right=1 of n] (s);
    \vertex[right=1 of o] (t);

    \vertex[below=.45 of p] (u);
    \vertex[below=1.15 of u] (v);

    \vertex[above=.9 of t] (tt);
    \vertex[right=2 of u] (ss);
    \vertex[left=1 of t, dot] (tt);
    \vertex[above=.04 of tt, dot] (ttt) {};
    \vertex[below=2 of p] (pp);
    \vertex[above=1.5 of ttt, dot] (t4) {};
    \vertex[right=1.1 of ttt] (electron1) {};
    \vertex[right=1.1 of t4] (electron2) {};
    \vertex[below=.4 of electron2] (electron3) {};
    \vertex[below=.3 of t4] (t5) {};
    \vertex[below=.4 of t4] (electron4);
    \vertex[right=1 of electron4] (electron5) {};
    
    \diagram*{
        (a) -- [] (b);
        (b) -- [] (c);
        (c) -- [] (d);

        (e) -- [] (l);
        (i) -- [] (pp);
        (pp) -- [double] (ttt);
        (ttt) -- [] (h);
        (ttt) -- [dashed] (t4);
        (t5) -- [anti majorana] (ttt);
        (ttt) -- [fermion] (electron1);
        (electron4) -- [fermion] (electron5);
        };

    \vertex[below=0.87 of p] (dot1) {};
    \vertex[right=-0.075 of p] (dot) {};
    \vertex[right=-0.075 of q] (dot) {};
    \vertex[right=-0.075 of u] (dot) {};
    \vertex[right=-0.075 of v] (dot) {};

    \vertex[below=1 of dot1, dot, black] (dot) {};
    \vertex[above=.04 of tt, dot] (t10101) {};
    \vertex[below=.91 of t4, square dot] (dot1) {};
    \vertex[above=1.1 of ttt,dot,black] (dot2) {};

    \end{feynman}
\end{tikzpicture}}
\end{center}
\end{minipage}\hspace{10pt}
\begin{minipage}{.22\textwidth}
    \begin{center}
    \scalebox{.75}{
    \begin{tikzpicture}[line width=1.2 pt]
    \begin{feynman}
    \vertex (a);
    \vertex[right=1 of a] (b);
    \vertex[right=2 of b] (c);
    \vertex[right=1 of c] (d);

    \vertex[below=1 of a] (e);
    \vertex[right=1 of e] (f);
    \vertex[right=2 of f] (g);
    \vertex[right=1 of g] (h);

    \vertex[below=2 of e] (i);
    \vertex[right=1 of i] (j);
    \vertex[right=2 of j] (k);
    \vertex[right=1 of k] (l);

    \vertex[below=1.2 of b] (m);

    \vertex[below=0.4 of c] (n);
    \vertex[below=1.6 of c] (o);

    \vertex[right=1 of b] (p);
    \vertex[below=2 of p] (q);
    \vertex[below=2 of q] (r);

    \vertex[right=1 of n] (s);
    \vertex[right=1 of o] (t);

    \vertex[below=.45 of p] (u);
    \vertex[below=1.15 of u] (v);

    \vertex[above=.9 of t] (tt);
    \vertex[right=2 of u] (ss);
    \vertex[left=1 of t, dot] (tt);
    \vertex[above=.04 of tt, dot] (ttt) {};
    \vertex[below=2 of p] (pp);
    \vertex[above=1.5 of ttt, dot] (t4) {};
    \vertex[right=1.1 of ttt] (electron1) {};
    \vertex[right=1.1 of t4] (electron2) {};
    \vertex[below=.4 of electron2] (electron3) {};
    \vertex[below=.3 of t4] (t5) {};
    \vertex[below=.4 of t4] (electron4);
    \vertex[right=1 of electron4] (electron5) {};
    \vertex[below=.32 of t4] (m1);
    \vertex[above=.32 of ttt] (m2);
    \vertex[below=.08 of m1] (electron5);
    \vertex[right=1 of electron5] (electron6);
    \vertex[below=.79 of m1] (electron7);
    \vertex[below=.29 of electron6] (electron8);
    
    \diagram*{
        (a) -- [] (b);
        (b) -- [] (c);
        (c) -- [] (d);

        (e) -- [] (l);
        (i) -- [] (pp);
        (pp) -- [double] (ttt);
        (ttt) -- [] (h);
        (t4) -- [dashed] (m1);
        (ttt) -- [dashed] (m2);
        (m1) -- [anti majorana] (m2);
        (electron5) -- [fermion] (electron6);
        (electron7) -- [fermion] (electron8);
        };

    \vertex[below=0.87 of p] (dot1) {};
    \vertex[right=-0.075 of p] (dot) {};
    \vertex[right=-0.075 of q] (dot) {};
    \vertex[right=-0.075 of u] (dot) {};
    \vertex[right=-0.075 of v] (dot) {};

    \vertex[below=1 of dot1, dot, black] (dot) {};
    \vertex[above=.04 of tt, dot] (t10101) {};
    \vertex[below=.725 of t4, square dot] (dot1) {};
    \vertex[below=.0 of m1,dot,black] {};
    \vertex[above=.01 of m2,dot,black] {};

    \end{feynman}
\end{tikzpicture}}
\end{center}
\end{minipage}\hspace{10pt}
\captionsetup{labelformat=default}
    \caption{Diagrams contributing to the neutrino potential which are proportional to $D_T$.}
    \label{fig:DT}
\captionsetup{labelformat=empty}
\end{figure}
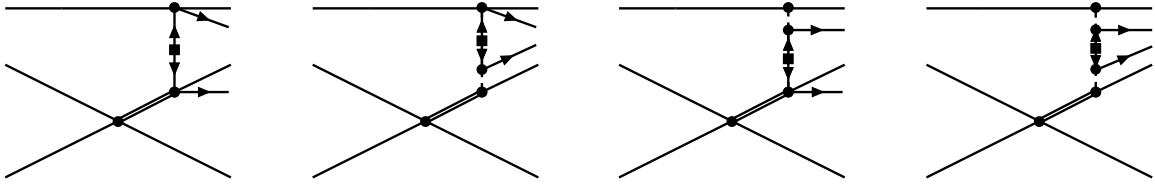

The operator structure given in Eq. \ref{eq:DT} is similar to the $c_D$ potential (\ref{eq:cD}) after taking into account all permutations of the diagrams shown in Fig. \ref{fig:DT}.

\begin{align}
\begin{split}
    V_{D_T}&=-\frac{4}{9}\frac{h_Ag_AD_T}{\Delta}\sum_{i\neq j\neq k} \frac{1}{\boldsymbol{q}_i^2}\left[\left(-2\boldsigma_i\cdot\boldsigma_k\tau_i^+\tau_k^++\frac{1}{2}\boldsymbol{\sigma}_i\cdot\boldsigma_j\times\boldsigma_k(\boldsymbol{\tau}_j\times\boldsymbol{\tau}_k)^+\tau_i^+\right)\right.
    \\
    &-\left.\frac{\boldsymbol{q}_i^2+2m_\pi^2}{\boldsymbol{q}_i^2+m_\pi^2}\left(-2\boldsigma_k\cdot\boldsymbol{q}_i\tau_i^+\tau_k^++\frac{1}{2}\boldsymbol{q}_i\cdot\boldsigma_j\times\boldsigma_k(\boldsymbol{\tau}_j\times\boldsymbol{\tau}_k)^+\tau_i^+\right)\right]=0
\end{split}
\label{eq:DT}
\end{align}

However, in the absence of a cutoff, the second terms in each parentheses in the expression above can be Fierz transformed to exactly cancel the first terms of the parentheses. This is analogous to situation in the three-nucleon forces whereby the three-nucleon diagram containing $c_D$ can be augmented with a $NN\rightarrow N\Delta$ vertex but results in a similar vanishing operator structure.

\bibliography{bibliography}

\end{document}